\documentclass[numberedappendix,apj,twocolumn]{emulateapj}
\usepackage[colorlinks,urlcolor=blue,citecolor=blue,linkcolor=blue]{hyperref}

\newcommand{\liwhite}{Li \& White}
\newcommand{\smf}{SMF}
\newcommand{\etal}{et~al.}

\newcommand{\isedfit}{{\tt iSEDfit}}
\newcommand{\kcorrect}{{\tt K-correct}}
\newcommand{\sex}{{\tt SExtractor}}

\newcommand{\vmax}{\ensuremath{V_{\rm max}}}
\newcommand{\oneovervmax}{\ensuremath{1/V_{\rm max}}}

\newcommand{\mass}{\ensuremath{\mathcal{M}}}
\newcommand{\masslim}{\ensuremath{\mathcal{M_{\rm lim}}}}
\newcommand{\mstar}{\ensuremath{\mathcal{M}^{\ast}}}
\newcommand{\msun}{\ensuremath{\mathcal{M}_{\sun}}}

\newcommand{\sfr}{\ensuremath{\psi}}

\newcommand{\sfrunits}{\ensuremath{\mathcal{M}_{\sun}~\textrm{yr}^{-1}}}

\newcommand{\tburst}{\ensuremath{t_{b}}}
\newcommand{\dtburst}{\ensuremath{\Delta t_{b}}}
\newcommand{\fburst}{\ensuremath{\mathcal{F}_{b}}}
\newcommand{\aburst}{\ensuremath{\mathcal{A}_{b}}}

\newcommand{\rhosfqq}{\ensuremath{\dot{\rho}_{{\rm SF}\rightarrow{\rm Q}}}}
\newcommand{\zmean}{\ensuremath{\langle z \rangle}}

\shorttitle{Evolution of the Stellar Mass Function}  
\shortauthors{Moustakas et~al.}
\journalinfo{The Astrophysical Journal, 767:50 (34pp), 2013 April 10
  \hspace{1.9in}\doi{10.1088/0004-637X/767/1/50}}
\slugcomment{ApJ, 767, 50}

\begin{document}

\title{PRIMUS: Constraints on Star Formation Quenching and Galaxy
  Merging, and the Evolution of the Stellar Mass Function From
  $\lowercase{z}=0-1$}

\author{
John Moustakas\altaffilmark{1}, 
Alison L. Coil\altaffilmark{2,11}, 
James Aird\altaffilmark{3},
Michael R. Blanton\altaffilmark{4}, 
Richard J. Cool\altaffilmark{5}, 
Daniel J. Eisenstein\altaffilmark{6},
Alexander J. Mendez\altaffilmark{2}, 
Kenneth C. Wong\altaffilmark{7}, 
Guangtun Zhu\altaffilmark{8}, and 
St\'{e}phane Arnouts\altaffilmark{9,10}
}

\altaffiltext{1}{Department of Physics and Astronomy, Siena College,
  515 Loudon Road, Loudonville, NY 12211; \email{jmoustakas@siena.edu}}
\altaffiltext{2}{Center for Astrophysics and Space Sciences,
  Department of Physics, University of California, 9500 Gilman Dr., La
  Jolla, CA 92093}
\altaffiltext{3}{Department of Physics, Durham University, Durham DH1
  3LE, UK}
\altaffiltext{4}{Center for Cosmology and Particle Physics, Department
  of Physics, New York University, 4 Washington Place, New York, NY
  10003} 
\altaffiltext{5}{MMT Observatory, University of Arizona, 1540 E Second
  Street, Tucson AZ 85721} 
\altaffiltext{6}{Harvard-Smithsonian Center for Astrophysics, 60
  Garden Street, Cambridge, MA 02138}
\altaffiltext{7}{Steward Observatory, University of Arizona, 933 North
  Cherry Avenue, Tucson, AZ 85721}
\altaffiltext{8}{Department of Physics and Astronomy, The Johns
  Hopkins University, 3400 North Charles Street, Baltimore, MD 21218} 
\altaffiltext{9}{Canada-France-Hawaii Telescope Corporation, 65-1238 
  Mamalahoa Hwy, Kamuela, HI 96743}
\altaffiltext{10}{Aix Marseille Universit\'{e}, CNRS, LAM
  (Laboratoire d'Astrophysique de Marseille) UMR 7326, 13388,
  Marseille, France}
\altaffiltext{11}{Alfred P. Sloan Foundation Fellow}
\setcounter{footnote}{11}

\begin{abstract}
We measure the evolution of the stellar mass function (SMF) from
$z=0-1$ using multi-wavelength imaging and spectroscopic redshifts
from the PRism MUlti-object Survey (PRIMUS) and the Sloan Digital Sky
Survey (SDSS).  From PRIMUS we construct an $i<23$ flux-limited sample
of $\sim40,000$ galaxies at $z=0.2-1.0$ over five fields totaling
$\approx5.5$~deg$^{2}$, and from the SDSS we select $\sim170,000$
galaxies at $z=0.01-0.2$ that we analyze consistently with respect to
PRIMUS to minimize systematic errors in our evolutionary measurements.
We find that the \smf{} of all galaxies evolves relatively little
since $z=1$, although we do find evidence for mass assembly
downsizing; we measure a $\approx30\%$ increase in the number density
of $\sim10^{10}$~\msun{} galaxies since $z\approx0.6$, and a
$\lesssim10\%$ change in the number density of all
$\gtrsim10^{11}$~\msun{} galaxies since $z\approx1$.  Dividing the
sample into star-forming and quiescent using an evolving cut in
specific star-formation rate, we find that the number density of
$\sim10^{10}$~\msun{} star-forming galaxies stays relatively constant
since $z\approx0.6$, whereas the space-density of
$\gtrsim10^{11}$~\msun{} star-forming galaxies decreases by
$\approx50\%$ between $z\approx1$ and $z\approx0$.  Meanwhile, the
number density of $\sim10^{10}$~\msun{} quiescent galaxies increases
steeply towards low redshift, by a factor of $\sim2-3$ since
$z\approx0.6$, while the number of massive quiescent galaxies remains
approximately constant since $z\approx1$.  These results suggest that
the rate at which star-forming galaxies are quenched increases with
decreasing stellar mass, but that the bulk of the stellar mass buildup
within the quiescent population occurs around $\sim10^{10.8}$~\msun.
In addition, we conclude that mergers do not appear to be a dominant
channel for the stellar mass buildup of galaxies at $z<1$, even among
massive ($\gtrsim10^{11}$~\msun) quiescent galaxies.
\end{abstract}


\section{Introduction}\label{sec:intro}

Surveys of the galaxy population in the nearby Universe such as the
Sloan Digital Sky Survey \citep[SDSS;][]{york00a} have found that the
distribution of many galaxy properties, including color, morphology,
and star formation rate (SFR), are bimodal, reflecting the existence
of two broad types of galaxies: blue, star-forming disk galaxies, and
red, quiescent (i.e., non star-forming) spheroidal or elliptical
galaxies \citep[e.g.,][]{kauffmann03b, blanton03b, baldry04a,
  wyder07a}.  At low redshift, quiescent galaxies tend to be luminous
and massive, and are prevalent in dense environments such as groups
and clusters, whereas star-forming galaxies typically have lower
stellar masses and are more commonly found in the field \citep[and
  references therein]{blanton09a}.  These broad empirical trends have
been shown to persist at least to $z\sim3$ \citep[e.g.,][]{bell04a,
  cooper07a, cassata08a, brammer09a, whitaker11a}.  Understanding how
these two populations come into existence and evolve with cosmic time,
therefore, is a fundamental outstanding problem in observational
cosmology.

Galaxy bimodality is likely a consequence of star formation in some
galaxies being \emph{quenched}---shut off---relatively quickly.  A
wide variety of quenching mechanisms have been proposed to match the
observed distributions of galaxy properties, including: major-merger
induced feedback from star formation and supermassive black holes
\citep{springel05a, di-matteo05a}; processes that prevent galaxies
from replenishing their cold-gas supply such as virial shock heating
\citep{keres05a, dekel06a}, often in concert with so-called radio-mode
feedback from an accreting active galactic nucleus
\citep[AGN;][]{croton06a, cattaneo06a, gabor11a}; and internal,
secular quenching due to disk or bar instabilities \citep{cole00a,
  martig09a}.  Late-type galaxies whose host halos are accreted by
larger dark-matter halos (i.e., by groups and clusters) are also
susceptible to \emph{environmental} quenching, such as ram-pressure
stripping \citep{gunn72a}, strangulation or starvation
\citep{larson80a, balogh00a}, and gravitational harassment
(\citealt{moore98a}; see \citealt{boselli06a} for a review).

By incorporating these quenching mechanisms into a coherent
cosmological framework, modern theoretical models have become
reasonably successful at reproducing a wide range of galaxy
properties, including galaxy bimodality \citep[e.g.,][]{hopkins08b,
  hopkins08a, somerville08a, bower12a}.  However, many discrepancies
between observations and predictions persist
\citep[e.g.,][]{fontana06a, fontanot09a, keres09b, dave11b, dave11a,
  lu12a, weinmann12a}.  The exact cause of these problems is hard to
determine because of the complexities of modeling the small-scale
physics of star formation, feedback, and black-hole accretion.
Consequently, empirical constraints on the relative fraction of
quiescent and star-forming galaxies as a function stellar mass,
redshift, environment, AGN content, and dark-matter halo mass are
critical for determining galaxy quenching, and the evolution of galaxy
bimodality \citep[e.g.,][]{weinmann06a, bell07a, kimm09a, peng10a,
  wetzel12a, woo12a, aird12a, knobel12a}.

In this paper, we focus on one key aspect of this problem by measuring
the stellar mass functions (\smf s) of quiescent and star-forming
galaxies from $z=0-1$, spanning the last $\sim8$~billion years of
cosmic time.  The \smf{} measures the comoving space density of
galaxies of a given stellar mass, making it a powerful observational
tracer of galaxy growth by {\em in situ} star formation, mergers, and
galaxy transformations due to star formation quenching
\citep[e.g.,][]{drory08a, peng10a}.  Measurements of the \smf{} are
also important for connecting the physics of galaxy formation to the
hierarchical assembly of dark matter halos, and large-scale structure
\citep[e.g.,][]{conroy09b, cattaneo11a, wang12a, leauthaud12a,
  behroozi12a}.

Deep extragalactic surveys in the last decade have begun to
characterize the evolution of galaxy bimodality over a significant
fraction of the age of the Universe.  At the highest redshifts,
$z\gtrsim2$, studies have shown that although quiescent galaxies
exist, they are outnumbered by star-forming galaxies at all stellar
masses \citep[e.g.,][]{whitaker10a, dominguez-sanchez11a}.  This early
epoch is followed by a period of rapid growth in the space density of
massive ($\gtrsim10^{11}$~\msun) quiescent galaxies between $z\sim2$
and $z\sim1$ \citep{arnouts07a, ilbert10a, nicol11a, brammer11a,
  mortlock11a}.  By $z\sim1$, the stellar mass dependence of galaxy
bimodality as observed locally is largely in place: star-forming
galaxies outnumber quiescent galaxies at the low-mass end of the \smf,
while quiescent galaxies dominate the massive galaxy population
\citep[e.g.,][]{bundy06a, borch06a}.  Subsequently, between $z\sim1$
and $z\sim0$, the transformation of star-forming galaxies into
quiescent, passively evolving galaxies continues, leading to an
approximately factor of two increase in the integrated stellar mass
density of quiescent galaxies \citep{bell04a, blanton06a, faber07a}.
The bulk of this stellar mass growth appears to be due to a rapidly
rising population of intermediate-mass ($\sim10^{10}$~\msun) quiescent
galaxies, although the extent to which massive galaxies also grow
through stellar accretion (i.e., mergers) remains controversial
\citep{cimatti06a, scarlata07a, brown07a, rudnick09a, stewart09a,
  ilbert10a, pozzetti10a, robaina10a, maraston12a}.  By the current
epoch, quiescent galaxies vastly outnumber star-forming galaxies above
$\sim3\times10^{10}$~\msun, and account for more than half of the
total stellar mass in the local Universe \citep{bell03b, baldry04a,
  driver06a}.
  
Despite significant progress, however, the detailed evolution of the
\smf s of quiescent and star-forming galaxies since $z\sim1$ remain
relatively uncertain, leaving several unresolved issues.  One
outstanding question is whether massive galaxies assemble their
stellar mass earlier (i.e., at higher redshift) relative to lower-mass
galaxies---that is, do galaxies undergo {\em mass assembly
  downsizing}?\footnote{Also called {\em downsizing in stellar mass}
  in the extensive study of the various manifestations of downsizing
  by \citet{fontanot09a}.}  Mass assembly downsizing poses a
significant challenge for theoretical models, which predict the
late-time assembly of massive galaxies \citep[i.e., {\em mass assembly
    upsizing}; e.g.,][]{de-lucia06a}; however, different observational
studies have reached different conclusions
\citep[e.g.,][]{perez-gonzalez08a, fontanot09a}.  

Another open question is the role of major and minor mergers for the
stellar mass growth of galaxies at $z<1$.  Because the merger rate is
notoriously difficult to measure directly \citep[e.g.,][]{lotz11a},
measurements of the \smf{} as a function of redshift can place
complementary constraints on the merger-driven growth of galaxies
\citep[e.g.,][]{drory08a, pozzetti10a}, in addition to constraining
the buildup of the diffuse stellar component (or intercluster light)
of groups and clusters \citep[e.g.,][]{murante07a, gonzalez07a}.

Finally, it remains poorly understood why the \smf{} of star-forming
galaxies evolves relatively little from $z=0-1$, despite vigorous
ongoing star formation.  In particular, it is not known why the
stellar mass growth by {\em in situ} star formation balances---almost
perfectly---the stellar mass growth of the quiescent galaxy population
due to quenching \citep[see, e.g.,][]{arnouts07a, martin07a, peng10a}.

Answers to these and related questions have remained elusive at
intermediate redshift because a combination of depth, area, and large
sample are needed to characterize the shape of the \smf{} over a large
dynamic range of stellar mass, while simultaneously minimizing the
effects of sample variance.\footnote{{\em Sample variance} refers to
  the variation in the number density of galaxies along a given
  line-of-sight due to large-scale clustering.  The more commonly
  adopted term {\em cosmic variance} should only strictly be used in
  the context of the existence of just one (observable) Universe.}
For example, \citet{moster11a} estimate a $\sim25\%$ uncertainty in
the number density of $\sim10^{11}$~\msun{} galaxies in a $\Delta
z=0.1$ wide redshift bin at $z=0.5$ due to sample variance in the
$\sim2$~deg$^{2}$ Cosmic Evolution Survey (COSMOS) field, which is
among the largest extragalactic deep fields with high-quality
spectroscopic and medium-band photometric redshifts
\citep{scoville07a, ilbert09a}.  Although broadband photometric
redshifts enable the \smf{} to be constructed over larger areas
\citep[e.g.,][]{matsuoka10a}, the redshift precision typically
achieved by these methods, $\sigma_{z}/(1+z)\approx1\%-5\%$, can
significantly bias the inferred shape of the \smf, and its evolution
with redshift.  Finally, previous studies have frequently relied on
published measurements of the local \smf{} (e.g., from the SDSS),
although the amount of evolution inferred can be significantly
affected by systematic differences in how stellar masses are derived
\citep{marchesini09a, bernardi10a}.

We alleviate many of these issues by measuring the evolution of the
\smf{} at intermediate redshift using data from the PRism MUlti-object
Survey \citep[PRIMUS;][]{coil11a, cool13a}.  From PRIMUS we select
$\sim40,000$ galaxies at $z=0.2-1$ with high-quality spectroscopic
redshifts and deep multi-wavelength imaging in the ultraviolet (UV)
from the {\em Galaxy Evolution Explorer} \citep[GALEX;][]{martin05a},
in the mid-infrared from the {\em Spitzer Space Telescope}
\citep{werner04a} Infrared Array Camera \citep[IRAC;][]{fazio04a}, and
in the optical and near-infrared from a variety of ground-based
surveys.  The broad wavelength coverage allows us to estimate stellar
masses and SFRs using detailed spectral energy distribution (SED)
modeling, and to robustly identify quiescent and star-forming galaxies
over the full redshift range.  The total area subtended by this sample
is $\approx5.5$~deg$^{2}$, making it the largest statistically complete
sample of faint galaxies with spectroscopic redshifts ever assembled.
Furthermore, we construct the \smf{} at $z\approx0.1$ using a sample of
$\sim170,000$ SDSS galaxies at $z=0.01-0.2$ with UV, optical, and
near-infrared photometry from \emph{GALEX}, the Two Micron All Sky
Survey \citep[2MASS;][]{skrutskie06a}, and the Wide-field Infrared
Survey Explorer \citep[WISE;][]{wright10a} that we analyze using the
same methodology as PRIMUS to minimize systematic errors in the
evolutionary trends we measure.

Using these data, we measure the evolution in the number and stellar
mass density of quiescent and star-forming galaxies since $z\approx1$,
in order to quantify their stellar mass growth by star formation and
mergers, and to constrain the physical mechanisms responsible for
quenching.  In Section~\ref{sec:obs} we present the \emph{GALEX},
optical, and IRAC photometry of galaxies in the PRIMUS fields that we
use, and we describe how we construct our local SDSS comparison
sample.  We select our parent sample of quiescent and star-forming
galaxies in Section~\ref{sec:sample}, and in
Section~\ref{sec:analysis} we describe the methodology we use to
construct a statistically complete \smf{} for both quiescent and
star-forming galaxies in seven redshift bins from $z=0-1$.  We present
our \smf s and quantify the number and stellar mass density evolution
of each galaxy type in Section~\ref{sec:results}, and quantify the
role of galaxy growth by mergers and star-formation quenching in
Section~\ref{sec:discussion}.  Finally, we summarize our results in
Section~\ref{sec:summary}.

Given the length of the paper, readers interested in our principal
results can skip ahead to Section~\ref{sec:results} (see especially
Figures~\ref{fig:mfqqsf_one} and \ref{fig:numbymass}), and to the
interpretation of our results in Section~\ref{sec:discussion}.

We adopt a concordance cosmology with $\Omega_{\rm m}=0.3$,
$\Omega_{\Lambda}=0.7$, and $h_{70} \equiv H_{0}/(70~{\rm km}~{\rm
  s}^{-1}~{\rm Mpc}^{-1})=1.0$, and the AB magnitude system
\citep{oke83a} throughout.  Unless otherwise indicated, all stellar
masses and SFRs assume a universal \citet{chabrier03a} initial mass
function (IMF) from $0.1-100~\mathcal{M}_{\sun}$.

\section{Observations}\label{sec:obs}

Our analysis of the \smf{} at intermediate redshift combines
multi-wavelength imaging of five distinct extragalactic deep fields
with spectroscopic redshifts from PRIMUS.  In Section~\ref{sec:primus}
we briefly describe PRIMUS and introduce the fields that we analyze.
We describe our analysis of the deep \emph{GALEX}/UV imaging of these
fields in Section~\ref{sec:galex}, and the optical and mid-infrared
photometric catalogs we use in Section~\ref{sec:optphot}.  Finally, in
Section~\ref{sec:sdss} we describe how we construct our $z\approx0.1$
SDSS galaxy sample.

\subsection{PRIMUS}\label{sec:primus}  

PRIMUS is a large faint-galaxy intermediate-redshift survey which
obtained precise ($\sigma_{z} / (1+z) \approx 0.5\%$) spectroscopic
redshifts for a statistically complete sample of $\sim120,000$
galaxies at $z\approx0-1.2$.  The survey targeted galaxies in seven
distinct extragalactic deep fields, totaling $\sim9$~deg$^{2}$, with a
wealth of ancillary multi-wavelength imaging from the X-ray to the
far-infrared.  PRIMUS was conducted using the IMACS spectrograph
\citep{bigelow03a} on the Magellan~I Baade 6.5~m telescope with a
slitmask and low-dispersion prism.  This novel experimental design
yielded low-resolution $(\lambda / \Delta \lambda \sim 40)$ spectra
for $\sim2000$ objects per slitmask, which is a factor of $\sim10$
higher multiplexing rate than traditional spectroscopic redshift
surveys \citep[see also][]{kelson12a}.  The full details of the survey
design, targeting, and data summary are in \citet{coil11a}, while the
details of the data reduction, redshift fitting and precision, and
survey completeness can be found in \citet{cool13a}.

In this paper we restrict our analysis to the fields targeted by
PRIMUS with \emph{GALEX} and \emph{Spitzer}/IRAC imaging.  Three of
these fields are part of the \emph{Spitzer} Wide-area Infrared
Extragalactic Survey
\citep[SWIRE\footnote{http://swire.ipac.caltech.edu/swire/swire.html};][]{lonsdale03a}:
the European Large Area ISO Survey - South~1 field
\citep[ELAIS-S1\footnote{http://dipastro.pd.astro.it/esis};][]{oliver00a};
the Chandra Deep Field South SWIRE field (hereafter, CDFS); and the
XMM Large Scale Structure Survey field \citep[XMM-LSS;][]{pierre04a}.
In detail, the XMM-LSS field in PRIMUS consists of two separate (but
spatially adjacent) subfields: the Subaru/XMM-Newton DEEP Survey field
\citep[XMM-SXDS\footnote{http://www.naoj.org/Science/SubaruProject/SDS};][]{furusawa08a},
and the Canada-France-Hawaii Telescope Legacy Survey (CFHTLS) field
(hereafter,
XMM-CFHTLS\footnote{http://www.cfht.hawaii.edu/Science/CFHLS}).  These
fields were targeted by PRIMUS using different photometric catalogs;
therefore, because of the slightly different selection functions we
treat them separately in our analysis of the \smf.  Finally, we
include in our analysis the well-studied
COSMOS\footnote{http://cosmos.astro.caltech.edu} field
\citep{scoville07a}, giving a total of five distinct fields.

\subsection{Ultraviolet Photometry}\label{sec:galex}


All five fields in our sample were observed in the near-UV (NUV) and
far-UV (FUV) as part of the \emph{GALEX} Deep Imaging Survey
\citep[DIS;][]{martin05a, morrissey05a}.  The mean exposure times of
these observations ranges from $58-78$~ks in our COSMOS field, to
between $15-42$~ks in our ELAIS-S1, CDFS (but one with $90$~ks), and
XMM-SXDS and XMM-CFHTLS (but one with $150$~ks) fields, making these
among the deepest UV observations ever obtained.

Given the $4\farcs2$ and $5\farcs3$ FWHM point-spread function (PSF)
of the \emph{GALEX} telescope in the FUV and NUV, respectively, in
$\gtrsim10$~ks exposures source confusion is a significant problem,
requiring great care when extracting a photometric catalog.
Therefore, to minimize contamination from neighboring sources, we use
the Bayesian photometric code {\sc EMphot}, which is based on the
expectation maximization (EM) algorithm of \citet{guillaume06a}.  {\sc
  EMphot} uses optical positional priors to measures the UV fluxes of
objects in the \emph{GALEX} images by adjusting a model of the \emph{GALEX} PSF.
The prior positions are based on deep, high-resolution optical imaging
of the same field in the bluest available band (see
Section~\ref{sec:optphot}).  {\sc EMphot} has been used successfully in
several previous studies to analyze deep \emph{GALEX} imaging \citep{xu05a,
  zamojski07a, salim09a, hammer10a}, so we refer the interested reader
to those papers for additional details.

\subsection{Optical \& Mid-Infrared Photometry}\label{sec:optphot} 

In this section we describe the ground-based optical and
\emph{Spitzer}/IRAC mid-infrared photometric catalogs that we utilize
in each field.  As most of these catalogs are publicly available, we
defer many of the finer details to the corresponding data release
documentation and papers cited below; additional details can also be
found in \citet{coil11a}.  

Our general strategy for obtaining integrated (total) fluxes in all
photometric bands is to use circular aperture photometry to constrain
the shape of the SED of each galaxy, and then to tie the overall
normalization of the SED to an estimate of the total magnitude in the
detection band (usually $i^{\prime}$ or $R$).  The strengths of this
procedure are that aperture colors are less affected by neighboring
sources, and they typically have higher signal-to-noise ratios than
total magnitudes measured in each band independently.  In three of our
fields---XMM-SXDS, XMM-CFHTLS, and COSMOS---aperture colors have been
measured from point spread function (PSF) matched images and so should
be very accurate; in the other two fields---CDFS and ELAIS-S1---the
aperture colors are not measured from PSF-matched mosaics, so for
these sources we adopt slightly higher minimum photometric
uncertainties when performing our SED modeling (see
Section~\ref{sec:mass}).

\subsubsection{CDFS, ELAIS-S1, XMM-SXDS, and
  XMM-CFHTLS}\label{sec:cdfsphot}  

In the CDFS field we use the ground-based optical photometric catalogs
distributed as part of the SWIRE Data Release 2/3, as described in the
SWIRE Data Delivery
Document.\footnote{\url{http://irsa.ipac.caltech.edu/data/SPITZER/SWIRE}}
The SWIRE team obtained $U g^{\prime} r^{\prime} i^{\prime}
z^{\prime}$ imaging of this field using the MOSAIC-II imager at the
CTIO/Blanco 4~m telescope, achieving a depth in each band of $25.2$,
$25.3$, $25.2$, $24.4$, and $23.8$~mag ($5\sigma$) for a point source
in a $2\arcsec$ diameter aperture
\citep{norris06a}.\footnote{\url{http://www.astro.caltech.edu/$\sim$bsiana/cdfs\_opt}}
Source catalogs were generated (by the SWIRE team) in each bandpass
individually using the Cambridge Astronomical Survey Unit
(CASU\footnote{\url{http://casu.ast.cam.ac.uk}}) pipeline, and then
merged using the \emph{Spitzer} Science Center {\tt
  bandmerge}\footnote{http://ssc.spitzer.caltech.edu/dataanalysistools/tools/bandmerge}
software package.  We adopt the fluxes measured in a fixed $2\farcs4$
diameter aperture, scaled to match CASU's estimate of the integrated
$i^{\prime}$ flux.

In the ELAIS-S1 field we use the $BVR$ and $Iz$ catalogs published by
\citet{berta06a, berta08a}.\footnote{http://www.astro.unipd.it/esis}
The $BVR$ imaging was obtained using the Wide Field Imager (WFI) at
the 2.2~m La Silla ESO-MPI telescope, achieving a depth of $24.9$,
$25.0$, and $24.7$~mag in $B$, $V$, and $R$, respectively.  The $I$-
and $z$-band observations were obtained using the VIsible Multi Object
Spectrograph \citep[VIMOS;][]{le-fevre03a} camera at the VLT 8.2~m
telescope, reaching a depth of $\sim23.3$ and $\sim22.8$~mag,
respectively.  All the quoted depths correspond to the $95\%$
completeness limit for point sources.  The publicly released
photometric catalogs were generated using \sex{} \citep{bertin96a}.
We adopt the aperture fluxes measured in a $3\farcs3$ diameter
aperture, scaled to the {\sc mag\_auto} (total) $R$-band magnitude.

Finally, in the XMM-SXDS and XMM-CFHTLS fields we rely on the
high-fidelity $u^{*} g^{\prime} r^{\prime} i^{\prime} z^{\prime}$
photometric catalogs generated as part of the CFHTLS Archive Research
Survey
\citep[CARS;][]{erben09a}.\footnote{ftp://marvinweb.astro.uni-bonn.de/data\_products/CARS\_catalogues}
These catalogs are based on deep imaging obtained as part of the
CFHTLS-Wide survey using the CFHT/Megacam camera
\citep{boulade03a}.\footnote{http://www.cfht.hawaii.edu/Science/CFHLS}
The CARS mosaics reach a $5\sigma$ depth of $25.2$, $25.3$, $24.4$,
$24.7$, and $23.2$~mag in a $2\arcsec$ diameter aperture in $u^{*}$,
$g^{\prime}$, $r^{\prime}$, $i^{\prime}$, and $z^{\prime}$,
respectively.  The CARS photometric catalogs were generated from
PSF-matched images using \sex{} in dual-image mode, with the
unconvolved $i^{\prime}$ mosaic serving as the detection image.  We
use the fluxes measured in a $3\arcsec$ diameter aperture, scaled to
the {\sc mag\_auto} magnitude measured from the unconvolved
$i^{\prime}$-band mosaic.

In addition to the ground-based optical observations described above,
all four of the preceding fields were also observed at $3.6$, $4.5$,
$5.8$, and $8$~\micron{} with \emph{Spitzer}/IRAC as part of SWIRE.
We use the SWIRE Data Release 2/3 IRAC catalogs matched to the
galaxies in PRIMUS using a $1\arcsec$ search radius.  These catalogs
are complete for point sources to a $5\sigma$ depth of $22.2$, $21.5$,
$19.8$, and $19.9$~mag at $3.6$, $4.5$, $5.8$, and $8$~\micron,
respectively.  Following the SWIRE Data Delivery document, we use the
fluxes measured in a $3\farcs8$ diameter circular aperture, multiplied
by an aperture correction derived in each band from isolated point
sources (see also \citealt{surace04a, ilbert09a}).\footnote{For
  reference, the aperture correction factors we use, as measured by
  the SWIRE team, are $0.736$, $0.716$, $0.606$, and $0.543$ at $3.6$,
  $4.5$, $5.8$, and $8$~\micron, respectively.  (The fluxes are
  \emph{divided} by these factors.)}  These aperture-corrected fluxes
optimize the trade-off between crowding, which favors a smaller
aperture, and signal-to-noise ratio, which favors a larger aperture,
and provide a reasonably accurate measurement of the total IRAC flux
(see the SWIRE Data Delivery Document).

\subsubsection{COSMOS}

In the COSMOS field we rely on the multi-wavelength photometric
catalog publicly released in April 2009 by the COSMOS team
\citep[see][]{capak07a}.\footnote{http://irsa.ipac.caltech.edu/data/COSMOS}
This catalog includes $V_{J}g^{+}r^{+}i^{+}z^{+}$ imaging obtained
using the Suprime-Cam instrument \citep{miyazaki02a} on the Subaru
8.2~m telescope; $u^{*}i^{*}$ imaging from the Megacam camera
\citep{boulade03a} on the 3.6~m Canada-France-Hawaii Telescope (CFHT);
and $K_{s}$-band imaging obtained using the Wide-field InfraRed Camera
\citep[WIRCam;][]{puget04a} on the CFHT
\citep{mccracken10a}.\footnote{We do not use the Subaru $B_{J}$- and
  $V_{J}$-band imaging of the COSMOS field, nor the UKIRT $J$-band
  imaging, because of their larger-than-average photometric zeropoint
  uncertainties \citep{ilbert09a}.}  In a $3\arcsec$ diameter
aperture, these images achieve a $5\sigma$ depth of $26.5$, $26.6$,
$26.1$, $23.5$, and $25.1$~mag in $u^{*}$, $g^{+}$, $V_{J}$, $r^{+}$,
$i^{+}$, $i^{*}$, and $z^{+}$, respectively; the near-infrared imaging
is more than $90\%$ complete for point sources to $K_{s}=23$.  The
COSMOS optical/near-infrared photometric catalog was generated from
PSF-matched mosaics using \sex{} in dual-image mode, with the
unconvolved $i^{+}$ mosaic as the detection image.  We use the fluxes
measured in a $3\arcsec$ diameter aperture, scaled to the $i^{+}$ {\sc
  mag\_auto} magnitude measured from the unconvolved $i^{+}$-band
mosaic.

We supplement these optical/near-infrared data with mid-infrared
photometry generated by \citet{mendez13a} using the public
\emph{Spitzer}-COSMOS\footnote{http://irsa.ipac.caltech.edu/data/SPITZER/S-COSMOS}
(S-COSMOS) IRAC mosaics \citep{sanders07a}.  The catalogs were
generated using the same tools utilized by the SWIRE team (see
Section~\ref{sec:cdfsphot}), thereby ensuring that our IRAC photometry
is consistent across all five fields; nevertheless, \citet{mendez13a}
demonstrate that their photometric measurements and uncertainties
agree in the mean with the S-COSMOS/IRAC catalogs publicly released by
the COSMOS team in June
2007.\footnote{http://irsa.ipac.caltech.edu/data/COSMOS/gator\_docs/\\ scosmos\_irac\_colDescriptions.html}
The catalogs are statistically complete for point sources brighter
than $24.0$, $23.3$, $21.3$, and $21.0$~mag ($5\sigma$) in each of the
four IRAC channels.  As for the other four fields, we use the
aperture-corrected $3\farcs8$ diameter aperture fluxes and match each
IRAC source to our PRIMUS sample using a $1\arcsec$ search radius. 

\subsection{SDSS-GALEX Sample}\label{sec:sdss}

An accurate measurement of the \smf{} at the current epoch is crucial
because it provides a low-redshift anchor against which we can
quantify the stellar mass buildup of galaxies through cosmic time.
Although many previous studies have measured the local \smf{} for the
global galaxy population (see Section~\ref{sec:mflocal}), few have
exploited the bimodality of the SFR versus stellar mass diagram to
separately measure the \smf s of quiescent and star-forming galaxies.
Moreover, given the susceptibility of stellar mass estimates to a
variety of model-dependent systematic uncertainties (see, e.g.,
\citealt{marchesini09a, bernardi10a}; Appendix~\ref{appendix:syst}),
it is important that we measure the local \smf{} using the same
assumptions and methodology used to generate the \smf{} of
intermediate-redshift galaxies.

With these considerations in mind, we select a sample of low-redshift
galaxies using the SDSS Data Release~7 \citep[DR7;][]{abazajian09a},
which provides high-fidelity $ugriz$ photometry and spectroscopic
redshifts for hundreds of thousands of galaxies in the nearby
Universe.  Specifically, we select $504,437$ galaxies from the New
York University Value-Added Galaxy Catalog
\citep[VAGC\footnote{http://sdss.physics.nyu.edu/vagc};][]{blanton05b}
that satisfy the main sample criteria defined by \citet{strauss02a},
and have Galactic extinction corrected \citep{schlegel98a} Petrosian
magnitudes $14.5<r<17.6$, and spectroscopic redshifts $0.01<z<0.2$.
Excluding areas of the survey that were masked because of bright stars
and other artifacts \citep{blanton05b}, this sample covers
$6956$~deg$^{2}$.  The VAGC also provides an estimate of the
statistical weight for each galaxy, which we use to correct the sample
for targeting incompleteness (generally due to fiber collisions) and
redshift failures \citep{blanton03d}.

Next, we restrict the parent SDSS sample to galaxies with medium-depth
observations from GALEX; UV photometry is needed to effectively divide
the sample into quiescent and star-forming galaxies (see
Section~\ref{sec:select}).  First, we retrieve the positions of all
the \emph{GALEX} tiles publicly available as part of the \emph{GALEX}
Release 6 (GR6) with a total exposure time greater than $1$~ks, which
consists of more than $5400$ tiles covering roughly $4450$~deg$^{2}$
of the sky.  The $1$~ks exposure time cut is the minimum time
necessary to obtain an accurate measurement of the total UV flux of
extended sources at the typical redshift of galaxies in the SDSS
\citep{wyder07a}.  Next, we use {\tt
  mangle}\footnote{http://space.mit.edu/$\sim$molly/mangle}
\citep[v2.2;][]{hamilton04a, swanson08a} to construct the angular
selection function of the joint SDSS-\emph{GALEX} sample.  For the
SDSS we use the {\tt mangle} polygons distributed as part of the VAGC,
while for each \emph{GALEX} pointing we adopt a simple $1\fdg1$
diameter circular field-of-view \citep{morrissey07a}.  The final
sample comprises $169,727$ SDSS galaxies with \emph{GALEX} imaging
distributed over $2505$~deg$^{2}$.

With the relevant list of objects in hand, we use the
MAST/CasJobs\footnote{http://galex.stsci.edu/casjobs} interface and a
$4\arcsec$ diameter search radius \citep{budavari09a} to retrieve the
NUV and FUV photometry of the galaxies in our sample.  We resolve
duplicate \emph{GALEX} observations due to overlapping tiles by selecting the
measurement with the highest signal-to-noise ratio, and adopt the
SExtractor {\sc mag\_auto} magnitude \citep{bertin96a} measured in the
NUV band as an estimate of the total NUV flux.  For the FUV photometry
we use the FUV flux measured inside the elliptical aperture defined in
the NUV band to ensure accurate galaxy colors.

For the $ugriz$ bands we use the SDSS {\tt model} magnitudes, which
provide reliable, high signal-to-noise measurements of the optical
colors of each galaxy \citep{stoughton02a}.  We scale the $ugriz$
photometry to the $r$-band {\tt cmodel} magnitude, which provides the
most reliable estimate of the total (integrated) galaxy flux
irrespective of galaxy type \citep{abazajian04a, bernardi10a,
  blanton11a}.  Finally, we supplement our UV and optical photometry
with integrated $JHK_{s}$ magnitudes from the 2MASS Extended Source
Catalog \citep[XSC;][]{jarrett00a}, and with integrated photometry (or
upper limits) at $3.4$ and $4.6$~\micron{} from the WISE All-Sky Data
Release.\footnote{\url{http://wise2.ipac.caltech.edu/docs/release/allsky}}

\section{Sample Selection}\label{sec:sample}

In Section~\ref{sec:parent} we define the PRIMUS parent sample, and in
Section~\ref{sec:select} we describe the criteria we use to select
quiescent and actively star-forming galaxies as a function of
redshift.  

\subsection{Parent Sample}\label{sec:parent}

We select our parent sample from the statistically complete
\emph{primary} sample of galaxies observed by PRIMUS (see
\citealt{coil11a} for full details).  In Table~\ref{table:sample} we
list the optical selection band and the corresponding magnitude limits
we impose to define the parent sample in each of our five fields.
These limits are $18<i^{\prime}<23$ in the CDFS, XMM-SXDS, and
XMM-CFHTLS fields, $18<I<23$ in COSMOS, and $18<R<23.2$ in ELAIS-S1.
Due to differences in the PRIMUS experimental design, the sample of
galaxies targeted for spectroscopy in the CDFS and ELAIS-S1 fields
were also required to be detected in IRAC imaging at $3.6$~\micron;
therefore, we further require our parent sample of galaxies in these
two fields to have $17<[3.6]<21$.  We emphasize that having our sample
flux-limited in both the optical and mid-infrared does not preclude us
from determining the \smf{} in these two fields, as we account for
both flux limits in our analysis (see Section~\ref{sec:masslim}).

Next, we construct the angular selection (window) function of our
sample by joining the PRIMUS, \emph{GALEX}, and IRAC window functions using
{\tt mangle}.  The final window function includes regions of the sky
with \emph{GALEX} and IRAC imaging, and ensures coverage from two or more
PRIMUS slitmasks, thereby minimizing targeting incompleteness due to
slit collisions.  The solid angle of each field ranges from
$0.80$~deg$^{2}$ in ELAIS-S1, to $1.70$~deg$^{2}$ in the XMM-CFHTLS
field, totaling $5.50$~deg$^{2}$ (see Table~\ref{table:sample}).

Finally, we select objects spectroscopically classified as galaxies by
the PRIMUS pipeline (thereby excluding stars and broad-line AGN; see
\citealt{cool13a} for details) with high-quality ($Q\ge3$)
spectroscopic redshifts in the range $z=0.2-1.0$.  Below $z=0.2$
PRIMUS is severely limited by sample variance, while our upper
redshift cut eliminates $<5\%$ of the primary sample.  The redshift
confidence cut $Q\ge3$ balances the need for a large sample while
minimizing the catastrophic outlier rate and maximizing the redshift
precision.  Based on a comparison with redshifts derived from
high-resolution spectroscopy, we estimate that the redshift precision
of our sample is $\sigma_{z}/(1+z)\approx0.4\%$, with a catastrophic
outlier rate of $\lesssim3\%$.  Table~\ref{table:sample} lists the
final number of galaxies in our sample.

To correct our sample for targeting incompleteness and redshift
failures, we use the statistical weights described by \citet{coil11a}
and \citet{cool13a}.  Briefly, we assign a statistically derived
weight, $w_i$, to each galaxy given by

\begin{equation}
w_i = (f_{\rm target}\times f_{\rm collision}\times f_{\rm success})^{-1},
\label{eq:totweight}
\end{equation}

\noindent where $f_{\rm target}$ is the fraction of galaxies that
passed the PRIMUS magnitude- and density-dependent target selection
criteria; $f_{\rm collision}$ is the fraction of potential targets
observed spectroscopically (i.e., whose spectrum would not collide, or
overlap with another potential target); and $f_{\rm success}$ is the
fraction of galaxies in a given bin of apparent magnitude and color
(typically $g-r$ or $B-R$) that yielded a robust (i.e., $Q\ge3$)
redshift.  Both the targeting fraction ($f_{\rm target}\approx80\%$;
\citealt{coil11a}) and the fraction of targets observed ($f_{\rm
  collision}\approx95\%$; \citealt{cool13a}) are high in PRIMUS due to
its novel experimental design and survey strategy.  The redshift
success rate decreases smoothly with apparent magnitude---essentially,
spectroscopic signal-to-noise ratio---and varies weakly with
observed-frame color.  For reference, $f_{\rm success}$ decreases from
$\gtrsim75\%$ at $i=21$, to $\approx45\%$ at $i=22.5$, and to
$\approx30\%$ at the limit of our survey, $i\approx23$
\citep{cool13a}.  We emphasize that the lack of a significant trend of
$f_{\rm success}$ with observed-frame color, as well as extensive
comparisons of the PRIMUS redshifts with redshifts derived from
high-resolution spectroscopic surveys for both intrinsically red and
blue (i.e., quiescent and star-forming) galaxies, indicate that the
redshift success in PRIMUS is not a strong function of galaxy type.

\subsection{Selecting Quiescent and Star-Forming
  Galaxies}\label{sec:select}

A variety of techniques have been proposed to separate passively
evolving galaxies from galaxies with ongoing star formation
\citep[see, e.g.,][]{williams09a, pozzetti10a}, but at a basic level
all methods exploit to varying degrees the existence of galaxy
bimodality (see Section~\ref{sec:intro}).

\begin{figure}
\centering
\includegraphics[scale=0.4]{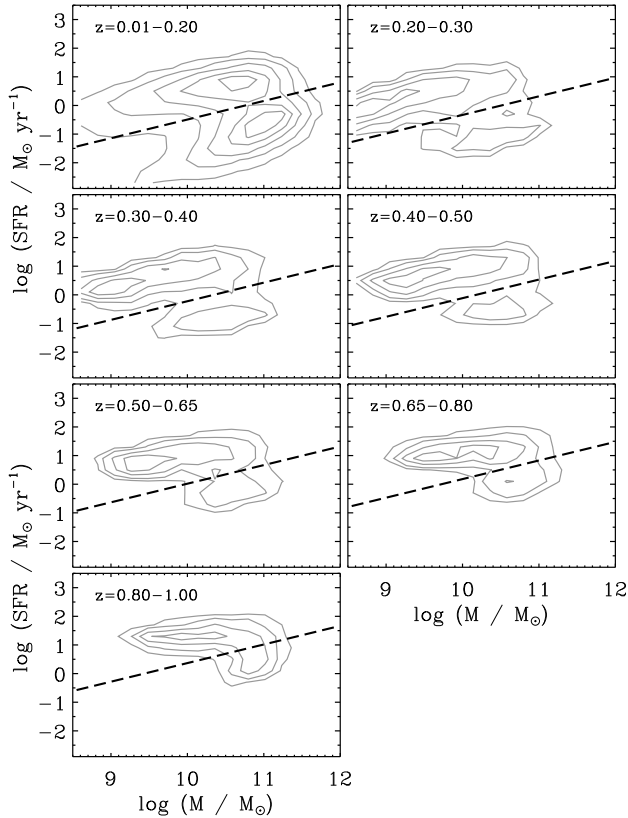}
\caption{Star formation rate (SFR) vs.~stellar mass in seven bins of
  redshift from $z=0-1$ based on our SDSS-\emph{GALEX} (upper-left
  panel) and PRIMUS (subsequent six panels) samples.  We divide our
  sample into star-forming or quiescent according to whether they lie
  above or below the dashed line, respectively; this line is parallel
  to the star formation (SF) sequence at $z\approx0.1$ and evolves
  with redshift according to equation~(\ref{eq:sfrmin}).
\label{fig:sfr}}
\end{figure}

Here, we leverage our broad wavelength coverage and precise
spectroscopic redshifts (see Section~\ref{sec:obs}) to measure
accurate stellar masses and SFRs for the galaxies in our sample using
\isedfit, a new Bayesian SED modeling code (see Section~\ref{sec:mass}
for details).  With these quantities in hand, we divide the galaxy
population into star-forming and quiescent based on whether they lie
on or below the so-called star formation (SF) sequence.  The SF
sequence (also called the {\em main sequence} of star formation;
\citealt{noeske07a}) is the correlation between SFR and stellar mass
exhibited by star-forming galaxies at least to $z\sim2$
\citep[e.g.,][]{oliver10a, karim11a}.  In Figure~\ref{fig:sfr} we plot
SFR versus stellar mass in seven redshift bins from $z=0-1$ for both
our SDSS-\emph{GALEX} and PRIMUS samples.  We find a well-defined SF
sequence whose amplitude increases smoothly toward higher redshift,
and a distinct population of quiescent galaxies that fall below the SF
sequence at a given stellar mass \citep[e.g.,][]{salim07a, elbaz07a}.
We postpone a more detailed discussion of the evolution of the SF
sequence to $z=1$ using PRIMUS to another paper.

To divide the galaxy population we use an evolving cut that traces the
lower envelope of the SF sequence in each redshift bin.  In detail, we
first rotate the SFR versus stellar mass diagram using the power-law
slope of the SF sequence derived by \citet{salim07a}; Salim~\etal{}
find ${\rm SFR}\propto\mass^{0.65}$ for galaxies at $z\approx0.1$,
which is also a good fit to our SDSS-\emph{GALEX} sample.  Next, we
construct the histogram distribution of ``rotated'' SFRs, given by
$\log\,({\rm SFR}_{\rm rot}) = \log\,({\rm
  SFR})-0.65(\log\,\mass-10)$, where SFR is in units of \sfrunits{}
and \mass{} is in \msun, and identify (by eye) the minimum of the
bimodality in each redshift bin.  Finally, we fit the minimum of the
bimodality versus redshift to obtain the following linear relation:
\begin{equation}
\log\, ({\rm SFR}_{\rm min}) = -0.49 + 0.65\,\log\,(\mass-10) + 
1.07\,(z-0.1).
\label{eq:sfrmin}
\end{equation}

\noindent We classify each galaxy in our sample into star-forming and
quiescent based on whether its SFR and stellar mass place it above or
below the SFR given by equation~(\ref{eq:sfrmin}), interpolated at the
redshift of the galaxy.

\section{Building the Stellar Mass Function}\label{sec:analysis}  

In this section we describe how we infer the stellar masses and SFRs
for the galaxies in our sample (Section~\ref{sec:mass}), present the
technique we use to construct a non-parametric estimate of the \smf{}
(Section~\ref{sec:method}), and calculate the stellar mass above which our
full (hereafter, the \emph{all} sample), quiescent, and star-forming
galaxy samples are statistically complete as a function of redshift
(Section~\ref{sec:masslim}).

\subsection{Stellar Masses and Star Formation Rates}\label{sec:mass} 

Modeling the broadband SEDs of galaxies using stellar population
synthesis models has become a powerful technique for inferring their
physical properties (see the recent review by \citealt{walcher11a},
and references therein).  We have developed \isedfit, a suite of
routines written in the {\sc idl} programming language to determine
within a simplified Bayesian framework the stellar masses, SFRs, and
other physical properties of galaxies from their observed broadband
SEDs \citep[e.g.,][]{kauffmann03a, salim07a, auger09a}.  We describe
\isedfit{} in more detail in Appendix~\ref{appendix:isedfit}, but in
essence the code uses the redshift and observed photometry of each
galaxy to compute the statistical likelihood of a large ensemble of
model SEDs---generated using population synthesis models---spanning a
wide range of observed colors and physical properties (stellar mass,
age, metallicity, star formation history, dust content, etc.).  Random
draws of the model parameters are chosen from user-defined prior
parameter distributions using a Monte Carlo technique.  Once the
posterior probability distribution function (PDF) has been computed,
the marginalized PDF of the quantity of interest, such as the stellar
mass, follows from the probability-weighted histogram of the
corresponding parameter values.  The median of the posterior PDF can
then be adopted as the best estimate of that parameter, and the
uncertainty can be derived from the cumulative distribution function
(see Appendix~\ref{appendix:isedfit} for more details).

Although SED modeling is conceptually straightforward, the inferred
physical properties depend on which population synthesis models and
prior parameters are adopted \citep[e.g.,][]{marchesini09a,
  kajisawa09a, muzzin09a}.  For example, differences in the stellar
libraries (e.g., theoretical vs.~empirical) among population synthesis
models, and the exact treatment of post-main sequence stellar
evolution can result in widely different predictions of the time- and
metallicity-dependent spectral evolution of even simple (i.e., coeval)
stellar populations \citep{conroy09a, conroy10b, mancone12a}.
Moreover, the exact choice of prior parameters (e.g., dust attenuation
curve, bursty vs.~smooth star formation histories, treatment of
metallicity evolution, etc.)  can also have a significant effect on
the derived properties \citep{kannappan07a, perez-gonzalez08a,
  carter09a, longhetti09a, muzzin09a, kajisawa09a, marchesini09a,
  maraston06a, maraston10a, conroy10a, pforr12a}.  Finally, SED
modeling typically includes an implicit assumption of a fixed,
universal IMF---an IMF that does not vary with redshift or galactic
physical conditions.  Throughout this paper we make the same
simplifying assumption, although recent observations suggest the IMF
may not be as universal as once was thought (\citealt{dave08a,
  treu10a, van-dokkum10a, cappellari12a}, but see \citealt{bastian10a}
for a critical review of the evidence).

An exhaustive investigation of the preceding issues is beyond the
scope of the current study.  Nevertheless, we would like to have a
qualitative and quantitative sense of which of our results are
(in)sensitive to the exact choice of population synthesis models and
priors.  Therefore, we proceed by presenting our principal results in
the main body of the paper using a fiducial set of SED modeling
assumptions, and in Appendix~\ref{appendix:syst} we examine the effect
of varying these assumptions on our conclusions.

Here, we briefly summarize the default population synthesis models and
prior parameters we use, but refer the interested reader to
Appendices~\ref{appendix:isedfit} and \ref{appendix:syst} for
additional parameter definitions and details.  Our fiducial stellar
masses and SFRs are derived using the Flexible Stellar Population
Synthesis
(FSPS\footnote{http://www.ucolick.org/$\sim$cconroy/FSPS.html}) models
\citep[v2.3;][]{conroy09a, conroy10b}, based on the
\citet{chabrier03a} IMF from $0.1-100$~\msun.  
We consider exponentially declining star formation histories with
stochastic bursts of varying onset, strength, and duration superposed
\citep{kauffmann03a, salim07a}, and allow a wide range of galaxy ages
and possible star formation histories.  Finally, we assume sensibly
distributed priors on stellar metallicity and dust attenuation, and
adopt the time-dependent attenuation curve of \citet{charlot00a}.

The photometric bands we use vary with the sample.  In our
SDSS-\emph{GALEX} sample we fit to $12$ bands of photometry: FUV and
NUV from \emph{GALEX}; $ugriz$ from the SDSS; $JHK_{s}$ from 2MASS;
and the $3.4$ and $4.6$~\micron{} bands from WISE (see
Section~\ref{sec:sdss}).  In our PRIMUS sample we fit to our
\emph{GALEX} FUV and NUV photometry (see Section~\ref{sec:galex}), the
two shortest IRAC bands at $3.6$ and $4.5$~\micron, and five optical
bands, except in COSMOS where we fit to seven optical and
near-infrared bands (see Section~\ref{sec:optphot}).  We do not fit to
the two longer-wavelength IRAC channels at $5.8$ and $8$~\micron{}
because of the potential contributions from hot dust and polycyclic
aromatic hydrocarbon (PAH) emission lines at these wavelengths
\citep[e.g.,][]{smith07a}, which \isedfit{} does not currently model.
When fitting, we assume minimum photometric uncertainties of $5.2\%$
and $2.6\%$ in the FUV and NUV, respectively \citep{morrissey07a},
$2\%-5\%$ in the optical/near-infrared bands, and $3\%$ in the 2MASS
and WISE near- and mid-infrared photometric bands.

\subsection{A Non-Parametric Estimate of the Stellar Mass
  Function}\label{sec:method} 

We build the \smf{} using the non-parametric \oneovervmax{} estimator
widely used in analyses of the galaxy luminosity function \cite[see
  the recent review by][and references therein]{johnston11a}.  We use
this technique in favor of complementary \emph{parametric} methods
because accurately fitting the observed \smf{} at $z=0-1$ requires
additional free parameters at the low-mass end \citep{baldry08a,
  drory09a, peng10a}, which cannot be reliably constrained at higher
redshift with PRIMUS.  Nevertheless, when calculating number and
stellar mass densities, we do rely on either a single standard
\citet{schechter76a} function, or a \emph{double} Schechter function
\citep[see, e.g.,][]{baldry08a}, depending on which model is a better
fit to the data, to extrapolate the observed binned \smf{} over small
intervals of stellar mass as needed.  We will also occasionally refer
to the `knee' of the \smf{} as \mstar, which marks the stellar mass
above which the \smf{} declines exponentially.

The differential, non-parametric \smf{} is given by 
\begin{equation}
\Phi(\log\mass)\,\Delta(\log\mass) = \sum_{i=1}^{N} \frac{w_{i}}{V_{\rm
    max,i}},   
\label{eq:mf}
\end{equation}

\noindent where \vmax{} is the maximum cosmological volume within
which each galaxy $i$ could have been observed given the apparent
magnitude limits of the survey, $w_{i}$ is the statistical weight for
each object (see Section~\ref{sec:parent}),
$\Phi(\log\mass)\times\Delta(\log\mass)$ is the number of galaxies
($N$) per unit volume with stellar masses in the range
$\log\mass\rightarrow$ $\log\mass+\Delta(\log\mass)$, and \mass{} has
units of \msun.

\begin{figure*}
\centering
\includegraphics[scale=0.55,angle=90]{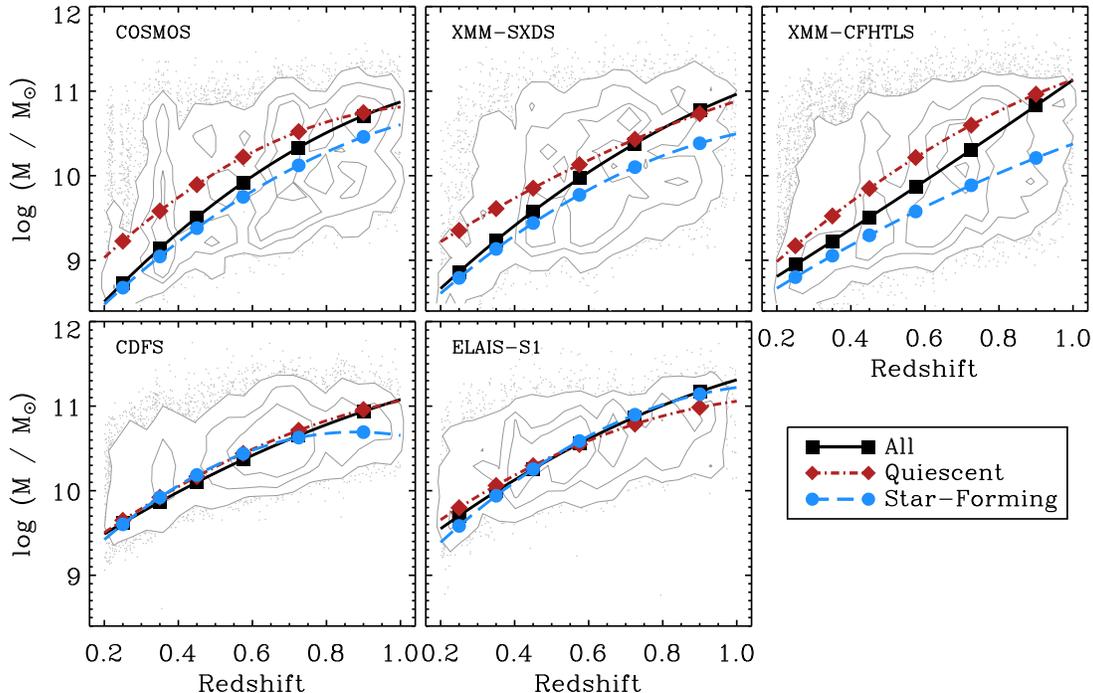}
\caption{Stellar mass vs.~redshift for all galaxies in the five
  individual PRIMUS fields.  The contours enclose $30\%$, $60\%$, and
  $90\%$ of the sample, while the grey points show galaxies lying
  outside the $90\%$ quantile.  The solid black, dot-dashed red, and
  dashed blue lines indicate the stellar mass completeness limit in
  each field, and the black squares, red diamonds, and blue points
  show the stellar mass limit at the center of each of our six adopted
  redshift bins (see Table~\ref{table:limits}).  In the CDFS and
  ELAIS-S1 fields the stellar mass limits among all, quiescent, and
  star-forming galaxies are comparable at all redshifts because these
  two samples are limited in both the optical and at $3.6$~\micron{}
  (see Section~\ref{sec:parent}).  Meanwhile, in the COSMOS, XMM-SXDS,
  and XMM-CFHTLS fields we find that the stellar mass limit for
  quiescent (star-forming) galaxies is higher (lower) at all
  redshifts, as expected given their typically larger (lower) stellar
  mass-to-light ratios.  Moreover, the stellar mass limit for the
  \emph{all} sample tracks low-mass star-forming galaxies at low
  redshift, and massive quiescent galaxies at high redshift.
\label{fig:limits}}
\end{figure*}

To estimate \vmax{} for each galaxy we first use
\kcorrect\footnote{http://howdy.physics.nyu.edu/index.php/Kcorrect}
\citep[v4.2;][]{blanton07a} to derive the redshift-dependent
$K$-correction, $K(z)$, from the observed SED.  We use \kcorrect{}
because of its speed and convenience, although we obtain similar
results if we instead use the best-fitting {\tt iSEDfit} model
(Section~\ref{sec:mass}).  Next, we write the apparent magnitude, $m$,
of a galaxy of absolute magnitude $M$ as
\begin{equation}
m = M + {\rm DM} + K - Qz,
\label{eq:appmag}
\end{equation}

\noindent where ${\rm DM}(z)$ is the distance modulus \citep{hogg99a}
and $Q$ is a (constant) luminosity evolution term which we discuss
below.  Given the observed apparent magnitude, $m_{\rm obj}$, of an
object at redshift $z_{\rm obj}$, we can use
equation~(\ref{eq:appmag}) to write $\Delta m(z)$, the change in the
apparent magnitude of the object as a function of redshift, as
\begin{eqnarray}
\Delta m(z) \equiv m_{\rm obj} & -& m(z) = {\rm DM}(z_{\rm obj})-{\rm
    DM}(z) \nonumber \\ & + & K(z_{\rm obj})-K(z) - Q(z_{\rm obj}-z).
\label{eq:deltaappmag}
\end{eqnarray}

\noindent By definition, $z_{\rm max}$ ($z_{\rm min}$) is the redshift
at which $\Delta m$ equals the faint (bright) apparent magnitude limit
of the survey.  Once $z_{\rm max}$ and $z_{\rm min}$ have been
computed for each galaxy with a measured redshift in the interval
$[z_{\rm lower},z_{\rm upper}]$, \vmax{} can be computed using
\begin{equation}
V_{\rm max} = \int_{\Omega}\int_{z_{1}}^{z_{2}} \frac{{\rm
    d}^{2}V_{c}}{{\rm d}z\,{\rm d}\Omega}\, {\rm d}z\,{\rm d}\Omega, 
\label{eq:vmax}
\end{equation}

\noindent where $z_{1} = {\tt max}\,(z_{\rm min}, z_{\rm lower})$,
$z_{2} = {\tt min}\,(z_{\rm max}, z_{\rm upper})$, $\Omega$ is the
solid angle of the survey, and $V_{c}$ is the comoving volume
\citep{hogg99a}.  

The coefficient $Q$ in equations~(\ref{eq:appmag}) and
(\ref{eq:deltaappmag}) allows us to include a simple luminosity
evolution model into our \vmax{} estimates.  Many recent measurements
of the optical luminosity function have shown that galaxies brighten
toward higher redshift \citep[e.g.,][]{blanton03c, faber07a,
  loveday12a, cool12a}; consequently, higher-redshift galaxies will be
observable over a larger cosmological volume (i.e., \vmax{} will be
larger) relative to the no-evolution case.  Using $Q>0$ in
equation~(\ref{eq:deltaappmag}) allows us to mimic the observed
luminosity evolution, at least in a statistical sense.

Of course, our luminosity evolution model is intentionally simplistic
and should not be over-interpreted.  For example, \citet{blanton03c}
find that among $z\approx0.1$ galaxies $Q$ varies by roughly a factor
of two at rest-frame optical wavelengths redward of the $4000$-\AA{}
break, whereas our simple model assumes that $Q$ is independent of
wavelength.  Nevertheless, the distribution of $V_{c}/\vmax$ for our
SDSS-\emph{GALEX} sample clearly shows the need to account for
luminosity evolution when estimating \vmax.  For example, without
luminosity evolution (i.e., $Q=0$), $V_{c}/\vmax$ for our
SDSS-\emph{GALEX} sample correlates with redshift, stellar mass, and
other intrinsic galaxy properties; adopting $Q=1-2$~mag~$z^{-1}$, on
the other hand, removes these first-order dependencies and results in
$\langle V_{c}/\vmax\rangle\approx0.5$, as expected for a homogenous,
statistically complete sample \citep{johnston11a}.  The distribution
of $V_c/\vmax$ for PRIMUS, on the other hand, is largely insensitive
to $Q$ because of the fairly narrow redshift bins we adopt (see, e.g.,
Figure~\ref{fig:sfr}).  Using $Q>0$ in place of $Q=0$ changes \vmax{}
for just $\approx10\%$ of PRIMUS galaxies in each redshift interval,
and has no significant effect on our derived \smf s.

With the preceding discussion in mind, we calculate \vmax{} for each
galaxy using equation~(\ref{eq:vmax}) and the solid angle and optical
apparent magnitude limits listed in Table~\ref{table:sample}.  We
adopt $Q=1.6$~mag~$z^{-1}$ \citep{blanton03c, cool12a} when
calculating \vmax{} based on our optical magnitude limits; in the CDFS
and ELAIS-S1 fields we separately estimate \vmax{} based on our
$3.6$~\micron{} flux limits assuming $Q=1.2$~mag~$z^{-1}$
\citep{dai09a}, and adopt the \emph{smaller} of the two (optical
vs.~mid-infrared) \vmax{} values.

Neglecting stellar mass uncertainties (see
Appendix~\ref{appendix:syst}), the two principal sources of
uncertainty in the \smf{} are due to sample size (i.e., Poisson
uncertainty) and sample variance.  In the limit $N\gg1$, the formal
Poisson uncertainty is given by
\begin{equation}
\sigma_{\Phi} = \frac{1}{\Delta(\log\mass)} \sqrt{\sum_{i=1}^{N}
  {\frac{w_{i}}{V_{{\rm max},i}^{2}}}}.
\label{eq:mferr}
\end{equation}

\noindent Equation~(\ref{eq:mferr}) becomes increasingly inaccurate,
however, as the number of galaxies approaches zero; therefore, we
calculate the effective number of galaxies in each mass bin following
\citet{zhu09a}, and use the analytic formulae of \citet{gehrels86a} to
compute the upper and lower statistical uncertainty of the \smf.

To estimate the uncertainty in the \smf{} due to sample variance we
use a standard jackknife technique.  We construct the \smf{} excluding
one field at a time, and calculate the uncertainty in the mean number
of galaxies in each stellar mass bin due to the field-to-field
variations.  Formally, we estimate the uncertainty, $\sigma_{\rm
  cv}^{j}$, in the $j^{\rm th}$ stellar mass bin due to sample
variance as
\begin{equation}
\sigma^{j}_{\rm cv} = \sqrt{\frac{M-1}{M} \sum_{k=1}^{M} (\Phi_{k}^{j}
  - \langle \Phi^{j} \rangle)^2}, 
\label{eq:cverr}
\end{equation}

\noindent where the sum extends over all $M$ individual fields, and
$\langle \Phi^{j} \rangle$ is the mean number density of galaxies in
that stellar mass bin measured from all the available data.  Note that
when computing the cumulative number and stellar mass densities (see
Section~\ref{sec:mfevol}), we integrate each jackknife realization of the
\smf, and compute the variance in these quantities using the same
formalism.  

This jackknife technique likely underestimates the level of sample
variance in PRIMUS, because two of the five fields (XMM-SXDS and
XMM-CFHTLS) are spatially adjacent, so there is covariance between the
fields due to structure on scales larger than the combined field.  For
simplicity, however, we ignore this covariance in our analysis, and
simply exclude each of the $M=5$ fields sequentially as if they were
independent.  Another potential issue is that the individual fields
are not the same size (angular area); therefore, one might expect that
the standard jackknife prefactor $\sqrt{(M-1)/M}$ is not strictly
correct.  However, we verified using a Monte Carlo calculation that
the correct prefactor for our sample differs by $<5\%$ from the
nominal value despite the more than factor of two variation in solid
angle among our five fields (see Table~\ref{table:sample}).

Finally, to estimate the level of sample variance in our
SDSS-\emph{GALEX} sample, we divide the sky into a $12\times9$
rectangular grid, and retain the $27$ $30\times60$~deg$^{2}$ regions
containing at least $1000$ galaxies.  We then compute the variance in
the SDSS-\emph{GALEX} \smf{} using the same methodology described
above, once again ignoring the potential covariance between adjacent
subfields.

\subsection{Stellar Mass Completeness Limits}\label{sec:masslim}

Before computing the \smf{} we need to determine the stellar mass
above which our sample is complete.  In a magnitude-limited survey
such as PRIMUS, the stellar mass completeness limit is a function of
redshift, the apparent magnitude limit of the survey, and the typical
stellar mass-to-light ratio of galaxies near this flux limit.  For
example, a quiescent galaxy of a given stellar mass will
preferentially fall below the survey flux limit compared to a
star-forming galaxy with the same stellar mass, because the stellar
mass-to-light ratios of quiescent galaxies are typically higher.

We empirically determine the stellar mass completeness limits of our
sample following \citet{pozzetti10a}.  First, we compute \masslim, the
stellar mass each galaxy would have if its apparent magnitude was
equal to the survey magnitude limit, $\log\,\masslim = \log\,\mass +
0.4\,(m-m_{\rm lim})$, where \mass{} is the stellar mass of the galaxy
in units of \msun, $m$ is the observed apparent magnitude in the
selection band, and $m_{\rm lim}$ is the corresponding magnitude limit
(see Table~\ref{table:sample}).  Next, we construct the cumulative
distribution of \masslim{} for the $15\%$ faintest galaxies in $\Delta
z=0.04$ wide bins of redshift, and calculate the minimum stellar mass
that includes $95\%$ of the objects.  We use the subset of galaxies
near the flux limit to account for the fact that in a flux-limited
sample the lowest-luminosity galaxies tend to have the lowest stellar
mass-to-light ratios; however, we did verify that using the whole
sample in each redshift slice changes the derived mass limits by
$\lesssim0.1$~dex.  Finally, we fit the limiting stellar mass versus
redshift with a quadratic polynomial separately for all, star-forming,
and quiescent galaxies.  We evaluate this fit at the center of each of
our six adopted redshift intervals (see Figure~\ref{fig:sfr}), and
list the results in Table~\ref{table:limits}.

In the CDFS and ELAIS-S1 fields we carry out the same procedure
described above except we consider the flux limits in both the
$i^{\prime}$ and $R$ selection bands, respectively, and in the IRAC
$3.6$~\micron{} band (see Section~\ref{sec:parent} and
Table~\ref{table:sample}).  At each redshift we then take the greater
of the two stellar mass limits implied by the two apparent magnitude
limits.  In detail, in these fields our sample is limited by the
$3.6$~\micron{} flux limit at low redshift, and by the optical flux
limit at higher redshift, with the transition redshift occurring
around $z\approx0.6$.

Figure~\ref{fig:limits} plots stellar mass versus redshift for the
galaxies in all five PRIMUS fields.  The black squares, red diamonds,
and blue points indicate the stellar mass completeness limits at the
center of each redshift bin for all, quiescent, and star-forming
galaxies, respectively, and the solid black, dot-dashed red, and
dashed blue lines show the corresponding polynomial fits.  As
expected, the completeness limits for the quiescent galaxies lie above
the star-forming galaxy limits at all redshifts, except in our CDFS
and ELAIS-S1 fields, which are flux-limited in both the optical and at
$3.6$~\micron.  Moreover, the completeness limits for all galaxies
overlap the star-forming galaxy limits at low redshift, and the
quiescent galaxy limits at high redshift.  This shift occurs because
at low redshift the combined sample is dominated by low-mass
star-forming galaxies, while at high redshift massive, quiescent
galaxies dominate, as we demonstrate in the next section.

Finally, for our SDSS-\emph{GALEX} sample we adopt a uniform stellar mass
limit of $10^{9}$~\msun, which is safely above the surface brightness
and stellar mass-to-light ratio completeness limits of the survey
\citep{blanton05a, baldry08a}.

\section{Evolution of the Stellar Mass Function from
  $\lowercase{z}=0-1$}\label{sec:results}   

We now have all the ingredients needed to compute the evolution of the
\smf{} from $z=0-1$.  We begin in Section~\ref{sec:mflocal} by
presenting the \smf{} at $z\approx0.1$ using our SDSS-\emph{GALEX}
sample.  Next, we combine the SDSS-\emph{GALEX} and PRIMUS samples in
Section~\ref{sec:mfevol} and show how the \smf{} of all, quiescent,
and star-forming galaxies has evolved since $z=1$.  The results we
present in this section are all based on our fiducial stellar mass
estimates (see Section~\ref{sec:mass}), although in
Appendix~\ref{appendix:syst} we present a detailed discussion of how
systematic uncertainties in our stellar mass estimates affect our
conclusions.

\subsection{Stellar Mass Function at
$\lowercase{z}\approx0.1$}\label{sec:mflocal}

\begin{figure}
\centering
\includegraphics[scale=0.4]{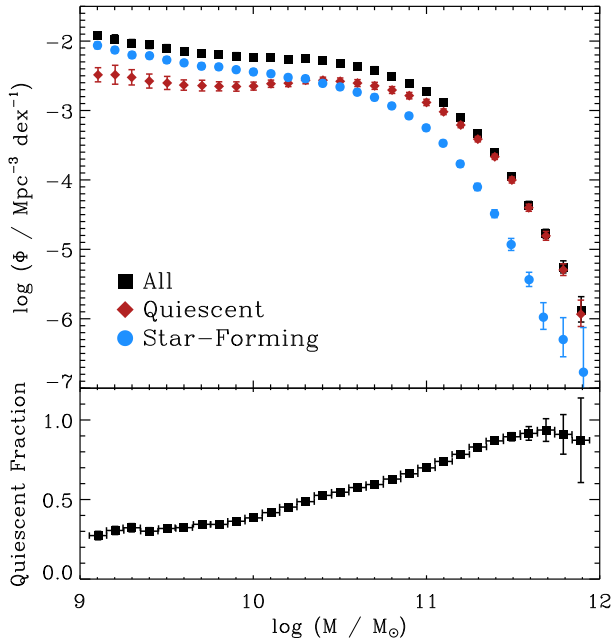}
\caption{(Upper panel) SDSS-\emph{GALEX} \smf{} for all (black squares),
  quiescent (red diamonds), and star-forming (blue points) galaxies at
  $z\approx0.1$.  (Lower panel) Fraction of quiescent galaxies as a
  function of stellar mass.  The massive end of the \smf{} is
  overwhelmingly comprised of quiescent galaxies, while below
  $\mass\sim3\times10^{10}$~\msun{} star-forming galaxies increasingly
  dominate the global galaxy population.  Quantitatively, the fraction
  of quiescent galaxies ranges from $\sim25\%$ around
  $\sim3\times10^{9}$~\msun{} to $\sim95\%$ around
  $\sim3\times10^{11}$~\msun.
\label{fig:mfsdss}}
\end{figure}

We begin by presenting in the upper panel of Figure~\ref{fig:mfsdss}
the \smf{} at $z\approx0.1$ based on the entire SDSS-\emph{GALEX} sample
(black squares), and separately for the star-forming (blue points) and
quiescent (red diamonds) galaxy subsamples.  Each point represents the
comoving number density of galaxies in $0.1$~dex wide bins of stellar
mass, and the vertical error bars indicate the quadrature sum of the
Poisson and sample variance uncertainties in each stellar mass bin.
We tabulate the \smf{} for each sample in Table~\ref{table:mflocal}.
The lower panel of this figure shows the variation in the fraction of
quiescent galaxies with stellar mass.

Figure~\ref{fig:mfsdss} conveys several striking (albeit well-known)
results.  First, the massive end of the \smf{} is almost entirely
comprised of quiescent galaxies, while star-forming galaxies vastly
outnumber quiescent galaxies at the low-mass end \citep[see,
  e.g.,][and references therein]{blanton09a}.  Above
$\sim2\times10^{11}$~\msun{} more than $\sim90\%$ of galaxies are
quiescent, whereas below $\sim10^{10}$~\msun{} star-forming galaxies
outnumber quiescent galaxies by more than a factor of three.  The
stellar mass at which each population begins to outnumber the other is
$\mass\sim3\times10^{10}~\msun$, in good agreement with previous
studies \citep[e.g.,][]{bell03b, kauffmann03b, baldry04a}.
Integrating the observed distributions above $\mass=10^{9}$~\msun{}
yields a total stellar mass density of
$2.36\times10^{8}$~\msun~Mpc$^{-3}$, of which approximately $60\%$
resides in quiescent galaxies.\footnote{Note that galaxies with
  $\mass<10^{9}~\msun$ contribute a negligible amount to the overall
  stellar mass budget of the nearby Universe \citep[see
    also][]{brinchmann04a}.}  For comparison, \citet{baldry04a} find
that $54\%-60\%$ of the stellar mass density at $z\approx0.1$ is in
red, early-type galaxies, where the precise result depends on the
method used to derive stellar masses.

\begin{figure}[b]
\centering \includegraphics[scale=0.4]{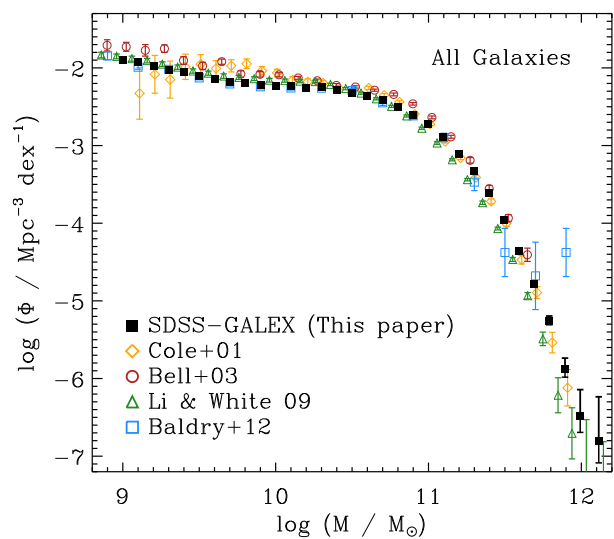}
\caption{Comparison of our measurement of the \smf{} at $z\approx0.1$
  for all galaxies against previous determinations from the
  literature, adjusted to our adopted cosmology and IMF where
  necessary.  Overall, our results agree well with these previous
  studies, albeit with some notable differences (see
  Section~\ref{sec:mflocal}).  \label{fig:mfsdss_lit}}
\end{figure}

We compare our results to previously published measurements of the
local \smf{} in Figure~\ref{fig:mfsdss_lit}, adjusting where necessary
for differences in the adopted IMF and cosmological parameters.  We
plot the SDSS-\emph{GALEX} \smf{} using filled black squares, and the results
from \citet{cole01a}, \citet{bell03b}, \citet{li09a}, and
\citet{baldry12a} using orange diamonds, red circles, green triangles,
and blue squares, respectively.  Overall, our results agree reasonably
well with these studies, although there are some notable differences.
The agreement between our \smf{} and the recent measurement by
\citet{baldry12a}, who analyzed a sample of $\sim10^{5}$ galaxies at
$z<0.06$ over $143$~deg$^{2}$ with spectroscopic redshifts from the
SDSS and GAMA \citep{driver11a} surveys, is especially good.
Unfortunately, the \citet{baldry12a} sample included too few galaxies
with stellar masses $\mass\gtrsim3\times10^{11}$~\msun{} for them to
reliably measure the massive end of the \smf.  

Compared to \citet{li09a}, the exponential tail of our \smf{} falls
off less steeply, which is somewhat surprising given that they
analyzed a comparably large sample of SDSS galaxies.  However,
\citet{bernardi10a} argue that \liwhite{} likely underestimated the
stellar masses of the most massive galaxies in their sample for two
reasons: first, \liwhite{} used Petrosian magnitudes, which are known
to underestimate the fluxes of galaxies with extended surface
brightness profiles such as the spheroidal galaxies that dominate the
massive end of the \smf{} \citep[see also][]{blanton11a}; and second,
\liwhite{} derived stellar masses using the standard set of
\kcorrect{} basis templates \citep{blanton07a}, which can
underestimate the stellar masses of massive early-type galaxies
dominated by very old stellar populations (see \citealt{bernardi10a}
and Appendix~\ref{appendix:syst}).

Finally, Figure~\ref{fig:mfsdss_lit} shows that the \smf{} measured by
\citet{bell03b} lies systematically above our \smf{} at all stellar
masses.  Bell~\etal{} constructed their \smf{} from a sample of
$\sim7000$ galaxies distributed over $\sim400$~deg$^{2}$ in the SDSS
Early Data Release (EDR) survey area \citep{stoughton02a}.  However,
this area of the sky is now known to contain one of the largest
structures ever mapped, the SDSS Great Wall at $z=0.078$
\citep{gott05a}.  Therefore, one possibility for the origin of the
discrepancy is that the Bell~\etal{} \smf{} may be more affected by
large-scale structure (i.e., sample variance) than originally
estimated.  Alternatively, a strong color-dependent difference in
stellar mass-to-light ratio that is negligible for massive quiescent
galaxies and $\sim0.3$~dex (factor of $\sim2$) for low-mass
star-forming galaxies could also explain the observed discrepancy
(E.~F. Bell 2012, private communication).  Whatever the reason, a
practical consequence of this result is that previous studies which
relied on the Bell~\etal{} \smf, and to a lesser degree the
Cole~\etal{} \smf{} as their low-redshift anchor, may have
overestimated the amount of number and stellar mass density evolution
\citep[e.g.,][]{brammer11a, mortlock11a}.

\begin{figure}
\centering
\includegraphics[scale=0.5]{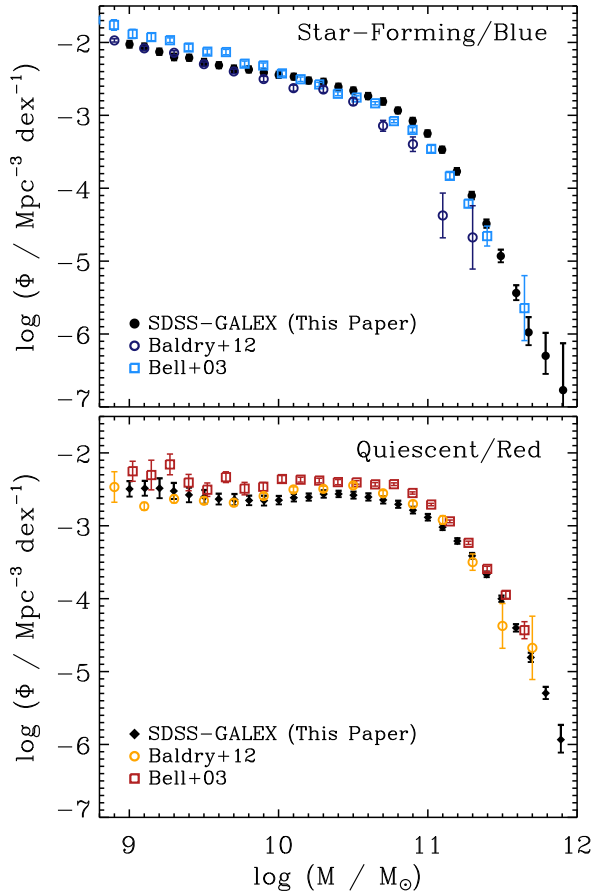}
\caption{Comparison of the SDSS-\emph{GALEX} \smf{} of star-forming
  (top) and quiescent (bottom) galaxies against the \smf s of
  optically-selected blue-cloud and red-sequence galaxies from
  \citet{bell03b} and \citet{baldry12a}, respectively.  We verified
  that when we divide the galaxy population using the $u-r$
  vs.~$M_{r}$ color-magnitude diagram we obtain excellent agreement
  with the corresponding \smf s from Baldry~\etal.  Therefore, we
  attribute the apparent differences between the \smf s to
  contamination of the red sequence by dusty star-forming galaxies.
\label{fig:mfsdss_lit_qq_sf}} 
\end{figure}

Next, we compare our \smf s of quiescent and star-forming galaxies
against previous measurements.  Most previous analyses have used the
optical color-magnitude diagram to identify ``quiescent'' and
``star-forming'' galaxies according to whether they lie on the {\em
  red sequence} or the {\em blue cloud} \citep[e.g.,][]{bell03b,
  taylor09b, baldry12a}; however, the optical red sequence is known to
contain both bona fide quiescent galaxies and dust-obscured
star-forming galaxies \citep[e.g.,][]{brand09a, maller09a, zhu11a}.
By contrast, we identify quiescent and star-forming galaxies according
to whether they lie on or below the star-forming sequence, which
allows us to select a purer sample of quiescent galaxies
\citep{salim07a, schiminovich07a}.  Therefore, we anticipate that the
overall normalization of our quiescent-galaxy \smf{} will be lower
relative to these previous studies.  Moreover, the differences are
likely to be stellar mass-dependent because the level of star
formation activity and amount of dust attenuation vary systematically
with stellar mass \citep{brinchmann04a, elbaz07a, garn10a}.

With the preceding ideas in mind, in Figure~\ref{fig:mfsdss_lit_qq_sf}
we compare our \smf s for star-forming and quiescent galaxies against
the \smf s published by \citet{bell03b} and \citet{baldry12a}.
Bell~\etal{} used the $g-r$ color-magnitude diagram to identify
red-sequence and blue-cloud galaxies, while Baldry~\etal{} leveraged
the bimodality in $u-r$ color.  As expected, our \smf{} of quiescent
galaxies agrees reasonably well with the \smf s from Bell~\etal{} and
Baldry~\etal{} at the massive end, where the amount of contamination
from dusty starburst galaxies is minimal, but is displaced
systematically below their \smf s below $\sim10^{11}$~\msun.  As
anticipated above, the reason for these differences is likely because
the $g-r$ and $u-r$ red sequences at intermediate mass contain an
admixture of both quiescent and dust-obscured star-forming galaxies.
In our analysis, these galaxies are (correctly) assigned to the
star-forming galaxy \smf, as illustrated in the upper panel of
Figure~\ref{fig:mfsdss_lit_qq_sf}.  We verified this interpretation by
constructing the \smf s for red-sequence and blue-cloud galaxies in
our SDSS-\emph{GALEX} sample using the same $u-r$ versus $M_{r}$
optical color-magnitude diagram as Baldry~\etal{}, and found
outstanding agreement.

\subsection{Evolution of the Stellar Mass Function \\ Since
  $\lowercase{z}\approx1$}\label{sec:mfevol} 

In the previous section we established the \smf{} at $z\approx0.1$
using our SDSS-\emph{GALEX} sample.  Here, we measure how the stellar
mass distribution of all, quiescent, and star-forming galaxies has
changed since $z\approx1$.

\subsubsection{All Galaxies}\label{sec:mfall}

\begin{figure*}
\centering \includegraphics[scale=0.6,angle=90]{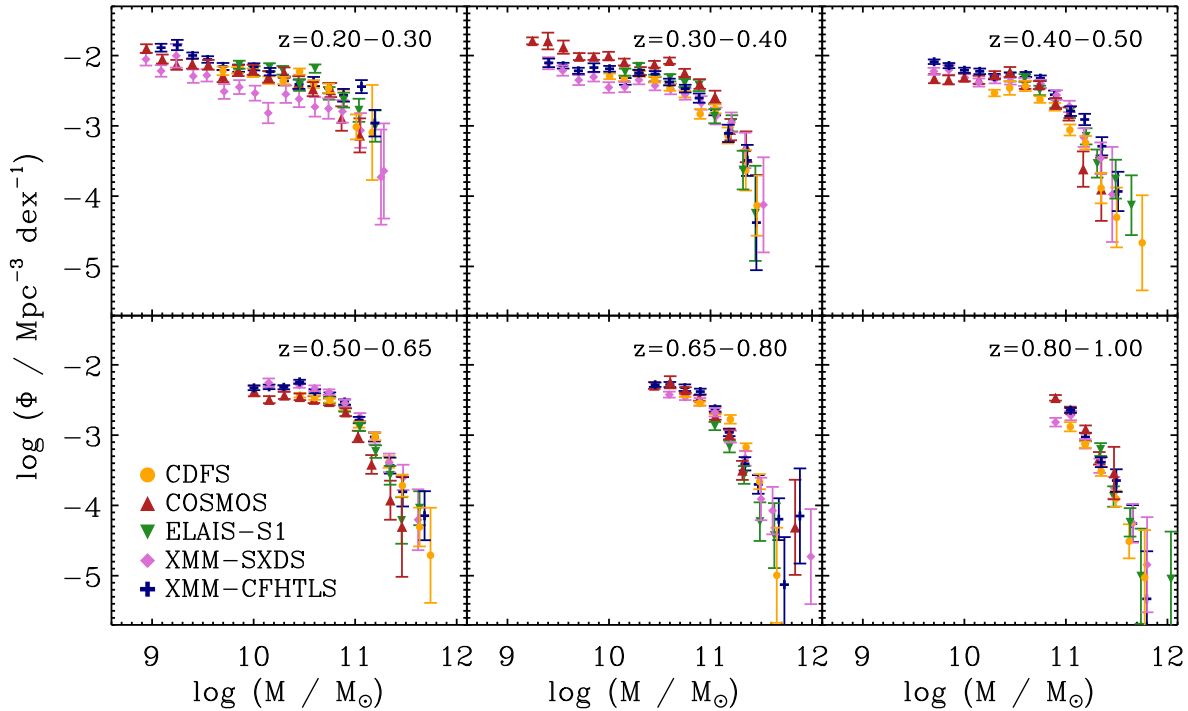}
\caption{Comparison of the \smf s in each of the five individual
  PRIMUS fields in six redshift bins from $z=0.2-1.0$.  The error bars
  reflect the statistical (Poisson) uncertainty in each stellar mass
  bin, and for clarity we only show the \smf{} in each field above our
  stellar mass completeness limit (see Section~\ref{sec:masslim}).
  This comparison demonstrates the overall consistency of the \smf s
  across the five fields, modulo expected differences due to sample
  variance.
\label{fig:mfbyfield}}
\end{figure*}

\begin{figure}[b]
\centering
\includegraphics[scale=0.4]{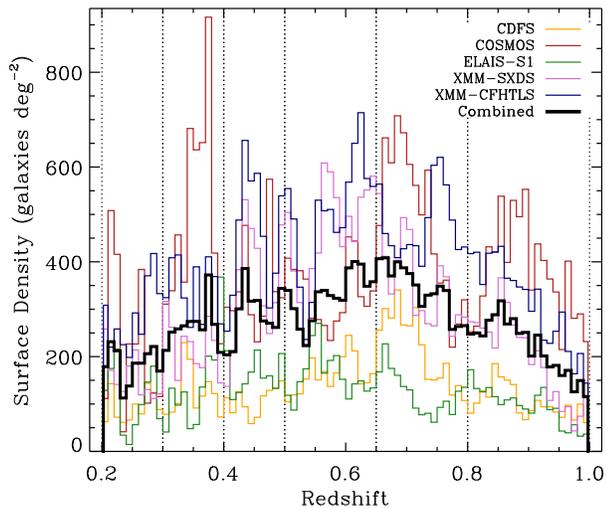}
\caption{Projected surface density of galaxies vs.~redshift in the
  five individual PRIMUS fields as indicated in the legend, and for
  the combined sample (thick black histogram).  The vertical dotted
  lines indicate the boundaries of the six redshift bins we have
  adopted.  The variation in the number density of galaxies in each
  field and redshift interval bin due to large-scale structure (sample
  variance) is striking.  The largest overdensities are in the COSMOS
  field at $z\approx0.35$, $\approx0.7$, and $\approx0.85$; in the
  XMM-CFHTLS field at $z\approx0.45$, $\approx0.6$, and $\approx0.75$;
  and in our CDFS field at $z\approx0.7$.  Note that the overall
  surface density in our CDFS and ELAIS-S1 fields is lower at all
  redshifts because of the additional $3.6$~\micron{} flux limit
  imposed in these fields (see
  Section~\ref{sec:parent}).  \label{fig:zhist}}
\end{figure}

We begin by comparing the \smf s in the five individual PRIMUS fields.
In Figure~\ref{fig:mfbyfield} we plot the \smf s in $0.15$~dex wide
bins of stellar mass divided into six redshift bins from $z=0.2-1.0$
centered on $\zmean=0.25$, $0.35$, $0.45$, $0.575$, $0.725$, and
$0.9$.  We choose these redshift bins because they correspond to
roughly equal $\sim0.9$~Gyr intervals of cosmic time.  For clarity we
only plot the portion of each \smf{} above our stellar mass
completeness limit in each field (see Section~\ref{sec:masslim}).  

We find good overall agreement among the individual \smf s, modulo
expected deviations due to sample variance.  In Figure~\ref{fig:zhist}
we illustrate the effects of large-scale structure explicitly by
plotting the differential surface density of galaxies brighter than
$i\approx23$ versus redshift in each of our five fields, and for our
combined PRIMUS sample.\footnote{Recall that an $[{\rm 3.6}]<21$ flux
  cut was also applied to the CDFS and ELAIS-S1 samples (see
  Section~\ref{sec:parent}).}  We find significant overdensities in
the COSMOS field at $z\approx0.35$, $\approx0.7$, and $\approx0.85$
\citep[see also][]{lilly09a, kovac10a}; in our XMM-CFHTLS field at
$z\approx0.45$, $\approx0.6$, and $\approx0.75$; and in our CDFS field
at $z\approx0.7$.  By constructing the area-weighted average of all
five fields (thick black histogram) we are able to reduce these
field-to-field variations significantly, although the effects of
sample variance are still apparent.  Indeed, in this and subsequent
sections we show that even with five independent fields covering
$\approx5.5$~deg$^{2}$, sample variance frequently limits the
precision with which we can constrain the evolution of the \smf{}
since $z=1$.  This conclusion is particularly sobering when one
considers that all previous analyses of the \smf{} at intermediate
redshift which utilized spectroscopic redshifts have been based on
samples covering at most $1-2$~deg$^{2}$.  In any case, in the
remainder of this paper we analyze the \smf{} constructed from the
area-weighted average of all five fields, and use the jackknife
technique described in Section~\ref{sec:method} to empirically
estimate the uncertainty in the \smf{} due to sample variance.

\begin{figure*}
\centering
\includegraphics[scale=0.65,angle=90]{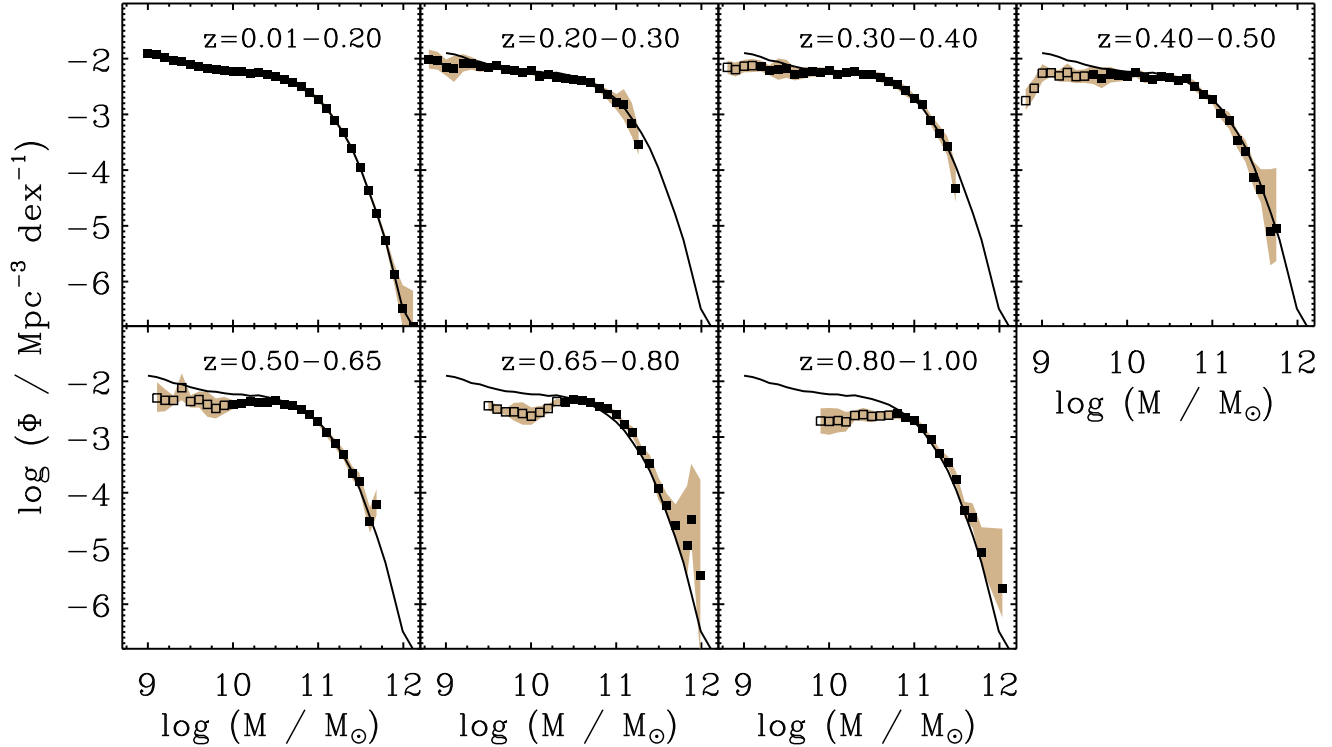}
\caption{Evolution of the \smf{} from $z=0-1$.  The black squares show
  the comoving number density of galaxies in $0.1$~dex wide bins of
  stellar mass based on our SDSS-\emph{GALEX} (upper-left panel) and
  PRIMUS samples (subsequent six panels), respectively.  Filled (open)
  squares indicate stellar mass bins above (below) the stellar mass
  completeness limit at the center of each redshift bin.  The shaded
  tan region in each panel reflects the quadrature sum of the Poisson
  and sample variance uncertainties in the \smf, and the solid curve,
  reproduced in every panel for reference, shows the SDSS-\emph{GALEX}
  \smf.  We find that the \smf{} for the ensemble population of
  galaxies has evolved remarkably little over the range of redshifts
  and stellar masses where PRIMUS is complete.  \label{fig:mfall}} 
\end{figure*}

\begin{figure*}
\centering
\includegraphics[scale=0.65,angle=90]{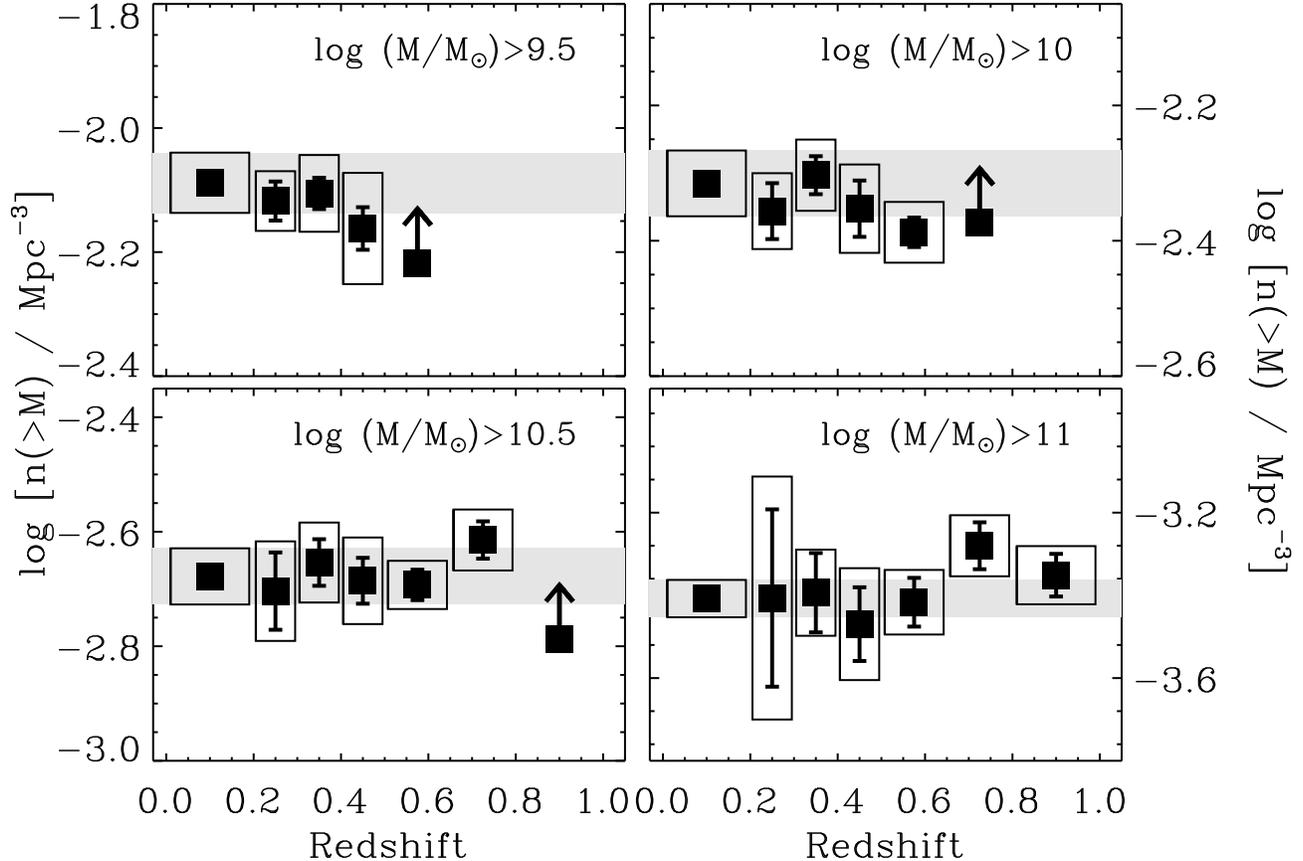}
\caption{Evolution of the cumulative comoving space density of
  galaxies more massive than (top-left) $10^{9.5}$~\msun, (top-right)
  $10^{10}$~\msun, (bottom-left) $10^{10.5}$~\msun, and (bottom-right)
  $10^{11}$~\msun{} from $z=0-1$ based on the \smf s presented in
  Figure~\ref{fig:mfall}.  The error bars reflect the Poisson
  uncertainty on each number density measurement, and the thin black
  boxes reflect the quadrature sum of the Poisson and sample variance
  uncertainties in the vertical direction, and the redshift bin width
  in the horizontal direction.  We designate lower limits using
  upward-pointing arrows.  The grey shaded region in each panel
  shows---as the null-evolution hypothesis---the number density of
  galaxies at $z\approx0.1$ based on our SDSS-\emph{GALEX} sample.  We
  find that the cumulative number density of galaxies with
  $\mass>10^{9.5}$~\msun{} and $\mass>10^{10}$~\msun{} has increased
  by $15\%\pm10\%$ and $21\%\pm19\%$ since $z=0.4$ and $z=0.6$,
  respectively, while the cumulative space density of
  $\mass>10^{10.5}$~\msun{} and $\mass>10^{11}$~\msun{} galaxies has
  changed by just $4.7\%\pm12\%$ and $11\%\pm17\%$ since $z=0.8$ and
  $z=1$, respectively.  
\label{fig:numden_all}}
\end{figure*}

\begin{figure*}
\centering
\includegraphics[scale=0.65,angle=90]{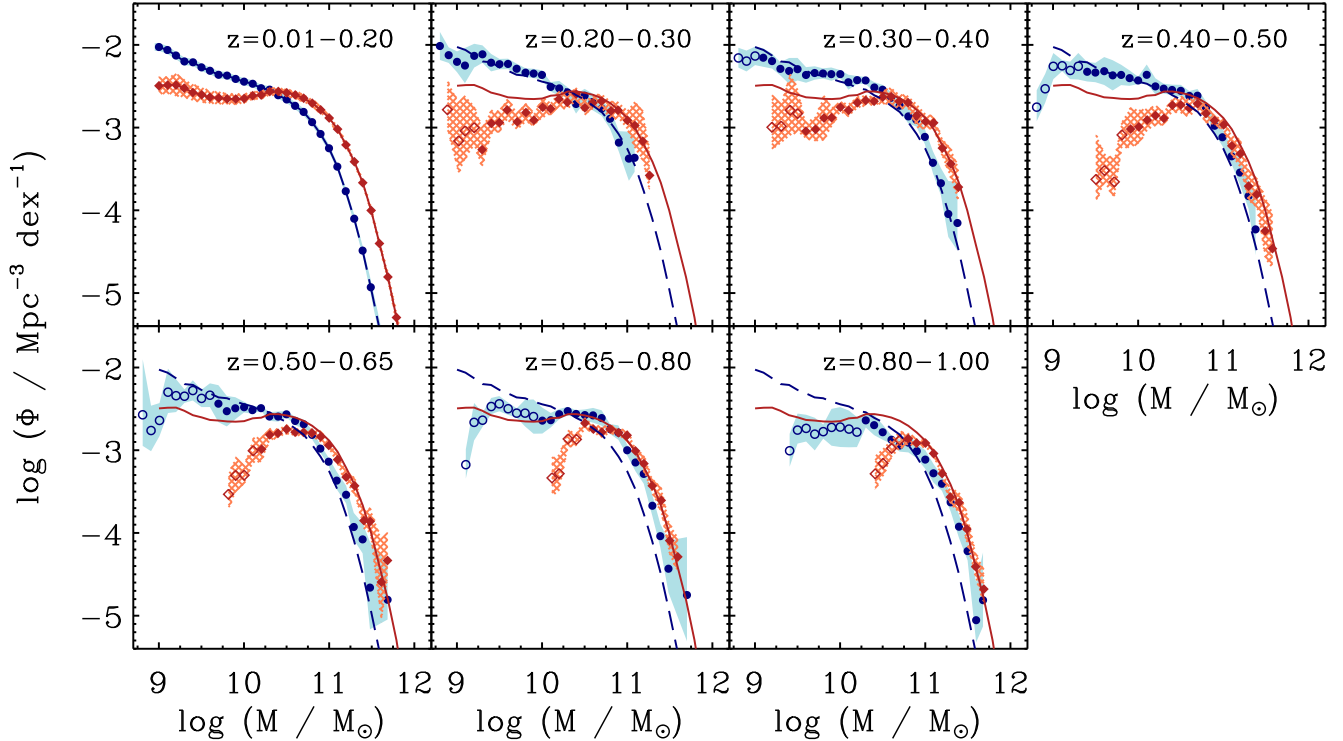}
\caption{Evolution of the \smf s of quiescent (dark red diamonds and
  light red hatched region) and star-forming (dark blue points and
  light blue shaded region) galaxies from $z=0-1$.  Filled (open)
  symbols correspond to stellar mass bins above (below) our stellar
  mass completeness limit in each redshift interval.  The dashed blue
  and solid red curves are the SDSS-\emph{GALEX} star-forming and
  quiescent-galaxy \smf s (upper-left panel), and have been reproduced
  in every panel for reference.  We find a significant increase in the
  number of intermediate-mass ($\sim10^{10}$~\msun) quiescent galaxies
  toward lower redshift, but essentially no change in the \smf{} of
  quiescent galaxies above $\sim10^{11}$~\msun.  Meanwhile, the \smf{}
  of star-forming galaxies is largely invariant below
  $\sim10^{11}$~\msun{} (at least where PRIMUS is complete), but
  exhibits significant evolution above $\sim10^{11}$~\msun{} by
  shifting toward lower stellar mass at fixed number density with
  decreasing redshift.  \label{fig:mfqqsf}}
\end{figure*}

In Figure~\ref{fig:mfall} we plot the \smf{} of all galaxies in seven
redshift bins from $z=0-1$ using our combined SDSS-\emph{GALEX} and
PRIMUS samples (see Table~\ref{table:mfevol}).  The black squares
reflect the comoving number density of galaxies in $0.1$~dex wide bins
of stellar mass, and the filled (open) symbols correspond to stellar
mass bins above (below) our completeness limit in each redshift
interval.  The tan shaded region shows the total (Poisson plus sample
variance) uncertainty in the \smf, and the solid curve in each panel
shows the SDSS-\emph{GALEX} \smf{} as the null-evolution hypothesis.

Examining Figure~\ref{fig:mfall}, we find strikingly little evolution
in the \smf{} for the global galaxy population since $z\approx1$, at
least over the range of stellar masses where PRIMUS is complete.  In
every redshift bin the observed \smf{} lies very close to the local
\smf, a result which we quantify below.  Given the expected stellar
mass growth of galaxies due to star formation (see, e.g.,
Figure~\ref{fig:sfr}) and galaxy mergers, this result may at first
appear surprising.  However, in Section~\ref{sec:mfqqsf}, we show that
the lack of significant evolution in the \smf{} for the ensemble
galaxy population is a consequence of how the \smf s of the
star-forming and quiescent galaxy populations separately evolve.
Moreover, in Section~\ref{sec:mergers} we show that the relative lack
of evolution in the global \smf, especially at the massive end,
suggests that mergers do not appear to play a significant role for the
stellar mass growth of galaxies at $z<1$.

By integrating the \smf{} above various stellar mass thresholds we can
quantify the observed (lack of) evolution in the \smf, and look for
evidence of mass assembly downsizing within the global galaxy
population (see Section~\ref{sec:intro}).  In
Figure~\ref{fig:numden_all} we plot versus redshift the cumulative
number density of galaxies with stellar masses greater than
$10^{9.5}$, $10^{10}$, $10^{10.5}$, and $10^{11}$~\msun.  We focus
here on the number density evolution, although the evolution in
stellar mass density leads to the same basic conclusions (see
Table~\ref{table:numden_all}).  We integrate the observed \smf{}
directly, but exclude stellar mass bins containing fewer than three
galaxies where the \smf{} is noisiest.  We use the best-fitting
Schechter model to extrapolate the observed \smf{} as needed over
small intervals of stellar mass.  We emphasize, however, that these
model-dependent corrections typically modify the measured number
densities by $\lesssim0.02$~dex, and therefore potential errors in the
extrapolations do not affect any of our conclusions.  The solid black
squares in Figure~\ref{fig:numden_all} show the mean number density,
while the vertical error bars indicate the Poisson uncertainty; the
thin black boxes around each point indicate the quadrature sum of the
Poisson and sample variance uncertainties in the vertical direction,
and the redshift bin width in the horizontal direction.  This
graphical representation shows that sample variance uncertainties are
frequently comparable to or larger than the Poisson uncertainties.
Finally, symbols with upward-pointing arrows represent lower limits,
and the grey shaded region shows for reference the comoving number
density of galaxies at $z\approx0.1$, to illustrate the case of no
evolution.

Figure~\ref{fig:numden_all} shows that the cumulative number density
of galaxies above all four stellar mass thresholds does not appear to
change significantly over the range of redshifts where PRIMUS is
complete.  To quantify this result, we fit a power-law function of
redshift, $n\propto(1+z)^{\gamma}$, to the measured densities,
excluding lower limits.  We scale the formal statistical uncertainties
by $\sqrt{\chi^{2}_{\nu}}$, the square root of the $\chi^{2}$
statistic divided by the number of degrees-of-freedom, in order to be
able to intercompare the significance of the evolutionary trends
across all four stellar mass thresholds.  We find
$\gamma=-0.43\pm0.3$, $-0.40\pm0.4$, $0.17\pm0.4$, and $0.32\pm0.4$
for the evolution in the cumulative number of galaxies more massive
than $10^{9.5}$, $10^{10}$, $10^{10.5}$, and $10^{11}$~\msun,
respectively.  Above the two highest stellar mass thresholds, the
measured number densities at $z\approx0.7$, and to a lesser extent at
$z\approx0.9$, are clearly affected by the large-scale overdensities
in several of the PRIMUS fields (see Figure~\ref{fig:zhist}).
Therefore, excluding the $z\approx0.7$ redshift bin, we obtain
$\gamma=-0.08\pm0.2$ and $0.16\pm0.3$ above $10^{10.5}$ and
$10^{11}$~\msun, respectively.  Stated another way, the cumulative
number of $\mass>10^{9.5}$~\msun{} and $\mass>10^{10}$~\msun{}
galaxies has increased by $15\%\pm10\%$ and $21\%\pm19\%$ since
$z=0.4$ and $z=0.6$, respectively.  Meanwhile,the cumulative space
density of $\mass>10^{10.5}$~\msun{} and $\mass>10^{11}$~\msun{}
galaxies has remained relatively constant, changing by just
$4.7\%\pm12\%$ and $11\%\pm17\%$ since $z=0.8$ and $z=1$,
respectively.

Thus, while we find hints of mass assembly downsizing---a more rapid
increase in the number of lower-mass galaxies toward low
redshift---the trends are only marginally significant.  By contrast,
previous studies have found much stronger evidence for downsizing
within the global galaxy population (see, e.g., \citealt{fontana06a,
  perez-gonzalez08a, pozzetti07a, pozzetti10a}).  For example,
\citet{pozzetti10a} report a $32\%\pm6\%$ increase in the cumulative
number of $\mass>10^{9.5}$~\msun{} galaxies since $z=0.44$, and no
statistically significant change ($7\%\pm17\%$) in the number density
of galaxies more massive than $10^{11}$~\msun{} since $z=1$ based on
an analysis of the COSMOS field.  While our findings are qualitatively
consistent with these studies---many of which pushed further down the
stellar mass function at higher redshift, and therefore had a larger
lever arm with which to detect downsizing---we also find that sample
variance can wash out the significance of the observed trends
\citep[see also][]{fontanot09a}.  In any case, we will show in
Section~\ref{sec:mfqqsf} that by measuring the change in the number
density of galaxies within fixed bins of stellar mass (as opposed to
using stellar mass thresholds) the signatures of downsizing will
become more apparent.

\subsubsection{Quiescent and Star-Forming Galaxies}\label{sec:mfqqsf}

In the previous section we measured the evolution of the \smf{} for
the ensemble population of galaxies from $z=0-1$.  We found only a
marginally significant increase ($16\%\pm9\%$) in the cumulative
number density of all $\mass\gtrsim10^{9.5}$~\msun{} galaxies since
$z\approx0.6$, and very little change ($8\%\pm10\%$) in the
space-density of $\mass\gtrsim10^{10.5}$~\msun{} galaxies since
$z\approx1$.  Here, we investigate these results in more detail by
dividing our sample into quiescent and star-forming galaxies based on
the criteria defined in Section~\ref{sec:select}.

\begin{figure*}
\centering
\includegraphics[scale=0.65,angle=90]{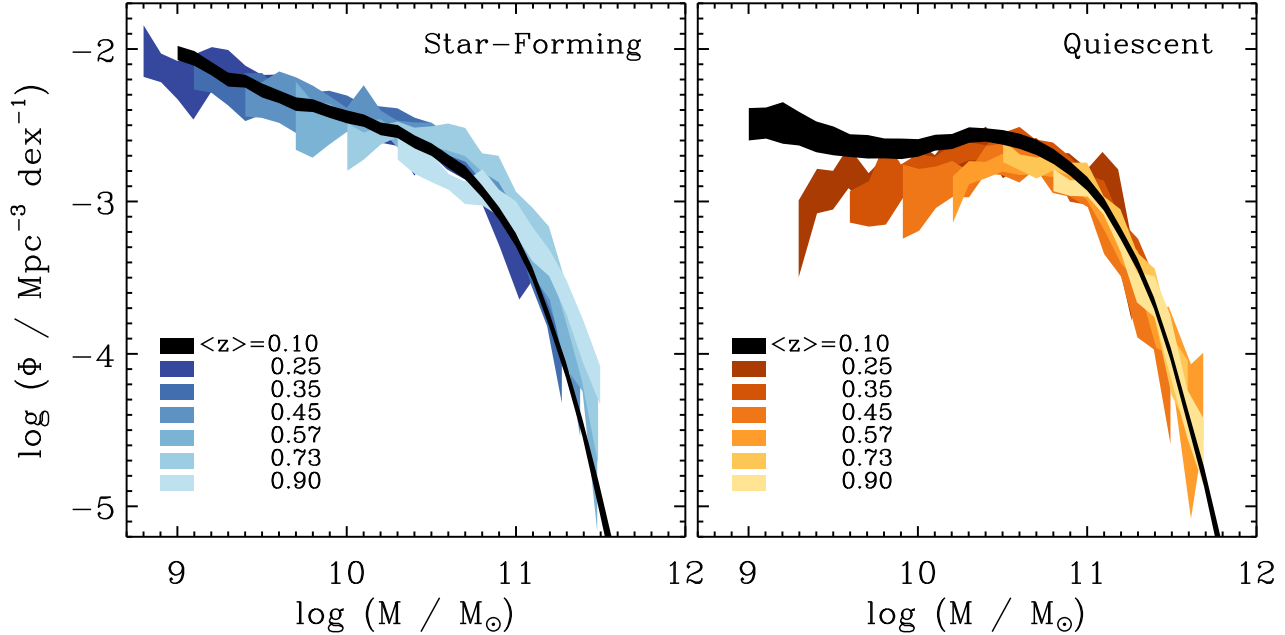}
\caption{Evolution of the \smf s of (left) star-forming and (right)
  quiescent galaxies from $z=0-1$ based on the data presented in
  Figure~\ref{fig:mfqqsf}.  We use progressively lighter shades of
  blue to show the evolution of the star-forming galaxy \smf, and
  shades of orange to show how the \smf{} of quiescent galaxies has
  evolved.  The black shaded region in each panel shows the
  corresponding SDSS-\emph{GALEX} \smf.  \label{fig:mfqqsf_one}}
\end{figure*}

\begin{figure*}
\centering
\includegraphics[scale=0.65,angle=90]{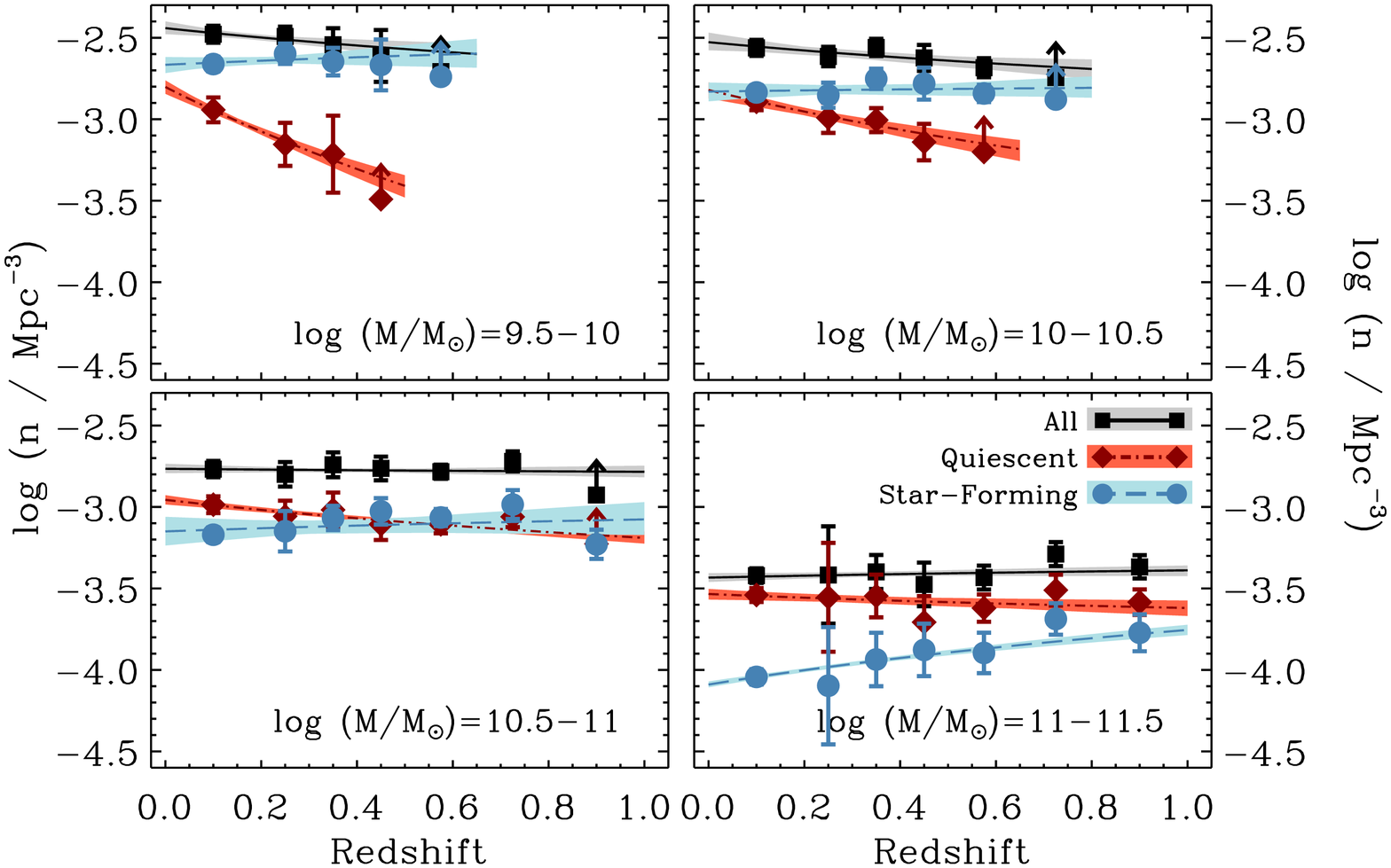}
\caption{Evolution in the number density of all (black squares),
  quiescent (red diamonds), and star-forming (blue points) galaxies in
  four $0.5$~dex wide intervals of stellar mass ranging from
  $10^{9.5}-10^{10}$~\msun{} in the upper-left panel, to
  $10^{11}-10^{11.5}$~\msun{} in the lower-right panel.  The error bar
  on each measurement is due to the quadrature sum of the Poisson and
  sample variance uncertainties in each redshift interval; we denote
  lower limits on the number density using upward-pointing arrows.
  The solid black, dot-dashed red, and dashed blue lines show weighted
  linear least-squares fits to the data, and the corresponding shaded
  regions show the $1\sigma$ range of power-law fits drawn from the
  full covariance matrix.  We find a factor of $\sim2-3$ increase in
  the number density of $\sim10^{9.5}-10^{10.5}$~\msun{} quiescent
  galaxies since $z\approx0.5$, and a remarkably little change
  ($-8\%\pm10\%$) in the space density of star-forming galaxies over
  the same range of stellar mass and redshift.  Between
  $10^{10.5}-10^{11}$~\msun, the space density of quiescent galaxies
  increases by $58\%\pm9\%$, while the number density of star-forming
  galaxies declines by $-13\%\pm23\%$.  Meanwhile, above
  $10^{11}$~\msun{} we find a steep $54\%\pm7\%$ decline in the number
  of massive star-forming galaxies since $z\approx1$, and a small
  increase ($22\%\pm12\%$) in the space-density of comparably massive
  quiescent galaxies.  The distinct evolutionary trends exhibited by
  star-forming and quiescent galaxies conspire to keep the number
  density of \emph{all} galaxies relatively constant over the range of
  stellar masses and redshifts probed by
  PRIMUS.  \label{fig:numbymass}}
\end{figure*}

In Figure~\ref{fig:mfqqsf} we plot the \smf s of quiescent and
star-forming galaxies in seven redshift bins from $z=0-1$.  Filled
(open) symbols indicate stellar mass bins above (below) our
completeness limit at the center of each redshift bin, and the solid
red and dashed blue curves, reproduced in every panel for reference,
show the SDSS-\emph{GALEX} quiescent and star-forming galaxy \smf s,
respectively.  In Section~\ref{sec:litcompare} we compare our
type-dependent \smf s against published measurements and show that
they are broadly consistent with previous studies.

Figure~\ref{fig:mfqqsf} shows that the strong stellar mass dependence
of galaxy bimodality observed among local galaxies persists over the
full range of redshifts probed by PRIMUS.  In other words, we find
that quiescent galaxies dominate the massive end
($\gtrsim10^{11}$~\msun) of the \smf, and star-forming galaxies
dominate among intermediate-mass ($\sim10^{10}$~\msun) galaxies at all
redshifts from $z=0-1$.  However, we also observe several striking
evolutionary trends.  Among quiescent galaxies, the number of
intermediate-mass galaxies increases dramatically toward the current
epoch, while the massive end of the \smf{} remains remarkably fixed.
Meanwhile, the largest changes in the \smf{} of star-forming galaxies
occur at the massive end.  We find a perceptible shift in the
star-forming galaxy \smf{} toward lower mass at fixed number density
with decreasing redshift, while the low-mass end of the \smf{} remains
relatively constant over the whole range of stellar masses and
redshifts where our sample is complete.  The so-called transition
mass---the stellar mass at which the quiescent and star-forming galaxy
\smf s cross---evolves roughly as $\propto(1+z)^{1.5}$, from
$\sim3\times10^{10}$~\msun{} at $z\approx0.1$ to
$\sim7\times10^{10}$~\msun{} at $z\approx0.9$, which agrees reasonably
well with previous measurements \citep[e.g.,][]{bundy06a, vergani08a,
  pozzetti10a}.  It is not clear, however, that the transition mass
has any physical interpretation, as Figure~\ref{fig:mfqqsf} shows that
its evolution is entirely driven by the rise in the number of
intermediate-mass quiescent galaxies \citep[e.g.,][]{borch06a,
  cattaneo08a}.

Another way to visualize these results is with
Figure~\ref{fig:mfqqsf_one}, which shows the individual \smf s from
all seven redshift bins on top of one another.  For clarity, we only
plot each \smf{} above our stellar mass completeness limit, and we
only show stellar mass bins containing three or more galaxies.  In the
left panel we use progressively lighter shades of blue to show the
evolution of the star-forming galaxy \smf, and in the right panel we
show the evolution of the quiescent-galaxy \smf{} using progressively
lighter shades of orange.  The black shaded region shows the
corresponding SDSS-\emph{GALEX} \smf, which we plot on top so that the
changes in the \smf{} with redshift can be more easily evaluated.
This figure clearly shows the significant steepening of the low-mass
end of the \smf{} of quiescent galaxies toward lower redshift, and the
simultaneous decline in the number of massive star-forming galaxies.

In Figure~\ref{fig:numbymass} we quantify the observed evolution by
plotting the integrated number density of galaxies measured in four
$0.5$~dex wide intervals of stellar mass between
$10^{9.5}-10^{11.5}$~\msun.  As in Section~\ref{sec:mfall}, we
calculate the number density by integrating the observed \smf s
directly, excluding stellar mass bins with fewer than three galaxies,
and use our Schechter model fits to extrapolate over small ranges of
stellar mass.  We plot the evolution in the number density of all,
quiescent, and star-forming galaxies using black squares, red
diamonds, and blue points, respectively, and indicate lower limits on
the number density in redshift bins where our \smf{} is partially
incomplete using upward-pointing arrows.  The error bars reflect the
quadrature sum of the Poisson and sample variance uncertainties.  We
list the derived number and stellar mass densities in
Table~\ref{table:numden_rhoden_bymass}.

We quantify the observed trends by fitting a power-law function of
redshift to the measured number densities, given by
\begin{equation}
n(z) = n_{0}(1+z)^{\gamma}.
\label{eq:numden_evol}
\end{equation}

\noindent In addition, we model the evolution in the stellar mass
density, $\rho(z)$, as
\begin{equation}
\rho(z) = \rho_{0}(1+z)^{\beta}.
\label{eq:rhoden_evol}
\end{equation}

\noindent In detail, we fit the data in $\log\,(n)-\log\,(1+z)$ and
$\log\,(\rho)-\log\,(1+z)$ space using weighted linear least-squares
minimization, and we only fit over the range of redshifts where our
measurements are complete (i.e., we ignore lower limits).  We also
exclude from the fits our measurements at $z\approx0.7$ because of the
above-average overdensity of galaxies in this redshift bin (see
Figure~\ref{fig:zhist}).  We emphasize, however, that including this
redshift bin would only strengthen our claim of minimal evolution in
the number density of massive galaxies; in other words, excluding the
$z\approx0.7$ measurements is a conservative choice.  The solid black,
dot-dashed red, and dashed blue lines in Figure~\ref{fig:numbymass}
show the results of fitting the number density of all, quiescent, and
star-forming galaxies, respectively, and the corresponding shaded
regions show the $1\sigma$ range of power-law fits drawn from the full
covariance matrix.  Table~\ref{table:numden_bymass_coeff} lists the
best-fitting coefficients and uncertainties, where the uncertainties
have been rescaled as in Section~\ref{sec:mfall} such that
$\chi^{2}_{\nu}=1$.

\begin{figure*}
\centering
\includegraphics[scale=0.65,angle=90]{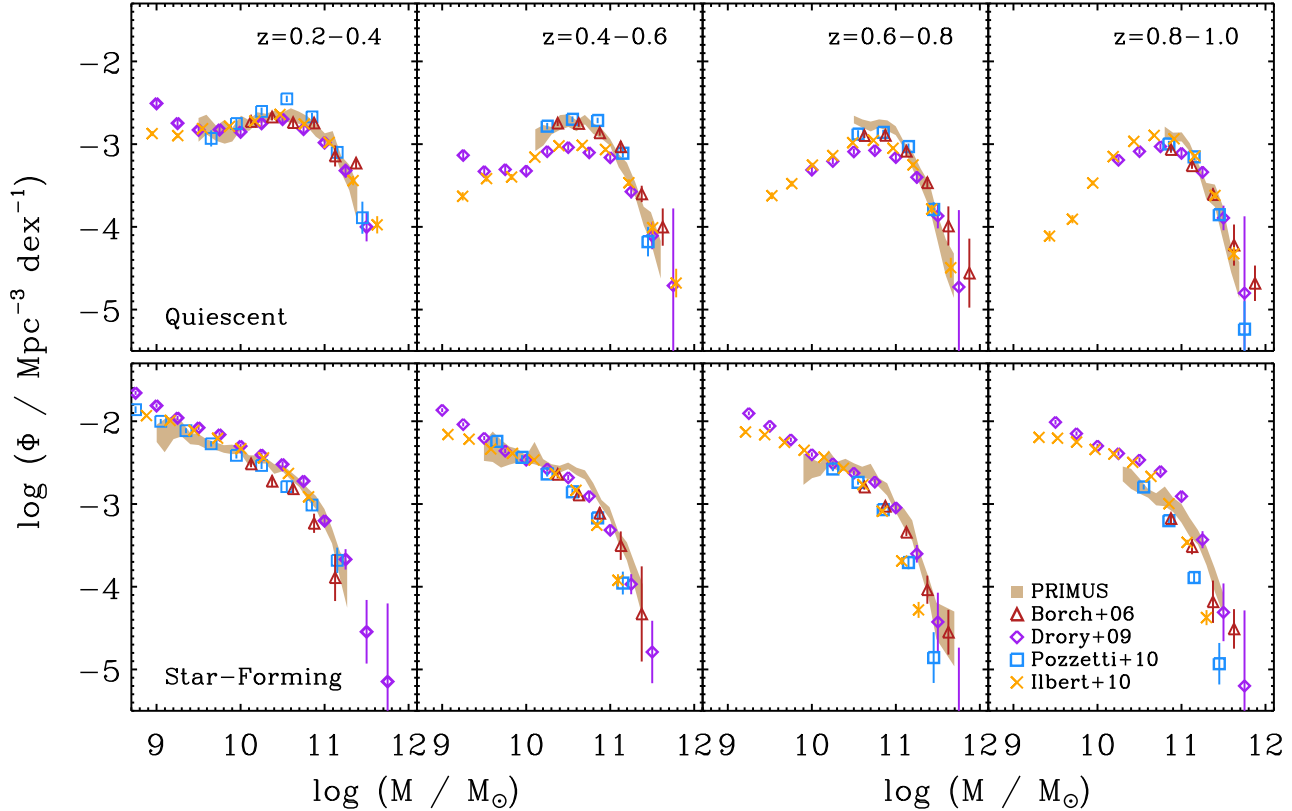}
\caption{Comparison of the \smf s for (top row) quiescent and (bottom
  row) star-forming galaxies in four redshift bins from $z=0.2-1.0$
  against previously measurements \smf s assembled from the
  literature.  We find that our \smf s are generally consistent with
  these previous studies.  \label{fig:mfqqsf_lit}} 
\end{figure*}

Figure~\ref{fig:numbymass} synthesizes nearly all the key results of
this section, and conveys many of the core conclusions of this paper.
First, we find that the number density of $10^{9.5}-10^{10}$~\msun{}
quiescent galaxies increases significantly toward lower redshift, by a
factor of $3.2\pm0.5$ since $z=0.4$, whereas the number density of
star-forming galaxies decreases marginally, by $-10\%\pm15\%$ over the
same redshift range.  Meanwhile, the number density of
$10^{10}-10^{10.5}$~\msun{} quiescent galaxies increases by a factor
of $2.2\pm0.4$ since $z=0.6$, while the number of comparably massive
star-forming galaxies changes by $-4\%\pm15\%$.  Finally, in the
$10^{10.5}-10^{11}$~\msun{} stellar masses bin we find a $58\%\pm9\%$
increase in the space density of quiescent galaxies since $z=0.8$, and
a $-13\%\pm23\%$ decrease in the number of star-forming galaxies over
the same redshift range.  Thus, we find remarkably little change
($-8\%\pm10\%$) in the number density of $10^{9.5}-10^{10.5}$~\msun{}
star-forming galaxies over the full range of redshifts where PRIMUS is
complete, and a gradual, but significant buildup in the population of
quiescent galaxies toward low redshift.  Moreover, we find that the
\emph{rate} at which the quiescent galaxy population builds up toward
low redshift increases steeply with \emph{decreasing} stellar mass.

Among the most massive galaxies in our sample, however,
Figure~\ref{fig:numbymass} reveals a striking inversion of the trends
seen at lower stellar mass.  Between $10^{11}-10^{11.5}$~\msun{} the
number density of quiescent galaxies increases by $22\%\pm12\%$ since
$z\approx1$, while the number of massive star-forming galaxies
\emph{declines} by $54\%\pm7\%$ over the same redshift range.  The
reason this destruction of massive star-forming galaxies (and presumed
transformation into quiescent systems) does not significantly affect
the space density of massive quiescent galaxies is because quiescent
galaxies vastly outnumber star-forming galaxies above
$\sim10^{11}$~\msun{} at all redshifts from $z=0-1$.  For example, at
$z=1$ quiescent galaxies outnumber $10^{11}-10^{11.5}$~\msun{}
star-forming galaxies by $\approx0.13$~dex ($\approx35\%$); therefore,
the $\approx55\%$ decline in the number of massive star-forming
galaxies can easily be subsumed into the quiescent galaxy population
by the current epoch.

Reviewing Figures~\ref{fig:mfqqsf}$-$\ref{fig:numbymass}, it is now
clear why the \smf{} for the global population of galaxies
(Figures~\ref{fig:mfall} and \ref{fig:numden_all}) evolves so little
since $z\approx1$ over the range of stellar masses where PRIMUS is
complete.  Between $10^{9.5}-10^{10.5}$~\msun{} the \smf{} is
dominated by star-forming galaxies, whose number density remains
relatively constant.  Meanwhile, among $\mass\gtrsim10^{10.5}$~\msun{}
galaxies the \smf{} becomes increasingly dominated by quiescent
galaxies, whose number density \emph{also} remains roughly constant
with redshift.  Consequently, the bimodal nature of the galaxy \smf{}
combined with the distinct evolutionary trends exhibited by
star-forming and quiescent galaxies conspire to keep the \smf{} for
the global population of galaxies from changing significantly at these
redshifts.

In Section~\ref{sec:mfall} we found hints of differential evolution in
the \smf{} of all galaxies based on stellar mass-threshold samples,
but the results were not very significant.  Do we find more
significant evidence for mass assembly downsizing based on the number
densities derived within fixed-interval bins of stellar mass?  Our
power-law fits to the black squares in Figure~\ref{fig:numbymass} (see
also Table~\ref{table:numden_bymass_coeff}) indicate a $28\%\pm11\%$
increase in the space density of all $10^{9.5}-10^{10}$~\msun{}
galaxies since $z=0.4$, and a $35\%\pm14\%$ increase in the number
density of $10^{10}-10^{10.5}$~\msun{} galaxies since $z=0.6$.  By
contrast, among $10^{10.5}-10^{11}$~\msun{} galaxies the number
density increases by $4\%\pm9\%$ since $z=0.8$, and declines by
$-10\%\pm9\%$ for $10^{11}-10^{11.5}$~\msun{} galaxies since $z=1$.
Thus, we \emph{do} find evidence for mass assembly downsizing---a
continued buildup of the low- and intermediate-mass galaxy population
toward low redshift, and no significant changes in the space density
of massive galaxies---within the global galaxy population.  However,
with the benefit of hindsight we now see that these relatively subtle
evolutionary trends are being driven entirely by the much more
significant evolutionary trends separately exhibited by the population
of quiescent and star-forming galaxies.

To summarize, we have shown that the evolution of the \smf s of both
quiescent and star-forming galaxies depends sensitively on stellar
mass.  Above $\mass\sim10^{11}$~\msun{} quiescent galaxies dominate
the galaxy population at all redshifts, and their number density
changes relatively little since $z=1$, whereas the number of
star-forming galaxies declines precipitously toward lower redshift.
Between $10^{9.5}-10^{10.5}$~\msun, on the other hand, star-forming
galaxies vastly outnumber quiescent galaxies, and their number density
changes by just $-8\%\pm10\%$ between $z\approx0.6$ and $z\approx0$.
Meanwhile, the number of $10^{9.5}-10^{10.5}$~\msun{} quiescent
galaxies increases significantly since $z\approx0.6$ at a rate that
accelerates with decreasing stellar mass.  Finally, the stellar mass
range between $10^{10.5}-10^{11}$~\msun{} marks the transition regime
between the dominance of star-forming galaxies and the rise of
quiescent galaxies at low mass, and the dominance of quiescent
galaxies and the decline of the star-forming population at large
stellar mass.

\subsection{Comparison With Previous Studies}\label{sec:litcompare}

A detailed quantitative comparison of our results against previous
studies is challenging for several reasons.  First, previous studies
have used a wide variety of techniques to divide the galaxy population
into ``quiescent'' and ``star-forming'' galaxies; however, many of
these techniques result in a quiescent population that is highly
contaminated by dusty star-forming galaxies \citep[e.g.,][]{maller09a,
  zhu11a}, which can severely bias the inferred \smf{} in
mass-dependent ways.  Second, systematic differences in stellar mass
estimates due to different prior assumptions and population synthesis
models can significantly affect the inferred \smf{} (e.g.,
\citealt{marchesini09a, kajisawa09a}; see also
Appendix~\ref{appendix:syst}) And finally, many previous studies have
neglected the effects of sample variance, and therefore have
underestimated the statistical uncertainties of their results.
Nevertheless, we can still perform a rudimentary comparison of our
quiescent and star-forming galaxy \smf s against previous measurements
assembled from the literature.

To facilitate this comparison, we recompute our \smf s using four
broader redshift bins with $\Delta z=0.2$ (due to the typically
smaller area and sample size of these analyses) centered on
$\zmean=0.3$, $0.5$, $0.7$, and $0.9$.  In Figure~\ref{fig:mfqqsf_lit}
we plot the \smf s for (top row) quiescent and (bottom row)
star-forming galaxies from PRIMUS as a tan shaded region, reflecting
the quadrature sum of the Poisson and sample variance uncertainties in
each redshift bin.  We compare these results to the \smf s published
by \citet{borch06a} (red triangles), \citet{drory09a} (purple
diamonds), \citet{pozzetti10a} (blue squares), and \citet{ilbert10a}
(orange crosses), accounting for differences in the adopted Hubble
constant and IMF.  For reference, the \smf s published by
\citet{drory09a}, \citet{ilbert10a}, and \citet{pozzetti10a} are all
based on the $\sim2$~deg$^{2}$ COSMOS field, while \citet{borch06a}
analyzed the three COMBO-17 \citep{wolf03a} fields, totaling
$\sim0.8$~deg$^{2}$.

Examining Figure~\ref{fig:mfqqsf_lit}, we find reasonably good
agreement between our \smf s and the literature, modulo expected
differences due to sample variance in PRIMUS (see, e.g.,
Figure~\ref{fig:zhist}) and the other reasons outlined above.  Among the
largest discrepancies are in the $\zmean=0.5$ and $\zmean=0.7$
redshift bins for quiescent galaxies, which shows that our \smf{}
agrees with the \smf s derived by \citet{borch06a} and
\citet{pozzetti10a}, but disagrees noticeably with the
\citet{drory09a} and \citet{ilbert10a} \smf s.  We also find a
somewhat higher number density of $\sim10^{10.5}-10^{11}$~\msun{}
star-forming galaxies at $\zmean=0.5$.  Overall, however, we conclude
that our results are consistent with previous measurements of the
\smf{} at intermediate redshift. 

\section{Discussion}\label{sec:discussion}  

We have measured the evolution of the \smf{} since $z\approx1$ of
quiescent and star-forming galaxies using PRIMUS, one of the largest
spectroscopic surveys of intermediate-redshift galaxies ever
conducted.  Our goals have been to characterize the stellar mass
growth of each population, and to measure the rate at which
star-forming galaxies are being quenched as a function of stellar mass
and redshift.  Compared to many previous studies, our analysis has
benefited from a large, statistically complete sample of faint
galaxies ($\sim40,000$ galaxies to $i\approx23$) spread across five
widely separated fields totaling $\approx5.5$~deg$^{2}$, and a
well-defined local SDSS-\emph{GALEX} comparison sample.  With these
data, we have been able to study the detailed evolution of the \smf{}
over a large dynamic range of stellar mass and redshift in a
consistent way, with well-quantified sample variance uncertainties.

We find that the evolution of the \smf s of both quiescent and
star-forming galaxies depends acutely on stellar mass, but in very
different ways (see Figure~\ref{fig:numbymass}).  Among quiescent
galaxies, the number of intermediate-mass ($\sim10^{10}$~\msun)
galaxies increases by a factor of $\sim2-3$ since $z\approx0.5$, but
remains approximately constant for massive ($\gtrsim10^{11}$~\msun)
galaxies since $z\approx1$.  By contrast, the most significant
evolutionary trends for star-forming galaxies occur above
$\sim10^{11}$~\msun.  Specifically, we find no significant change in
the number density of intermediate-mass galaxies, and a $\approx55\%$
\emph{decrease} in the number of massive star-forming galaxies since
$z\approx1$.

These galaxy-type dependent trends conspire rather remarkably to make
the \smf{} for the global galaxy population \emph{appear} to not have
changed significantly since $z\approx1$, at least over the range of
stellar masses and redshifts probed by PRIMUS
(Figure~\ref{fig:mfall}).  One implication of these results is that an
analysis of the global galaxy population by itself would yield a
highly incomplete view of galaxy evolution because it would mask the
rich interplay between the coevolution of star-forming and quiescent
galaxies.  

In the next two sections we synthesize our results to investigate the
effect of mergers on the stellar mass growth of galaxies, and to
quantify the stellar mass dependence of star formation quenching from
$z=0-1$.

\subsection{Constraints on the Stellar Mass Growth of Galaxies by
  Mergers}\label{sec:mergers}

The stellar mass of an individual galaxy can change by forming new
stars, or by merging with other galaxies.  By accounting for the
stellar mass growth of galaxies by {\em in situ} star formation, the
redshift evolution of the \smf{} (i.e., the derivative of the \smf{}
with respect to cosmic time) in principal can be used to constrain the
growth of galaxies by mergers \citep[e.g.,][]{drory08a, walcher08a,
  conroy09b, pozzetti10a}.

Our finding that the \smf{} of the global galaxy population evolves
relatively little between $z\approx1$ and $z\approx0$ (see
Figure~\ref{fig:mfall}) suggests that mergers play a subdominant role
for the stellar mass growth of galaxies at these redshifts.  To
quantify this result, we use the measured SFR of each galaxy (see
Section~\ref{sec:mass}) to estimate how much their stellar mass will
increase by {\em in situ} star formation.  Specifically, we compute
for each galaxy of a given stellar mass, $\mass$, a new stellar mass,
$\mass^{\prime}$, given by
\begin{equation}
\mass^{\prime}(z^{\prime}) = \mass(z) +
(1-\mathcal{R})\times(\sfr\Delta t),
\end{equation}

\noindent where $\mathcal{R}$ is the {\em return fraction}, the
stellar mass returned (assumed instantaneously) to the interstellar
medium by supernovae and stellar winds, \sfr{} is the SFR in
\sfrunits, and $\Delta t\equiv t(z^{\prime})-t(z)$ is the elapsed
cosmic time (increase in the age of the Universe) in years between
redshift $z$ and $z^{\prime}$, where $z^{\prime}<z$.  We adopt
$\mathcal{R}\approx0.5$, which is appropriate for the
\citet{chabrier03a} IMF \citep{leitner11a}, and make the simplifying
assumption that the SFR is constant over the time interval $\Delta t$.
Although the SFRs of most star-forming galaxies at these redshifts are
declining with decreasing redshift (see, e.g., Figure~\ref{fig:sfr};
\citealt{noeske07a}), $\Delta t$ in our analysis is sufficiently short
($\lesssim1$~Gyr) for all but the last redshift bin (where $\Delta t
\approx1.6$~Gyr), that incorporating a more detailed star formation
history into our calculation would not significantly change our
results.  Moreover, recall that our UV-based SFRs trace star formation
over the last $\sim100$~Myr, which is reasonably well matched to the
$\Delta t$ timescale.

Using this formalism, we can use the \emph{observed} \smf{} at
redshift $z$ to \emph{predict} what the \smf{} will look like at a
(lower) redshift $z^{\prime}$ if galaxies grow by star formation alone
(i.e., assuming mergers do not occur).  Note that although {\em in
  situ} star formation conserves the \emph{total} number of galaxies,
the number of galaxies of a given stellar mass can increase or
decrease because the SFR varies with stellar mass (see, e.g.,
Figure~\ref{fig:sfr}).  After controlling for star formation growth,
any residual change in the number density of galaxies of a given
stellar mass must be due to merging.  We emphasize that our
measurement complements, but is only implicitly related to
measurements of the major and minor {\em merger rate} \citep[see,
  e.g.,][and references therein]{lotz11a}.  For example, the technique
we use only reveals whether mergers have a \emph{net} effect on the
\smf, but cannot be used to infer the underlying stellar mass
distribution of galaxies that are merging (see \citealt{drory08a} for
more details).  Another caveat regarding this technique is its
implicit assumption that mergers retain \emph{all} the stellar mass
involved in the merger, when it is likely that a non-negligible
fraction of that mass is dispersed to large radii to form the diffuse
stellar component (DSC) of groups and clusters
\citep[e.g.,][]{murante07a}.  In fact, our precise measurement of the
evolution of the \smf{} above $\sim10^{11}$~\msun{} could be turned
around to help constrain this fraction in the context of the
cosmological growth of dark-matter halos \citep[e.g.,][]{monaco06a,
  behroozi12a}.  

With the preceding discussion in mind, we define the fractional {\em
  merger growth rate}, $\mathcal{G}(\mass,z)$, as
\begin{equation}
\mathcal{G}(\mass,z) \equiv \left. \frac{1}{n}\frac{\Delta n}{\Delta
  t}\right|_{\rm mergers} = \frac{1}{\Delta t(z,z^{\prime})} \left[1-\frac{n^{\rm
    pred}(\mass,z)}{n^{\rm obs}(\mass,z)}\right],  
\label{eq:mergerrate}
\end{equation}

\noindent where $n^{\rm obs}$ and $n^{\rm pred}$ are the observed
(measured) and predicted number density of galaxies at redshift $z$,
respectively, and $\Delta t$ is in Gyr.  We divide by the observed
number density to account for the shape of the \smf; for example, a
merger-induced absolute increase in number density of
$10^{-3}$~galaxies~Mpc$^{-3}$~Gyr$^{-1}$ will have a much larger
fractional effect on the exponential tail of the \smf{} relative to
the low-mass end, where such a change would be negligible.  We derive
the number densities in equation~(\ref{eq:mergerrate}) as in
Section~\ref{sec:mfevol} by numerically integrating the observed and
predicted \smf s over chosen intervals of stellar mass, but use the
best-fitting single or double Schechter model to extrapolate to lower
or higher mass as needed.  Note that $\mathcal{G}$ can be either
positive or negative depending on whether mergers result in a net
increase or decrease of galaxies of a certain stellar mass.  Moreover,
$\mathcal{G}\approx0$ does not \emph{necessarily} indicate that
mergers are not occuring or important; mergers could result in a small
value of $\mathcal{G}$ if the growth and destruction of galaxies into
and out of a certain stellar mass bin on average balanced one another.

\begin{figure}
\centering
\includegraphics[scale=0.4]{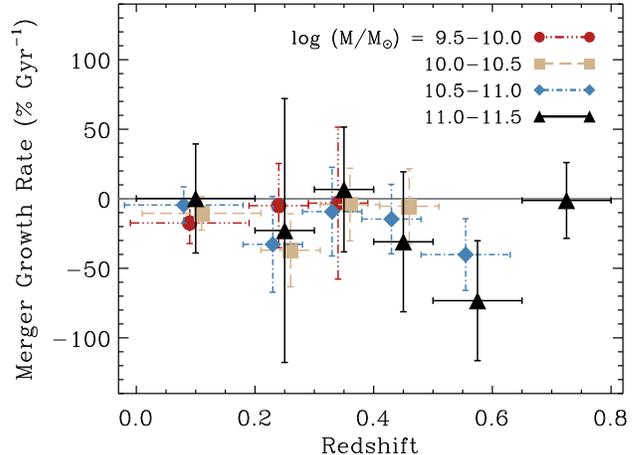}
\caption{Merger growth rate, $\mathcal{G}(\mass,z)$, the fractional
  change in the number density of galaxies due to mergers after
  accounting for stellar mass growth by {\em in situ} star formation,
  vs.~redshift in four $0.5$~dex wide intervals of stellar mass
  between $10^{9.5}$~\msun{} and $10^{11.5}$~\msun.  Note that
  $\mathcal{G}$ as defined by equation~(\ref{eq:mergerrate}) can be
  either negative or positive depending on whether mergers
  preferentially destroy or create galaxies of a given stellar mass.
  We find a net change in the number density of
  $10^{9.5}-10^{10.5}$~\msun{} galaxies of $-13\%\pm9\%$~Gyr$^{-1}$
  from $z\approx0$ to $z\approx0.5$, and a change of
  $-14\%\pm18\%$~Gyr$^{-1}$ in the number density of
  $10^{10.5}-10^{11.5}$~\msun{} galaxies since $z\approx0.8$.  In
  other words, although mergers are almost certainly occurring, they
  do not have a large \emph{net} effect on the \smf{} over this range
  of stellar masses and redshifts.  \label{fig:mergerrate}}
\end{figure}

In Figure~\ref{fig:mergerrate} we plot $\mathcal{G}$ versus redshift
in four intervals of stellar mass between $10^{9.5}-10^{11.5}$~\msun.
Qualitatively, we find no significant variations of $\mathcal{G}$ with
either stellar mass or redshift (where PRIMUS is complete).
Quantitatively, we find a net change in the number density of
$10^{9.5}-10^{10.5}$~\msun{} galaxies of $-13\%\pm9\%$~Gyr$^{-1}$ from
$z\approx0$ to $z\approx0.5$, and a change of
$-14\%\pm18\%$~Gyr$^{-1}$ in the number density of
$10^{10.5}-10^{11.5}$~\msun{} galaxies since $z\approx0.8$.  Although
the uncertainties are significant, we conclude, therefore, that
mergers do not appear to be an important channel for stellar mass
growth at late cosmic times, even among massive
($\gtrsim10^{11}$~\msun) galaxies \citep[see also][]{pozzetti10a}.
Although beyond the scope of the present study, a detailed comparison
of these results with theoretical galaxy formation models, which
generally find that massive galaxies grow much more substantially
through mergers at $z<1$ \citep[e.g.,][]{de-lucia06a, cattaneo08a,
  cattaneo11a}, would be very interesting.

We extend our analysis further by separately considering the stellar
mass growth of quiescent and star-forming galaxies.  One particularly
important question we would like to address is whether quiescent
galaxies grow significantly by dissipationless mergers (also known as
{\em dry mergers}) at $z<1$.  To investigate this question we measure
the redshift evolution of $\mass_{c}$, the stellar mass at fixed
cumulative number density, $n(\mass>\mass_{c})$.  To a good
approximation, both star formation and mergers will increase
$\mass_{c}$ without changing the number density, while the
transformation of one galaxy type into another (e.g., due to star
formation quenching) will tend to decrease $\mass_{c}$ at a given
cumulative number density \citep{van-dokkum10a, brammer11a}.  Because
the amount of {\em in situ} star formation in our sample of quiescent
galaxies is by construction negligible (see Figure~\ref{fig:sfr}), an
increase in $\mass_{c}$ with decreasing redshift can be attributed to
dissipationless mergers.

\begin{figure}
\centering
\includegraphics[scale=0.4]{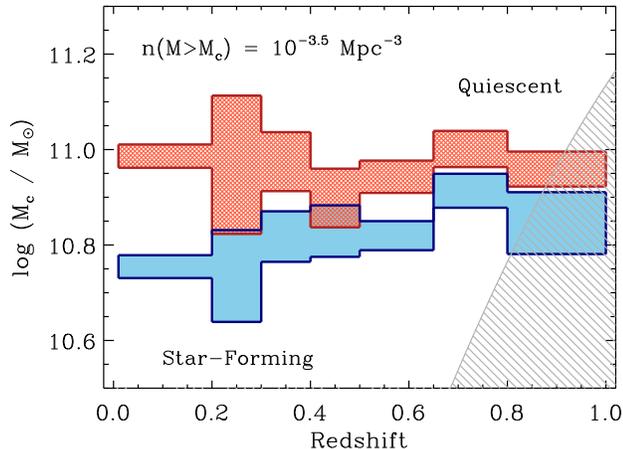}
\caption{Stellar mass, $\mass_{c}$, at which the cumulative number
  density, $n(\mass>\mass_{c})$, equals $10^{-3.5}$~Mpc$^{-3}$
  vs.~redshift.  The red and blue shaded regions correspond to
  quiescent and star-forming galaxies, respectively, and the grey
  hatched area indicates for reference the stellar mass completeness
  limit for quiescent galaxies; the completeness limit for
  star-forming galaxies extends to much lower stellar mass.  The
  constancy of $\mass_{c}$ for quiescent galaxies indicates negligible
  growth of this population due to mergers, while the decline in
  $\mass_{c}$ with decreasing redshift reflects the progressive
  quenching and transformation of massive star-forming galaxies into
  quiescent galaxies from $z=0-1$.  \label{fig:cumuphi}}
\end{figure}

In Figure~\ref{fig:cumuphi} we plot $\mass_{c}$ versus redshift for
quiescent (hatched red shading) and star-forming (light blue shading)
galaxies corresponding to $n(\mass>\mass_{c})=10^{-3.5}$~Mpc$^{-3}$.
We choose this number density threshold because it samples a
significant fraction of the exponential tail of the \smf{} at each
redshift, and because PRIMUS is complete to both galaxy types all the
way to $z\approx1$.  For reference, the grey shaded region in the
lower-right corner of Figure~\ref{fig:cumuphi} shows the stellar mass
completeness limit for quiescent galaxies (the stellar mass limit for
star-forming galaxies is much lower).  In each redshift bin, our
measurement of $\mass_{c}$ reflects the quadrature sum of both the
Poisson and sample variance uncertainties.

We find no notable change in $\mass_{c}$ for quiescent galaxies from
$z=0-1$, and a significant decrease for star-forming galaxies with
decreasing redshift.  Modeling the observed trends as a power-law
function of redshift, $\mass_{c} \propto (1+z)^{q}$, we find
$q=-0.09\pm0.16$ for quiescent galaxies, and $q=0.60\pm0.17$ for
star-forming galaxies.  In other words, $\mass_{c}$ for star-forming
galaxies decreases on average by $0.18\pm0.05$~dex ($51\%\pm12\%$)
since $z=1$, whereas $\mass_{c}$ for quiescent galaxies is constant to
within $\pm0.05$~dex ($\pm12\%$) over the same redshift range.

The decline in $\mass_{c}$ for star-forming galaxies is most likely
due to the progressive transformation of massive star-forming galaxies
into quiescent, passively evolving systems toward low redshift (see
Figure~\ref{fig:numbymass} and Section~\ref{sec:quenchrate}).  In
essence, $\mass_{c}$ decreases toward low redshift because we have to
integrate further down the \smf{} to count the same total number of
galaxies.  Meanwhile, the constancy of $\mass_{c}$ for quiescent
galaxies follows directly from the lack of evolution at the massive
end of the \smf.  Note that $\mass_{c}$ for quiescent galaxies remains
approximately constant even though star-forming galaxies are being
quenched because massive star-forming galaxies constitute a very small
fraction of the total number of massive galaxies at $z=0-1$ (see
Section~\ref{sec:mfqqsf}).  We conclude, therefore, that most massive,
quiescent galaxies are fully assembled by $z\approx1$, and do not appear
to grow significantly by dissipationless mergers over the last
$\sim8$~billion years of cosmic time.

\subsection{Buildup of the Quiescent Galaxy Population by
  Star Formation Quenching}\label{sec:quenchrate}

Our analysis of the \smf{} has revealed significant changes in the
population of both quiescent and star-forming galaxies.  The two key
results we focus on in this section are the rapid rise in the number
of intermediate-mass ($\sim10^{10}$~\msun) quiescent galaxies since
$z\approx0.5$, and the steep decline in the population of massive
($\gtrsim10^{11}$~\msun) star-forming galaxies since $z\approx1$.  Taken
together, these results indicate that quenching---the rapid cessation
of star formation in galaxies---is an important driver of galaxy
evolution at $z<1$.  Moreover, whatever mechanism or mechanisms are
responsible for quenching must affect galaxies spanning a wide range
of stellar mass at these redshifts.

\begin{figure*}
\centering
\includegraphics[scale=0.65]{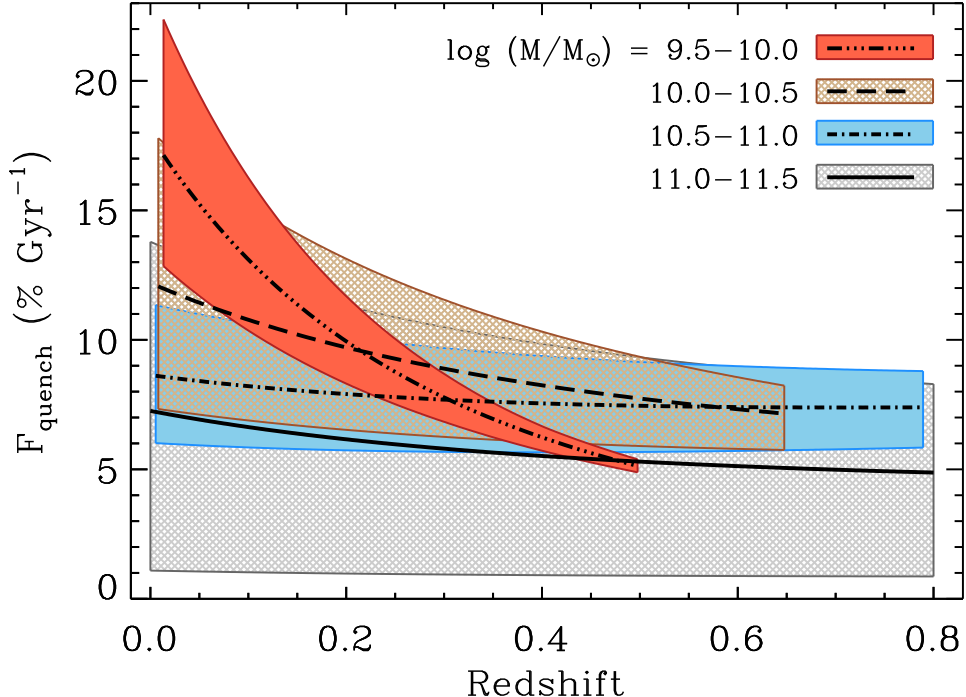}
\caption{Fractional quenching rate $F_{\rm quench}$ vs.~redshift in
  four intervals of stellar mass between $10^{9.5}-10^{11.5}$~\msun.
  $F_{\rm quench}$ is the proportion of star-forming galaxies in a
  given stellar mass interval that must be quenched per Gyr in order
  to match the measured evolution in the number density of quiescent
  galaxies.  Although we implicitly ignore the effects of mergers in
  calculating $F_{\rm quench}$, that does not necessarily mean that
  mergers are not responsible for quenching star formation in some
  galaxies (see the discussion in Section~\ref{sec:quenchrate}).
  Although the uncertainties are large, we find that $F_{\rm quench}$
  is typically a factor of $\sim2-3$ higher for
  $10^{9.5}-10^{10.5}$~\msun{} galaxies compared to more massive
  galaxies $\gtrsim10^{11}$~\msun.  Moreover, the fractional quenching
  rate in lower-mass systems appears to be increasing toward the
  current epoch. \label{fig:fquench}}  
\end{figure*}

We can use the results presented in Section~\ref{sec:results} to
quantify both the {\em quenching rate}---the frequency with which
star-forming galaxies are being transformed into quiescent
galaxies---and the stellar mass growth of the quiescent galaxy
population due to the addition of newly quenched galaxies.  The
evolution with redshift of both these quantities should place
important constraints on the broad range of proposed quenching
processes, and their implementation into theoretical galaxy formation
models.

\begin{figure*}
\centering \includegraphics[scale=0.65]{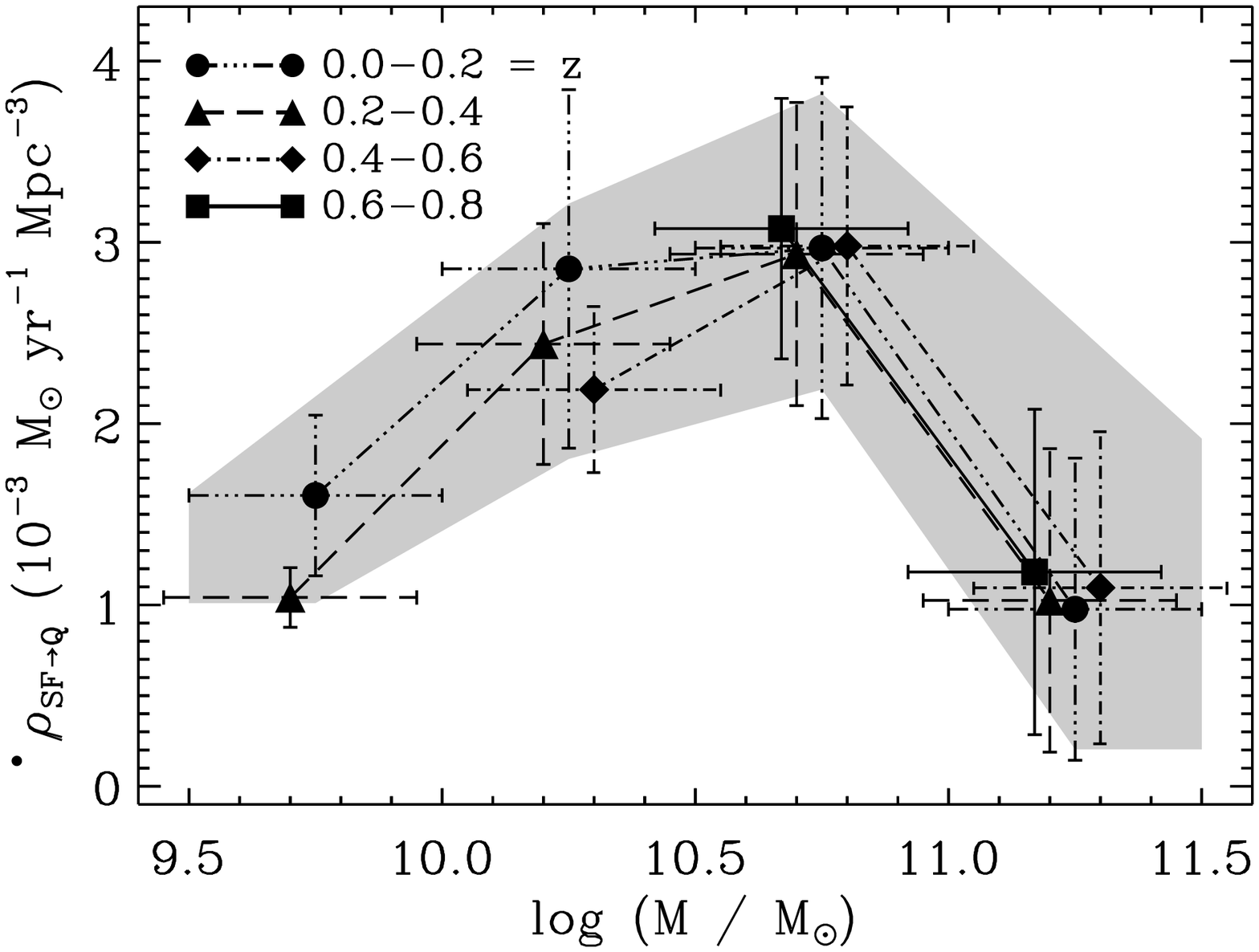}
\caption{Rate at which stellar mass is being transferred from the
  star-forming to the quiescent galaxy population, \rhosfqq,
  vs.~stellar mass in four intervals of redshift between $z=0-0.8$.
  The individual symbols with error bars (offset slightly horizontally
  for clarity), correspond to different redshift bins, and the grey
  shaded region shows the mean trend and $1\sigma$ confidence region
  over all redshifts.  We find that \rhosfqq{} is largely independent
  of redshift, and peaks around $\sim10^{10.8}$~\msun.  Consequently,
  although a larger proportion of lower-mass galaxies are being
  quenched (see Figure~\ref{fig:fquench}), the bulk of the stellar
  mass buildup of the quiescent galaxy population is occuring near the
  `knee' of the \smf.  \label{fig:rhosfqq}}
\end{figure*}

In the subsequent analysis we make the simplifying but well-motivated
assumption (see Section~\ref{sec:mergers}) that we can neglect the
effects of mergers.  More specifically, we implicitly assume that if
mergers are taking place, they do not cause a significant number of
galaxies to shift from one $0.5$~dex wide stellar mass bin to another
(e.g., from the $10^{10}-10^{10.5}$~\msun{} to the
$10^{10.5}-10^{11}$~\msun{} stellar mass bin).  Consequently, although
we do not explicitly include their potential contribution, mergers
could still be a viable means of quenching star formation in some
galaxies.  In addition, we implicitly ignore {\em in situ} star
formation within the quiescent galaxy population, which by design is
negligible (see Figure~\ref{fig:sfr}).

With the preceding caveats in mind, we define the fractional quenching
rate as $F_{\rm quench} \equiv ({\rm d}n_{\textsc{q}}/{\rm
  d}t)/n_{\textsc{sf}}$, or the fraction of star-forming galaxies that
must be quenched per Gyr as a function of stellar mass and redshift in
order to account for the measured evolution of the quiescent-galaxy
population with decreasing redshift.  Using
equation~(\ref{eq:numden_evol}), we obtain
\begin{equation}
F_{\rm quench} = \gamma_{\textsc{q}}
\left(\frac{n_{0,\textsc{q}}}{n_{0,\textsc{sf}}}\right)
(1+z)^{\gamma_{\textsc{q}}-\gamma_{\textsc{sf}}-1} 
\left(\frac{{\rm d}t}{{\rm d}z}\right)^{-1},
\label{eq:fquench}
\end{equation}

\noindent where the {\sc q} and {\sc sf} subscripts refer to quiescent
and star-forming galaxies, respectively, and the relevant coefficients
and uncertainties are listed in Table~\ref{table:numden_bymass_coeff}.
The $({\rm d}t/{\rm d}z)$ term is the derivative of the age-redshift
function, $t(z)$, where $t$ is the age of the Universe
\citep{hogg99a}.  For convenience, we note that in our adopted
cosmology $({\rm d}t/{\rm d}z)$ can be approximated from $0<z<1$ to an
accuracy of better than $0.5\%$ by a third-order polynomial of the
form $({\rm d}t/{\rm d}z) \approx -13.8835 + 19.3598z -13.621z^2 +
4.2141z^3$~Gyr$^{-1}$.

In Figure~\ref{fig:fquench} we plot $F_{\rm quench}$ versus redshift
in four bins of stellar mass between $10^{9.5}-10^{11.5}$~\msun.  This
figure reveals several interesting results.  First, we find that
$F_{\rm quench}$ is relatively small at all redshifts and stellar
masses; it varies between $\approx5\%-10\%$~Gyr$^{-1}$, except in the
lowest stellar mass bin at $z\lesssim0.2$, where it rises to
$\approx12\%-18\%$~Gyr$^{-1}$.  For comparison, \citet{blanton06a}
estimate that roughly $25\%$ of blue star-forming galaxies must be
quenched since $z\approx1$ in order to match the measured buildup of
the optical luminosity function for red-sequence galaxies.  Second,
although the uncertainties are large, our measurements suggest that
$F_{\rm quench}$ varies systematically with stellar mass.  Among
$>10^{11}$~\msun{} galaxies, we find a typical quenching rate of
$\approx5\%$~Gyr$^{-1}$, whereas among $\lesssim10^{10.5}$~\msun{}
galaxies the rate is around $\approx7\%-15\%$~Gyr$^{-1}$, a factor of
$\sim2-3$ higher.  Finally, Figure~\ref{fig:fquench} suggests that
$F_{\rm quench}$ \emph{increases} toward low redshift among lower-mass
galaxies.  Specifically, we find a weak trend of an increasing $F_{\rm
  quench}$ toward $z\approx0$ among $10^{10}-10^{10.5}$~\msun{}
galaxies, and a much more significant and rapid rise toward low
redshift among $10^{9.5}-10^{10}$~\msun{} galaxies.  Although the
uncertainties are large, these results suggest that star formation
quenching is more prevalent among lower-mass galaxies, and that the
fraction of low-mass star-forming galaxies that are being quenched is
increasing toward the current epoch.

We can extend this analysis one step further and calculate
$\rhosfqq\equiv {\rm d}\rho_{\textsc{q}}/{\rm d}t$, the rate of
stellar mass transfer from the population of star-forming to quiescent
galaxies, using the measured stellar mass growth of the
quiescent-galaxy population.  Once again, we implicitly assume that
the stellar mass growth of the quiescent population is entirely due to
the addition of newly quenched (i.e., previously star-forming)
galaxies, and that dissipationless mergers between two quiescent
galaxies does not cause a significant number of galaxies to change
their stellar mass by more than a factor of three.  With these caveats
in mind, we use equation~(\ref{eq:rhoden_evol}) to write \rhosfqq{} in
units of \msun~yr$^{-1}$~Mpc$^{-3}$ as
\begin{equation}
\rhosfqq = \rho_{0, {\textsc{q}}}\ \beta_{\textsc{q}}\,
(1+z)^{\beta_{\textsc{q}}-1} \left(\frac{{\rm d}t}{{\rm d}z}\right)^{-1},
\label{eq:rhosfqq}
\end{equation}

\noindent where the relevant coefficients are given in
Table~\ref{table:numden_bymass_coeff}.  

In Figure~\ref{fig:rhosfqq} we plot $\rhosfqq$ versus stellar mass in
four equal-sized bins of redshift from $z=0-0.8$.  The individual
symbols with error bars, which have been offset slightly in the
horizontal direction for clarity, correspond to different redshift
bins, and the grey shaded region reflects the broad trend we deduce.
We find that \rhosfqq{} depends weakly on redshift, but has a very
strong stellar mass dependence.  Quantitatively, \rhosfqq{} rises from
$\approx1.5\times10^{-3}$~\msun~yr$^{-1}$~Mpc$^{-3}$ around
$\sim10^{9.8}$~\msun{} to a peak value of
$\approx3\times10^{-3}$~\msun~yr$^{-1}$~Mpc$^{-3}$ around
$\sim10^{10.8}$~\msun.  Above $\sim10^{11}$~\msun, \rhosfqq{} declines
sharply to a mean value of
$\approx1\times10^{-3}$~\msun~yr$^{-1}$~Mpc$^{-3}$.

These results reveal that although fractionally more low- and
intermediate-mass ($\sim10^{9.5}-10^{10.5}$~\msun) galaxies are being
quenched (see Figure~\ref{fig:fquench}), the bulk of the stellar-mass
buildup within the quiescent galaxy population occurs around
$\sim10^{10.8}$~\msun, near the `knee' of the \smf.  As emphasized by
\citet{bell07a}, this stellar mass scale appears to be important
because both the stellar mass-weighted star-formation rate density in
galaxies, and the stellar mass growth of the quiescent galaxy
population peak around this stellar mass at $z<1$; however, the
underlying physical cause of this coincidence remains unknown.

\section{Summary}\label{sec:summary}

We have measured the evolution of the \smf s of quiescent and
star-forming galaxies from $z=0-1$ using two large, statistically
complete, spectroscopic samples.  At low redshift we use a sample of
$\sim170,000$ SDSS galaxies with \emph{GALEX}, 2MASS, and WISE
photometry, and at intermediate redshift we use a sample of
$\sim40,000$ galaxies brighter than $i\approx23$ drawn from PRIMUS
with deep \emph{GALEX} and IRAC imaging.  Our PRIMUS sample is notable
for its depth, sample size, and area, which totals
$\approx5.5$~deg$^2$ over five widely-separated fields, while our
SDSS-\emph{GALEX} sample comprises one of the largest statistical
samples of local galaxies with SDSS and \emph{GALEX} photometry ever
assembled.

The exceptional multi-wavelength coverage of both datasets provides
deep UV to mid-infrared imaging over the entire spectroscopic survey
area, allowing us to robustly estimate SFRs and stellar masses using a
new Bayesian SED-modeling code (\isedfit; see
Appendix~\ref{appendix:isedfit}).  We use these measurements to
separate the galaxy population into quiescent and star-forming based
on their position in the SFR-stellar mass diagram, and to measure the
evolution of the \smf{} over a large dynamic range in stellar mass and
redshift with relatively small sample variance and Poisson
uncertainties.  In addition, we carefully assess the effect of
systematic errors in our stellar mass and SFR estimates, and find that
the evolutionary trends we measure are broadly insensitive to the
exact choice of priors and population synthesis models.

Our principal quantitative results are as follows:

\begin{enumerate}

\item[1.]{We find for the global galaxy population that the \smf{} has
  evolved relatively little since $z=1$, although we do find evidence
  for differential evolution---mass assembly downsizing.  We measure a
  $31\%\pm9\%$ increase in the integrated number density of
  $\sim10^{10}$~\msun{} galaxies since $z\approx0.6$, and a
  $-3\%\pm6\%$ change in the number density of all
  $\sim10^{11}$~\msun{} galaxies since $z\approx0.9$.  Most massive
  galaxies, therefore, appear to be largely in place by $z=1$, while
  lower-mass galaxies continue to assemble toward the present epoch.}

\item[2.]{The relatively subtle changes in the \smf{} of the global
  population, however, mask much more dramatic evolution in the \smf s
  of star-forming and quiescent galaxies.  Within the star-forming
  population the most rapid evolution occurs among massive galaxies,
  whereas the low-mass end of the star-forming galaxy \smf{} does not
  change significantly.  We find that the comoving number density of
  $10^{9.5}-10^{11}$~\msun{} star-forming galaxies changes by
  $-8\%\pm10\%$ between $z\approx0.8$ and $z\approx0.0$, whereas the
  space-density of massive ($10^{11}-10^{11.5}$~\msun) star-forming
  galaxies \emph{declines} by $54\%\pm7\%$ since $z\approx1$.}

\item[3.]{Meanwhile, among quiescent galaxies the most significant
  evolutionary changes occur among low- and intermediate-mass
  ($10^{9.5}-10^{10.5}$~\msun) galaxies, whereas most massive
  ($\gtrsim10^{11}$~\msun) quiescent galaxies are largely in place
  from $z=0-1$.  Quantitatively, we find a factor of $\sim2-3$
  increase in the number-density of $10^{9.5}-10^{10}$~\msun{}
  galaxies since $z\approx0.5$, and a marginally significant increase,
  $22\%\pm12\%$, in the space-density of $>10^{11}$~\msun{} quiescent
  galaxies since $z\approx1$.}
\end{enumerate}

We use these measurements to place new constraints on the growth of
galaxies by mergers, and to quantify the buildup of the quiescent
galaxy population due to star formation quenching as a function of
redshift and stellar mass:

\begin{enumerate}

\item[1.]{Using a simple model to account for the expected growth of
  galaxies due to star formation, we find that mergers do not appear
  to be a dominant channel for the stellar mass buildup of galaxies at
  $z<1$, even among massive ($\gtrsim10^{11}$~\msun) systems.
  Quantitatively, we find that mergers are responsible for a net
  change in the number density of $10^{9.5}-10^{10.5}$~\msun{}
  galaxies of $-13\%\pm9\%$~Gyr$^{-1}$ from $z\approx0$ to
  $z\approx0.5$, and a change of $-14\%\pm18\%$~Gyr$^{-1}$ in the
  number density of $10^{10.5}-10^{11.5}$~\msun{} galaxies since
  $z\approx0.8$.  These results do not imply that mergers are not
  occurring, only that they do not have a large \emph{net} effect on
  the \smf{} over the range of stellar masses and redshifts probed by
  PRIMUS.}

\item[2.]{Our results also imply that the rate at which star formation
  is quenched in galaxies depends both on stellar mass and redshift,
  with a peak around $\sim10^{10.8}$~\msun, and an increase in the
  quenching rate at lower redshift for lower mass galaxies.  In
  particular, we find that the quenching rate for massive galaxies
  with $>10^{11}$~\msun{} is consistently low
  ($\approx5\%$~Gyr$^{-1}$) at all redshifts to $z\approx1$.}
\end{enumerate}

To fully characterize the build up of stellar mass for all galaxies to
$z=1$, additional measurements at the lower mass end of the \smf{} are
needed, which requires both deeper multi-wavelength imaging and
spectroscopy.  In addition, while our results at intermediate redshift
use a large sample of $\sim40,000$ galaxies across five separate
fields covering $\approx5.5$~deg$^2$, they are still dominated by
sample variance.  This result argues that even more wide-field imaging
and spectroscopy to at least $i=23$ are needed to precisely measure
the \smf{} from $z=0-1$.

\acknowledgements

We gratefully acknowledge feedback on the manuscript and insightful
conversations with Peter Behroozi, Eric Bell, Aaron Bray, Charlie
Conroy, Aleks Diamond-Stanic, Du\v{s}an Kere\v{s}, Leonidas Moustakas,
Gregory Rudnick, Samir Salim, Ramin Skibba, and Risa Wechsler, and we
thank the anonymous referee for their careful report.  We also
acknowledge Rebecca Bernstein, Adam Bolton, Scott Burles, Douglas
Finkbeiner, David W. Hogg, Timothy McKay, Sam Roweis, Wiphu
Rujopakarn, and Stephen Smith for their contributions to the PRIMUS
project.  In addition, we extend our appreciation to Mariangela
Bernardi, Olivier Ilbert, and Lucia Pozzetti for providing their
published stellar mass functions in electronic format, Peter Capak and
Thomas Erben for answering questions regarding the COSMOS and CARS
photometric catalogs, respectively, Brian Siana for assistance with
the CDFS/SWIRE optical photometry, and Samir Salim for sending an
electronic catalog of the stellar masses and SFRs published in
\citet{salim07a}, and for enlightening conversations regarding the
intricacies of SED modeling.  We would also like to thank the CFHTLS,
COSMOS, DLS, and SWIRE teams for their public data releases and/or
access to early releases. This paper includes data gathered with the
6.5 meter Magellan Telescopes located at Las Campanas Observatory,
Chile; we thank the support staff at LCO for their help during our
observations, and we acknowledge the use of community access through
NOAO observing time.

Funding for PRIMUS has been provided by NSF grants AST-0607701,
0908246, 0908442, 0908354, and NASA grant 08-ADP08-0019.  ALC
acknowledges support from the Alfred P. Sloan Foundation and NSF
CAREER award AST-1055081, and MRB acknowledges financial support
through NASA grant 08-ADP08-0072.

This research has made use of the NASA/IPAC Infrared Science Archive,
which is operated by the Jet Propulsion Laboratory, California
Institute of Technology, under contract with the National Aeronautics
and Space Administration; data products from the Wide-field Infrared
Survey Explorer, which is a joint project of the University of
California, Los Angeles, and the Jet Propulsion Laboratory/California
Institute of Technology, funded by the National Aeronautics and Space
Administration; data products from the Two Micron All Sky Survey,
which is a joint project of the University of Massachusetts and the
Infrared Processing and Analysis Center/California Institute of
Technology, funded by the National Aeronautics and Space
Administration and the National Science Foundation; observations
obtained with MegaPrime/MegaCam, a joint project of CFHT and
CEA/DAPNIA, at the Canada-France-Hawaii Telescope (CFHT) which is
operated by the National Research Council (NRC) of Canada, the
Institut National des Science de l'Univers of the Centre National de
la Recherche Scientifique (CNRS) of France, and the University of
Hawaii; and data products produced at TERAPIX and the Canadian
Astronomy Data Centre as part of the Canada-France-Hawaii Telescope
Legacy Survey, a collaborative project of NRC and CNRS.  The Galaxy
Evolution Explorer (\emph{GALEX}) is a NASA Small Explorer, whose
mission was developed in cooperation with the Centre National d'Etudes
Spatiales of France and the Korean Ministry of Science and Technology.
Funding for the SDSS and SDSS-II has been provided by the Alfred
P. Sloan Foundation, the Participating Institutions, the National
Science Foundation, the U.S. Department of Energy, the National
Aeronautics and Space Administration, the Japanese Monbukagakusho, the
Max Planck Society, and the Higher Education Funding Council for
England. The SDSS Web Site is \url{http://www.sdss.org}.  The SDSS is
managed by the Astrophysical Research Consortium for the Participating
Institutions. The Participating Institutions are the American Museum
of Natural History, Astrophysical Institute Potsdam, University of
Basel, University of Cambridge, Case Western Reserve University,
University of Chicago, Drexel University, Fermilab, the Institute for
Advanced Study, the Japan Participation Group, Johns Hopkins
University, the Joint Institute for Nuclear Astrophysics, the Kavli
Institute for Particle Astrophysics and Cosmology, the Korean
Scientist Group, the Chinese Academy of Sciences (LAMOST), Los Alamos
National Laboratory, the Max-Planck-Institute for Astronomy (MPIA),
the Max-Planck-Institute for Astrophysics (MPA), New Mexico State
University, Ohio State University, University of Pittsburgh,
University of Portsmouth, Princeton University, the United States
Naval Observatory, and the University of Washington.

\begin{appendix}

\setcounter{figure}{0}
\renewcommand\thefigure{A\arabic{figure}}

\section{\isedfit{} Spectral Energy Distribution Modeling
  Code}\label{appendix:isedfit}

\subsection{Background}

\isedfit\footnote{\url{https://github.com/moustakas/iSEDfit}} was
developed in the {\sc idl} programming language to be a fast and
flexible tool to extract the physical properties of both nearby and
high-redshift galaxies from their broadband UV, optical, and
near-infrared SEDs.  It builds on the Bayesian formalism pioneered by
\citet{kauffmann03a} to model the optical spectral features of SDSS
galaxies, and its subsequent extension to broadband photometry
\citep[e.g.,][]{bundy05a, salim05a, salim07a, da-cunha08a, auger09a,
  mcgee11a, taylor11a}.  To date, stellar masses and other physical
parameters derived using \isedfit{} have been used to measure the
evolution of the stellar mass-metallicity relation
\citep{moustakas11a}, the stellar mass dependence of AGN accretion
\citep{aird12a}, the stellar mass and SFR surface densities of compact
starbursts with high-velocity outflows \citep{diamond-stanic12a}, and
the ages, SFRs, and stellar masses of galaxies at $z=6-10$
\citep{zitrin12a, zheng12a}, with many subsequent applications
forthcoming.

Like all SED-modeling codes designed to extract the physical
properties of galaxies\footnote{See \url{http://www.sedfitting.org}
  and \citet{walcher11a} for extensive references.}, \isedfit{} relies
on stellar population synthesis (SPS) models to provide as input
$\mathcal{S}(\lambda,t,Z)$, the spectral evolution of a simple stellar
population (SSP) of a given stellar metallicity, $Z$.  An SSP is an
idealized stellar population formed in an instantaneous burst of star
formation which evolves passively thereafter with time $t$.  The basic
ingredients of an SSP are \citep[e.g.,][]{tinsley68a, bruzual83a}: (1)
stellar evolution calculations for stars spanning the full range of
initial mass; (2) a stellar library that provides the emergent
spectrum of a star at each position in the Hertzsprung-Russell (HR)
diagram; and (3) an assumed IMF, which specifies the relative number
of stars of a given stellar mass.

Unfortunately, even the relatively ``simple'' goal of modeling SSPs is
limited by uncertainties in calculating particular phases of stellar
evolution, inadequecies in the stellar libraries (e.g., non-solar
abundance ratios, spectral coverage and resolution, etc.), and other
simplifying assumptions (see \citealt{conroy09a} and
\citealt{conroy10b} for recent in-depth discussions).  For example,
among the least well understood phases of stellar evolution are the
thermally pulsating asymptotic giant branch (TP-AGB) stars, blue
stragglers (BS), and horizontal branch (HB) stars, all of which are
relatively luminous and can therefore affect the integrated spectrum
of the stellar population \citep[e.g.,][]{maraston05a, melbourne12a}.
SSP calculations also implicitly assume a well-sampled (i.e., fully
populated), IMF, which may not always be satisfied
\citep[e.g.,][]{fumagalli11a}.

Differences in how these issues are addressed (or ignored) for a given
IMF can lead to significant systematic discrepancies among SSPs
derived using different SPS models.  In principal, the uncertainties
affecting SSPs should be incorporated into the SED modeling in order
to obtain reliable parameter estimates and realistic confidence
intervals \citep[e.g.,][]{conroy10a}.  In practice, however, this
procedure is both cumbersome and computationally challenging.
Instead, \isedfit{} adopts the simplified approach of allowing the
user to select from among many different SPS models, thereby allowing
the effects of choosing one set of SSPs over another to be
systematically investigated.

\subsection{From Simple Stellar Populations to Spectral Energy
  Distributions of ``Galaxies''} 

Given $\mathcal{S}(\lambda,t,Z)$, \isedfit{} computes the integrated
SED of a ``galaxy'' (a composite stellar population) as a function of
time $t$ using the following convolution integral:

\begin{equation}
\mathcal{C}(\lambda,t,Z) = \int_{0}^{t} \sfr(t-t^{\prime})\,
\mathcal{S}[\lambda,t^{\prime},Z(t-t^{\prime})]\,10^{-0.4A(\lambda,t^{\prime})}\, 
        {\rm d}t^{\prime},
\label{eq:csp}
\end{equation}

\noindent where $\sfr(t)$ is the star-formation history (SFH), and
$A(\lambda)$ is the wavelength-dependent attenuation, which in general
may depend on time \citep[e.g.,][]{charlot00a}.  The current version
of \isedfit{} only handles mono-metallic stellar populations (i.e., it
does not treat the chemical evolution of the system self-consistently)
in which case equation~(\ref{eq:csp}) reduces to

\begin{equation}
\mathcal{C}(\lambda,t,Z) = \int_{0}^{t} \sfr(t-t^{\prime})\,
\mathcal{S}(\lambda,t^{\prime},Z)\,10^{-0.4A(\lambda,t^{\prime})}\, 
        {\rm d}t^{\prime}.
\label{eq:simplecsp}
\end{equation}

In order to solve this integral, \isedfit{} requires several
additional inputs (implicit prior assumptions) from the user.  First,
an extinction or attenuation\footnote{Recall that attenuation includes
  the effects of both absorption and scattering, whereas extinction
  describes the absorption of light by a homogenous foreground dust
  screen \citep{witt92a, witt00a}.}  curve, $k(\lambda)\equiv
A(\lambda)/E(B-V)$, must be chosen among several different
possibilities, where $A(\lambda)$ is the total wavelength-dependent
attenuation and $E(B-V)$ is the color excess \citep{calzetti01a}.  The
currently supported possibilities are the \citet{calzetti00a}
starburst galaxy attenuation curve, the \citet{charlot00a} attenuation
law, the \citet{odonnell94a} Milky Way extinction curve, the
extinction curve of the Small Magellanic Cloud \citep{gordon03a}, or
no attenuation.  Note that among these only the \citet{charlot00a}
attenuation curve is time-dependent.

\begin{figure}
\centering
\includegraphics[scale=0.6,angle=90]{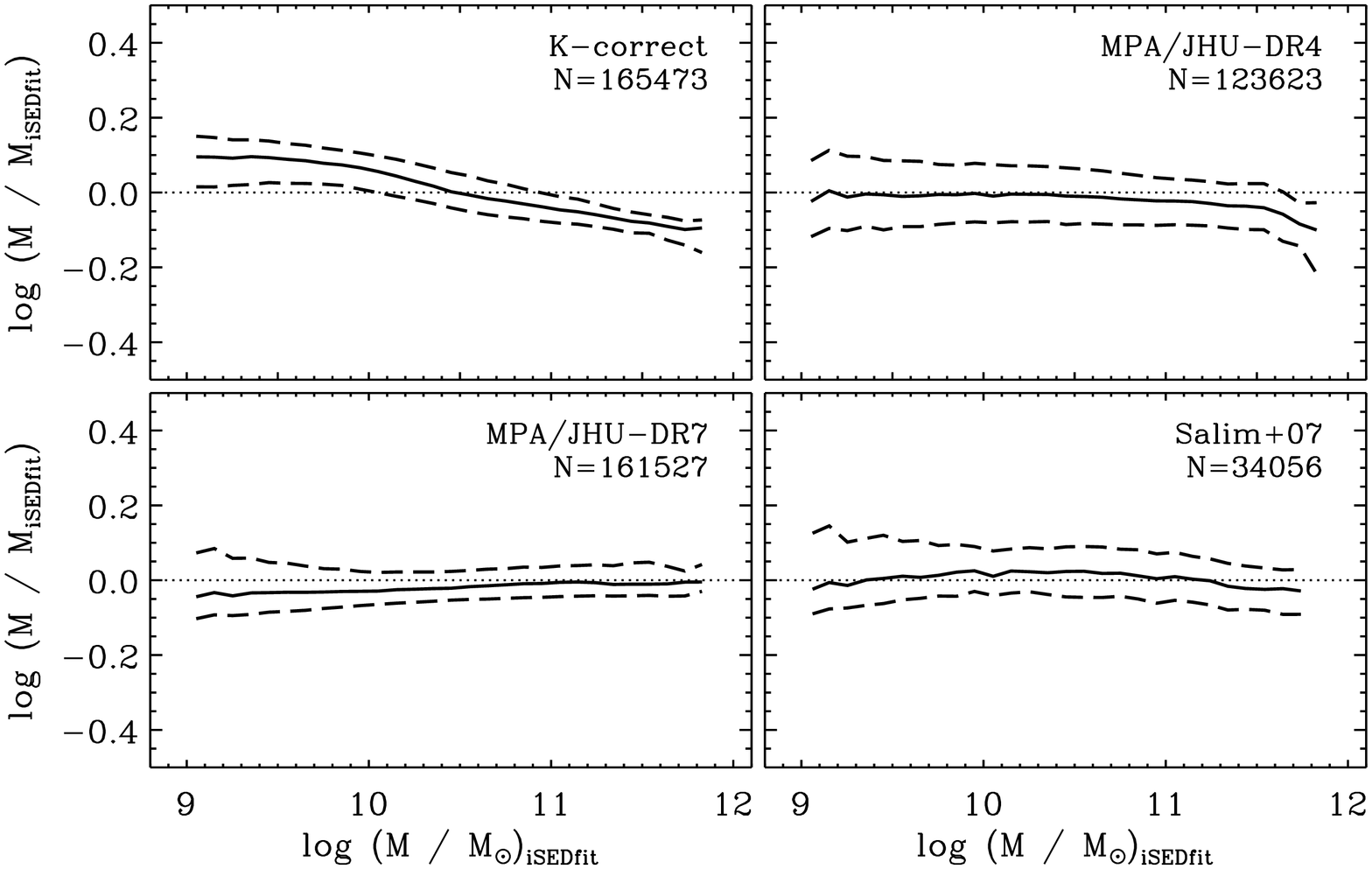}
\caption{Comparison of stellar masses derived using \isedfit{} (see
  Section~\ref{sec:mass}) for our SDSS-\emph{GALEX} sample vs.~stellar masses
  derived using the four independent techniques described in
  Appendix~\ref{sec:checks}.  The solid line in each panel indicates
  the median residual trend, and the dashed lines show the
  interquartile range of the residuals.  The dotted horizontal line
  shows for reference the one-to-one relation.  Overall, we find very
  good agreement between the stellar masses derived using \isedfit{}
  and these various other methods. \label{fig:masscompare}}  
\end{figure}

Next, a parametric form for the SFH must be specified.  In principal,
$\sfr(t)$ could be arbitrarily complex, or even non-parametric
\citep[e.g.,][]{cid-fernandes05a, tojeiro07a, panter08a}.  In
practice, however, it is challenging if not impossible to constrain
the detailed SFHs of individual galaxies from broadband photometry
alone (i.e., without high-resolution spectroscopy, although see
\citealt{dye08a}).  On the other hand, the integrated spectra of many
star-forming galaxies may be poorly fitted by simple (e.g.,
exponentially declining) SFHs.  Therefore, \isedfit{} optionally
allows stochastic bursts to be superposed on a backbone of smooth SFHs
\citep[e.g.,][]{kauffmann03a}.  For the underlying smooth component,
the user can choose either exponentially declining SFHs (so-called
simple $\tau$-models; \citealt{sandage86a}):

\begin{equation}
\sfr_{s}(t) = \frac{\mass_{\rm tot}}{\tau} e^{-t/\tau}; 
\label{eq:tau}
\end{equation}

\noindent or ``delayed'' $\tau$-models:

\begin{equation}
\sfr_{s}(t) = \frac{\mass_{\rm tot}}{\tau^2} t\,e^{-t/\tau},
\label{eq:delayed}
\end{equation}

\noindent where the subscript ``s'' indicates that these are
``smooth'' SFHs, $t$ is the age of the stellar population (the time
since the onset of star formation), $\tau$ is the characteristic time
for star formation, and the normalization is defined to be $\mass_{\rm
  tot}=1$~\msun.  The delayed $\tau$-models are advantageous because
they allow for both exponentially declining ($t/\tau\gg1$) and
linearly rising ($t/\tau\ll1$) SFHs to be explored, the latter of
which are needed to accurately reproduce the colors of high-redshift
($z\gtrsim2$) galaxies \citep[e.g.,][]{maraston10a, papovich11a,
  behroozi12a}.  

We characterize each burst by three independent parameters: the time
the burst begins, \tburst, its duration, \dtburst, and the {\em burst
  fraction}, the relative strength of the burst, \fburst.  The SFH of
each burst, $\sfr_{b}(t)$, is a Gaussian function given by

\begin{equation}
\sfr_{b}(t)
= \frac{\aburst}{\sqrt{2\pi}} \,e^{-(t-\tburst)^2/2\dtburst^2},
\label{eq:sfrburst}
\end{equation}

\noindent where \aburst{} is the {\em burst amplitude}.  Defining
\fburst{} to be the mass formed in the burst divided by the total mass
formed by the underlying $\tau$ model until the peak of burst, we
obtain

\begin{equation}
\fburst \equiv \frac{\mass_{b}}{\mass_{\tau}(\tburst)} =
\frac{\dtburst} {\mass_{\rm tot} (1-e^{-\tburst/\tau})}\,
\aburst, 
\label{eq:fburstdef}
\end{equation}

\noindent where $\mass_{\tau}(\tburst)\equiv \int_{0}^{\tburst}
\sfr_{\tau}(t)\,dt$, and $\mass_{\rm tot}=1$~\msun.
Equation~(\ref{eq:fburstdef}) assumes a simple $\tau$-model, but a
similar expression can be derived for the delayed $\tau$-model.  The
final composite SFH is given by

\begin{equation}
\sfr(t) = \sfr_{s}(t) + \sum_{j=1}^{N_{\rm burst}} \sfr_{b_{j}}(t),
\label{eq:sfhfinal}
\end{equation}

\noindent where $N_{\rm burst}$ is the total number of bursts
experienced by each model galaxy.  The number of bursts is determined
by specifying the cumulative probability $P_{\rm burst}$ for a burst
to occur within a $\Delta P_{\rm burst}$ time interval.  Finally, we
note that the current version of \isedfit{} additionally allows the
final burst to be truncated exponentially with a characteristic time
$\tau_{\rm trunc}$, thereby allowing the SFHs and physical properties
of post-starburst galaxies to be investigated
\citep[e.g.,][]{tremonti07a}.

\subsection{Extracting the Physical Properties of Galaxies from
  Broadband Photometry}

Based on the large number of free parameters needed to model the
integrated SEDs of galaxies, it would be far too computationally
expensive to explore all possible parameter combinations (e.g., on a
uniform grid).  Moreover, traditional best-fitting (maximum
likelihood) techniques are limited because they only account for
photometric uncertainties, but not physical degeneracies among
different models (parameter combinations).  Alternatively, Markov
Chain Monte Carlo (MCMC) algorithms may be more suitable for exploring
the multi-dimensional parameter space \citep[e.g.,][]{acquaviva11a}.
However, MCMC methods are typically too slow to enable tens, hundreds,
or even millions of galaxies to be fitted in a timely manner with
multiple independent prior parameter combinations or SPS models.

Given the limitations of these other techniques, \isedfit{} extracts
the physical parameters of interest using a simplified Bayesian
approach \citep[and references therein]{walcher11a}.  First, the model
parameters are drawn from a user-specified prior probability
distribution using a Monte Carlo technique.  Next, given the broadband
fluxes $F_{i}$ of a galaxy at redshift $z$ in $i=1,N$ filters, and the
corresponding $\sigma_{i}$ uncertainties, \isedfit{} uses Bayes'
theorem to compute the posterior probability distribution function
(PDF)

\begin{equation}
p({\mathbf Q}|F_{i},z) = p({\mathbf Q})\times p(F_{i},z|{\mathbf Q}),
\label{eq:posterior}
\end{equation}

\noindent where ${\mathbf Q}$ represents the set of model parameters
(stellar mass, age, metallicity, etc.).  Here, $p({\mathbf Q})$ is the
prior probability of the model parameters, $p(F_{i},z|{\mathbf Q})$ is
the likelihood $\mathcal{L}\propto \exp[-\chi^2(F_{i},z | {\mathbf
    Q})/2]$ of the data given the model, and $\chi^{2}$ is the usual
goodness-of-fit statistic appropriate for normally distributed
photometric uncertainties, given by

\begin{equation}
\chi^{2}(F_{i}, z | {\mathbf Q}) = \sum_{i=1}^{N} \frac{[F_{i} -
    \mathcal{A}\,\mathcal{C}_{i}({\mathbf Q},z)]^2}{\sigma_{i}^{2}},
\label{eq:chi2}
\end{equation}

\noindent where $\mathcal{A}$ is a normalization factor, and the
$\mathcal{C}_{i}({\mathbf Q},z)$ are the broadband fluxes of each
model SED given the redshift and parameter combination ${\mathbf Q}$.
Once $\chi^{2}$ has been computed for every model, the marginalized
posterior PDF of the parameter of interest, for example $p(\mass)$ for
the stellar mass, can be derived by randomly drawing each parameter
value with probability given by equation~(\ref{eq:posterior}), thereby
effectively integrating (using histogram binning) over the other
``nuisance'' parameters.  Although \isedfit{} is capable of
reconstructing the full posterior distributions in post-processing
after the computationally intensive fitting has been completed, the
code by default provides the median of the posterior PDF as the best
estimate of each parameter, and estimates the uncertainty in each
parameter as $1/4$ of the $2.3-97.7$ percentile range of the posterior
distribution, which would be equivalent to $1\sigma$ for a Gaussian
distribution.

\subsection{Consistency Checks}\label{sec:checks}

In Figure~\ref{fig:masscompare} we verify that the stellar masses we
derive using \isedfit{} for our SDSS-\emph{GALEX} sample based on our
fiducial prior parameters (see Section~\ref{sec:mass} and
Appendix~\ref{appendix:syst}) are reasonable, by comparing them
against the stellar masses of the same objects derived using four
other independent techniques.  In each panel the solid line indicates
the median residual trend, and the dashed lines show the interquartile
range of the residuals.  

The upper-left panel compares our stellar masses against the stellar
masses derived using \kcorrect{} \citep{blanton07a}, fitted to the
same $12$-band UV to mid-infrared photometry as \isedfit{} (see
Section~\ref{sec:sdss}).  We find a weak residual trend with stellar
mass, in the sense that \kcorrect{} yields slightly higher (lower)
stellar masses for lower-mass (higher-mass) galaxies, which for the
massive galaxies at least is similar to the residual trend reported by
\citet{bernardi10a}.  The upper-right and lower-left panels compares
our masses against the stellar masses publicly released by the MPA/JHU
team based on the SDSS
DR4\footnote{\url{http://www.mpa-garching.mpg.de/SDSS/DR4}} and
DR7\footnote{\url{http://www.mpa-garching.mpg.de/SDSS/DR7}} data
releases.  The MPA/JHU-DR4 masses are based on fitting the
H$\delta_{A}$ and $D_{\rm n}(4000)$ optical spectral indices
\citep{kauffmann03a}, and the DR7 stellar masses are derived using a
similar technique as \isedfit, but only fitting to the SDSS $ugriz$
photometry.  In both cases the agreement between the various
independent determinations is outstanding.  Finally, in the
lower-right panel we compare our mass estimates against the stellar
masses derived by \citet{salim07a} using a similar SED-modeling
technique as \isedfit, but just fitting to the \emph{GALEX} plus $ugriz$
photometry.  Once again, the agreement between the two stellar mass
estimates is very good.

\setcounter{figure}{0}
\renewcommand\thefigure{B\arabic{figure}}

\section{Effect of Prior Parameter Choices and Population Synthesis
  Models on our Results}\label{appendix:syst}

\begin{figure*}
\centering
\includegraphics[scale=0.65,angle=90]{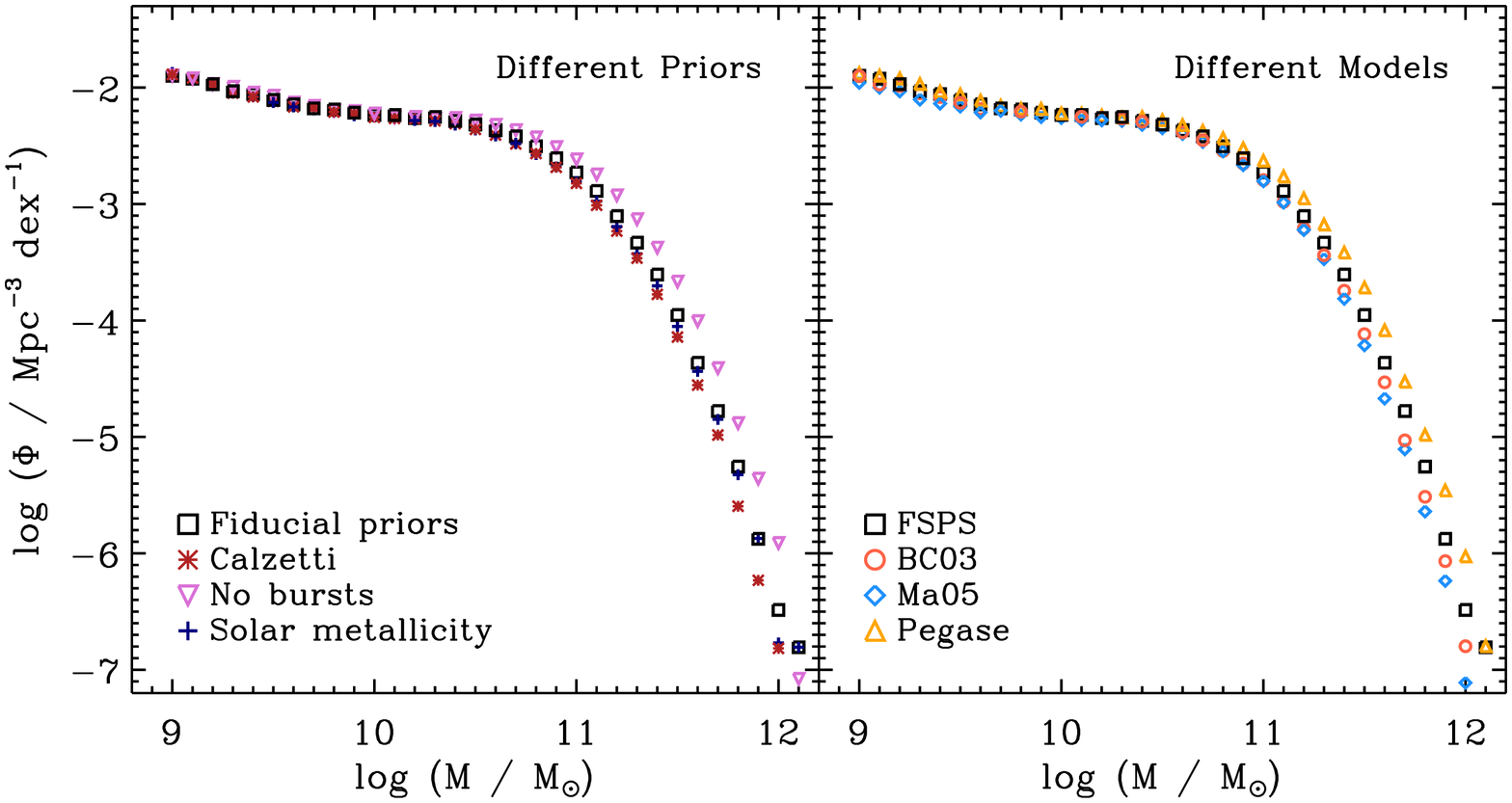}
\caption{Effect on the SDSS-\emph{GALEX} \smf{} of (left) different prior
  parameter combinations and (right) different population synthesis
  models (see the text in Appendix~\ref{appendix:syst} for acronym
  definitions and details).  We find that assuming that galaxies do
  not experience stochastic bursts (``no bursts'') has a significant
  effect on the massive end ($\mass\gtrsim10^{11}$~\msun) of the \smf,
  whereas for all other combinations of priors and population
  synthesis models the effects are relatively small
  ($\lesssim0.1$~dex).  \label{fig:sdsspriors}}
\end{figure*}

In this appendix we examine the effect of varying the SPS models and
prior parameters we use to derive stellar masses and SFRs on our
results (see also Section~\ref{sec:mass}).  We consider four distinct
SPS models: FSPS \citep[v2.3;][]{conroy09a, conroy10b};
\citet[hereafter BC03]{bruzual03a}; the SPS models of \citet[hereafter
  Ma05]{maraston05a}\footnote{http://www-astro.physics.ox.ac.uk/$\sim$maraston/Claudia's\_Stellar\_Population\_Models.html};
and {\sc pegase}\footnote{\url{http://www2.iap.fr/pegase/pegasehr}}
\citep{fioc97a, fioc99a, le-borgne04a}.  For the FSPS and BC03 models
we adopt the \citet{chabrier03a} IMF from $0.1-100$~\msun, and for
Ma05 and {\sc pegase} we use the \citet{kroupa01a} IMF from
$0.1-100$~\msun.  We neglect the $\sim0.03$~dex systematic difference
between the two IMFs.  Each of these models relies on a different
combination of stellar evolution calculations and stellar libraries
and therefore differ in their predictions of the integrated spectra of
galaxies.

\begin{figure*}
\centering
\includegraphics[scale=0.65,angle=90]{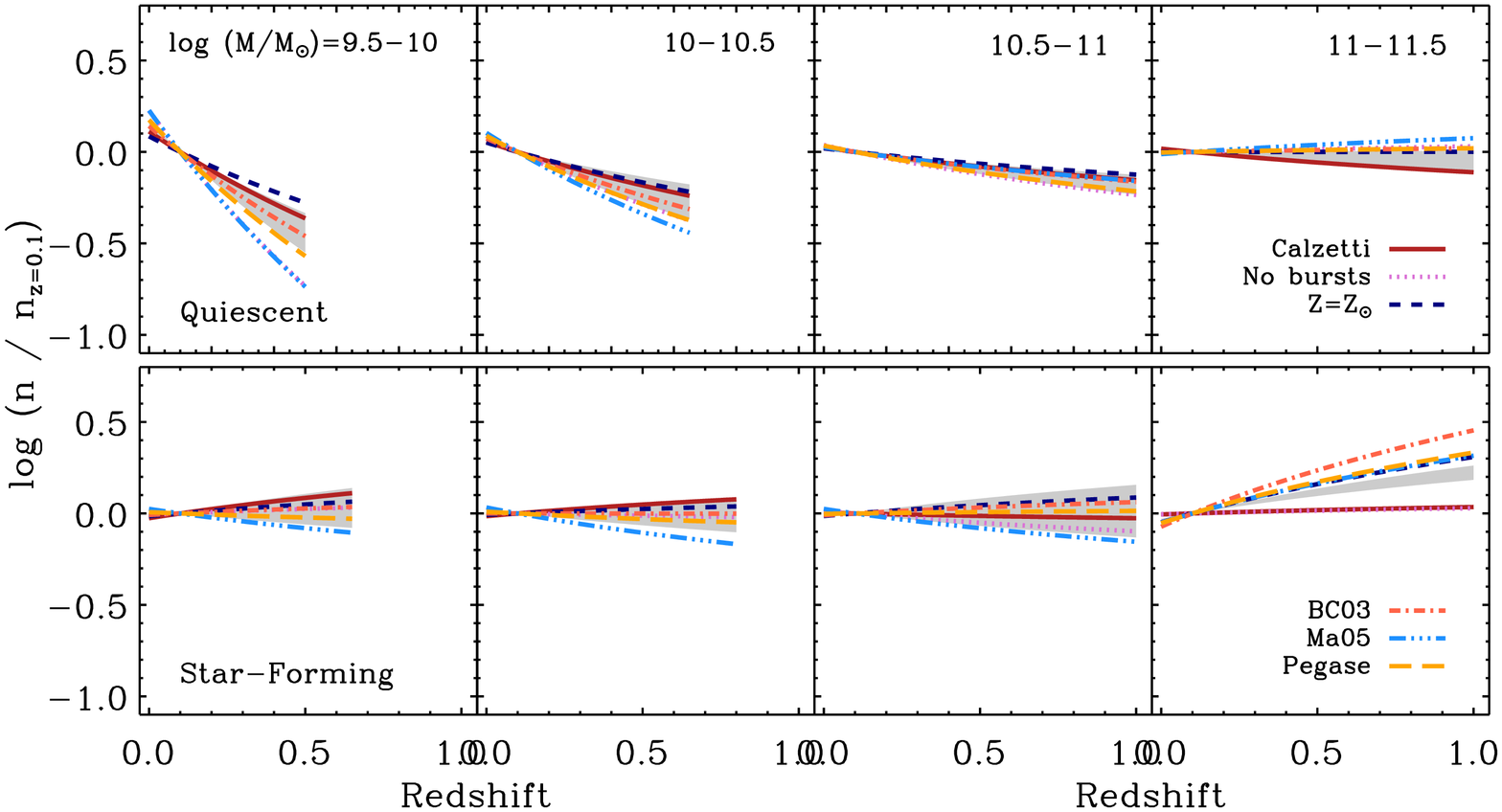}
\caption{Relative number density vs.~redshift for (top) quiescent and
  (bottom) star-forming galaxies in four bins of stellar mass between
  $10^{9.5}-10^{11.5}$~\msun{} based on different SPS models and prior
  assumptions (see text).  Each relation has been normalized by the
  number density at $z=0.1$ so that the relative evolutionary trends
  can be compared.  The grey shaded region reflects the $1\sigma$
  confidence region of the mean fitted relation based on our fiducial
  SPS models and prior assumptions (see Figure~\ref{fig:numbymass}).
  We find that the individual evolutionary trends are generally
  consistent with one another at the $\pm1\sigma$ level except for the
  stellar masses and SFRs derived using the \citet{calzetti00a} dust
  law, which exhibits a shallower (steeper) decline (increase) in the
  number of massive star-forming (quiescent) galaxies.  Overall,
  however, we conclude that our results are broadly insensitive to our
  choice of SPS models and priors.  \label{fig:syst}} 
\end{figure*}

For completeness, we briefly summarize the salient features of each of
these SPS models.  Our fiducial stellar masses and SFRs are based on
the empirically calibrated version of the FSPS models described by
\citet{conroy10b}.  We couple these models to the
Padova\footnote{http://stev.oapd.inaf.it/cgi-bin/cmd} stellar
evolutionary isochrones \citep{girardi00a, marigo07a, marigo08a},
which have been supplemented with the \citet{vassiliadis94a} models
for post-AGB stars.  Integrated spectra are generated using the
low-resolution, semi-empirical BaSeL~3.1 library \citep{lejeune97a,
  lejeune98a, westera02a}, which extends from the UV to the rest-frame
near-infrared, except for the TP-AGB stars, for which the empirical
spectra of \citet{lancon02a} are used over the full wavelength range.
The BC03 models we use are based on the Padova (1994) isochrones
\citep{alongi93a, bressan93a, fagotto94a}, supplemented with the
\citet{vassiliadis93a} and \citet{vassiliadis94a} models for TP-AGB
and post-AGB stars, respectively.  Integrated spectra are synthesized
using the empirical STELIB stellar library \citep{le-borgne03a} in the
optical ($3200-9500$~\AA) and extended into the UV and near-IR at
lower resolution using the BaSeL~3.1 library.  Finally, the Ma05
models are based on the stellar tracks and isochrones through the
main-sequence turnoff published by \citet{schaller92a} and
\citet{cassisi97a, cassisi00a}.  The fuel consumption theorem
\citep{maraston98a, maraston05a} is used to calculate the post-main
sequence phases of stellar evolution, including the TP-AGB phase.
These evolutionary calculations are then tied to the \citet{lancon02a}
empirical spectra for TP-AGB stars and to the BaSeL~3.1 spectral
library for other stellar populations, resulting in a set of
integrated spectra from the UV to the near-infrared.  In our
calculations we adopt the version of the models computed using the
``red horizontal branch'' morphology (see \citealt{maraston05a} for
details).  Finally, {\sc pegase} utilizes the early-1990s version of
the Padova stellar isochrones and couples those to the BaSeL stellar
library.

Our fiducial prior parameters were briefly described in
Section~\ref{sec:mass}, but here we provide more details (see
Appendix~\ref{appendix:isedfit} for additional salient details and
parameter definitions).  We assume exponentially declining SFHs with
Gaussian bursts of varying onset, strength, and duration randomly
superposed.  Following \citet{kauffmann03a} and \citet{salim07a}, we
draw $\tau^{-1}$ from a uniform distribution in the range
$0.01-10$~Gyr$^{-1}$ and allow bursts to occur with a cumulative
probability $P_{\rm burst}=0.5$ every $\Delta P_{\rm burst}=2$~Gyr.
We draw \dtburst{} from a logarithmic distribution in the range
$30-300$~Myr (i.e., shorter-duration bursts are preferred) and
\fburst{} from a logarithmic distribution spanning $0.03-4$
\citep{salim07a, wild09a}.  We allow the age $t$ (time for the onset
of star formation) of each model to range with equal probability
between $0.1-13$~Gyr, although we disallow ages older than the age of
the Universe at the redshift of each galaxy.  We assume a uniform
prior on stellar metallicity $Z$ in the range $0.004-0.03$
\citep[roughly $20\%-150\%$ times the solar
  metallicity;][]{asplund09a}.  Because the SSP models are generally
only available for a small number of tabulated values of $Z$, we
linearly interpolate between these values to obtain an SSP with an
arbitrary metallicity.  Finally, we adopt the time-dependent
attenuation curve of \citet{charlot00a}, in which stellar populations
older than $10$~Myr are attenuated by a factor $\mu$ times less than
younger stellar populations.  We draw $\mu$ from an order four Gamma
distribution that ranges from zero to unity centered on a typical
value $\langle\mu\rangle=0.3$ \citep{charlot01a, wild11a} and the
$V$-band optical depth from an order two Gamma distribution that peaks
around $A_{V}\approx1.2$~mag, with a tail to $A_{V}\approx6$~mag.

We consider the effect of varying a small number of these priors and
SPS models on our results.  Specifically, we consider stellar masses
and SFRs derived assuming: (1) the \citet{calzetti00a} starburst
galaxy attenuation curve (``Calzetti''); (2) that galaxies do not
experience stochastic bursts of star formation (``no bursts''); and
(3) fixed solar metallicity (``solar metallicity'').  Although these
parameter combinations are not exhaustive, they have been chosen to
reasonably span the range of priors commonly adopted in other studies
of the \smf{} \citep[see, e.g.,][]{marchesini09a}.

In Figure~\ref{fig:sdsspriors} we plot the SDSS-\emph{GALEX} \smf{}
derived using these different priors and SPS models.  In the left
panel we use our fiducial FSPS models and vary the prior assumptions,
and in the right panel we use the same fiducial priors and vary the
SPS models.  Overall we find that these variations have a relatively
small systematic effect on the \smf.  The most significant differences
result when we do not include the effects of bursts, which leads to
typically higher stellar masses for massive galaxies.  In
Figure~\ref{fig:syst} we plot number density versus redshift for (top)
quiescent and (bottom) star-forming galaxies in four bins of stellar
mass between $10^{9.5}-10^{11.5}$~\msun{} (see
Section~\ref{sec:mfevol}).  For clarity we only show the slope of the
line [in log-log space; see equation~(\ref{eq:numden_evol})] fitted to
the mean number density of galaxies measured in each redshift bin,
normalized by the number density at $z=0.1$ so that the
\emph{relative} evolutionary trends can be compared.  In each panel,
the grey shaded region reflects the $1\sigma$ confidence region of the
mean fitted relation based on our fiducial SPS models and prior
assumptions (see Figure~\ref{fig:numbymass} and
Table~\ref{table:numden_bymass_coeff}).  We find that the relative
evolutionary trends we infer are generally within $\pm1\sigma$ of the
trends inferred using these other SPS models and prior assumptions.

\end{appendix}

\clearpage
\begin{deluxetable}{lcccc}
\tablecaption{Sample Properties\label{table:sample}}
\tablewidth{0pt}
\tablehead{
\colhead{} & 
\colhead{Selection} & 
\colhead{Magnitude} & 
\colhead{$\Omega$\tablenotemark{b}} & 
\colhead{} \\
\colhead{Field} & 
\colhead{Band} & 
\colhead{Limit\tablenotemark{a}} & 
\colhead{(deg$^{2}$)} & 
\colhead{$N$\tablenotemark{c}} 
}
\startdata
CDFS\tablenotemark{d} & 
$i^{^{\prime}}$ & 
$23.0$ & 
$1.496$ & 
$8050$ \\
COSMOS & 
$I$ & 
$23.0$ & 
$0.856$ & 
$7290$ \\
ELAIS-S1\tablenotemark{d} & 
$R$ & 
$23.2$ & 
$0.800$ & 
$4140$ \\
XMM-SXDS & 
$i^{^{\prime}}$ & 
$23.0$ & 
$0.646$ & 
$6403$ \\
XMM-CFHTLS & 
$i^{^{\prime}}$ & 
$23.0$ & 
$1.700$ & 
$14547$ \\
\cline{1-5}
PRIMUS\tablenotemark{e} & 
 & 
 & 
$5.499$ & 
$40430$ \\
SDSS-\emph{GALEX} & 
$r$ & 
$17.6$ & 
$2504$ & 
$169727$
\enddata
\tablenotetext{a}{Corrected for foreground Galactic extinction.}
\tablenotetext{b}{Angular area surveyed.}
\tablenotetext{c}{Number of galaxies satisfying the selection criteria given in \S\ref{sec:parent}.}
\tablenotetext{d}{Due to differences in the PRIMUS experimental design, our CDFS and ELAIS-S1 samples are also flux-limited in the $3.6$~\micron{} IRAC band between $17<{\rm [3.6]}<21$.}
\tablenotetext{e}{Combination of all five fields.}
\end{deluxetable}

\begin{deluxetable*}{cccccc}
\tablecaption{Stellar Mass Completeness Limits\tablenotemark{a}\label{table:limits}}
\tablewidth{0pt}
\tablehead{
\colhead{} & 
\colhead{COSMOS} & 
\colhead{XMM-SXDS} & 
\colhead{XMM-CFHTLS} & 
\colhead{CDFS} & 
\colhead{ELAIS-S1} \\
\cline{1-6}
\\
\colhead{Redshift Range} & 
\multicolumn{5}{c}{$\log\,(\mass_{\rm lim}/\msun)$} 
}
\startdata
\multicolumn{6}{c}{All} \\
\cline{1-6}
$0.20-0.30$ & 
 8.73 &  8.86 &  8.95 &  9.62 &  9.70 \\
$0.30-0.40$ & 
 9.14 &  9.23 &  9.23 &  9.87 &  9.99 \\
$0.40-0.50$ & 
 9.51 &  9.58 &  9.51 & 10.10 & 10.26 \\
$0.50-0.65$ & 
 9.92 &  9.97 &  9.87 & 10.37 & 10.56 \\
$0.65-0.80$ & 
10.33 & 10.38 & 10.31 & 10.65 & 10.87 \\
$0.80-1.00$ & 
10.71 & 10.78 & 10.83 & 10.94 & 11.17 \\
\cutinhead{Star-Forming}
$0.20-0.30$ & 
 8.68 &  8.79 &  8.80 &  9.60 &  9.58 \\
$0.30-0.40$ & 
 9.05 &  9.13 &  9.06 &  9.92 &  9.94 \\
$0.40-0.50$ & 
 9.38 &  9.44 &  9.30 & 10.19 & 10.25 \\
$0.50-0.65$ & 
 9.75 &  9.77 &  9.58 & 10.44 & 10.59 \\
$0.65-0.80$ & 
10.12 & 10.10 &  9.89 & 10.63 & 10.90 \\
$0.80-1.00$ & 
10.46 & 10.38 & 10.21 & 10.69 & 11.14 \\
\cutinhead{Quiescent}
$0.20-0.30$ & 
 9.23 &  9.35 &  9.17 &  9.65 &  9.80 \\
$0.30-0.40$ & 
 9.58 &  9.61 &  9.52 &  9.92 & 10.06 \\
$0.40-0.50$ & 
 9.89 &  9.85 &  9.85 & 10.17 & 10.30 \\
$0.50-0.65$ & 
10.22 & 10.13 & 10.22 & 10.44 & 10.55 \\
$0.65-0.80$ & 
10.52 & 10.43 & 10.60 & 10.71 & 10.79 \\
$0.80-1.00$ & 
10.75 & 10.73 & 10.96 & 10.96 & 10.99
\enddata
\tablenotetext{a}{Stellar mass completeness limits among all, quiescent, and star-forming galaxies as a function of redshift.  Above these limits our sample includes more than $95\%$ of all types of galaxies, accounting for the flux limit in each field and mass-to-light ratio variations.  For comparison, in our SDSS-\emph{GALEX} sample the completeness limit is $10^{9}$~\msun{} for all three samples.}
\end{deluxetable*}

\begin{deluxetable*}{cccccccccccc}[!h]
\tabletypesize{\small}
\tablecaption{SDSS-\emph{GALEX} Stellar Mass Function\label{table:mflocal}}
\tablewidth{0pt}
\tablehead{
\colhead{} & 
\multicolumn{3}{c}{All} & 
\colhead{} & 
\multicolumn{3}{c}{Star-Forming} & 
\colhead{} & 
\multicolumn{3}{c}{Quiescent} \\
\cline{2-4}\cline{6-8}\cline{10-12} \\
\colhead{$\log\,(\mass)$} & 
\colhead{$\log\,(\Phi)$} & 
\colhead{$\sigma_{\rm SV}$} & 
\colhead{} & 
\colhead{} & 
\colhead{$\log\,(\Phi)$} & 
\colhead{$\sigma_{\rm SV}$} & 
\colhead{} & 
\colhead{} & 
\colhead{$\log\,(\Phi)$} & 
\colhead{$\sigma_{\rm SV}$} & 
\colhead{} \\
\colhead{$(h_{70}^{-2}\,\msun)$} & 
\multicolumn{2}{c}{$(h_{70}^{3}\,{\rm Mpc}^{-3}\,{\rm dex}^{-1})$} & 
\colhead{$N$} & 
\colhead{} & 
\multicolumn{2}{c}{$(h_{70}^{3}\,{\rm Mpc}^{-3}\,{\rm dex}^{-1})$} & 
\colhead{$N$} & 
\colhead{} & 
\multicolumn{2}{c}{$(h_{70}^{3}\,{\rm Mpc}^{-3}\,{\rm dex}^{-1})$} & 
\colhead{$N$} 
}
\startdata
$9.0$ & 
$-1.899^{+0.017}_{-0.017}$ & 
$0.052$ & 
$1040$ & 
 & 
$-2.026^{+0.018}_{-0.017}$ & 
$0.043$ & 
$854$ & 
 & 
$-2.495^{+0.048}_{-0.043}$ & 
$0.096$ & 
$186$ \\
$9.1$ & 
$-1.923^{+0.017}_{-0.016}$ & 
$0.048$ & 
$1239$ & 
 & 
$-2.062^{+0.017}_{-0.016}$ & 
$0.045$ & 
$1030$ & 
 & 
$-2.486^{+0.044}_{-0.041}$ & 
$0.093$ & 
$209$ \\
$9.2$ & 
$-1.970^{+0.015}_{-0.015}$ & 
$0.059$ & 
$1397$ & 
 & 
$-2.129^{+0.015}_{-0.015}$ & 
$0.041$ & 
$1162$ & 
 & 
$-2.485^{+0.038}_{-0.035}$ & 
$0.10$ & 
$235$ \\
$9.3$ & 
$-2.031^{+0.015}_{-0.014}$ & 
$0.052$ & 
$1594$ & 
 & 
$-2.201^{+0.014}_{-0.014}$ & 
$0.044$ & 
$1328$ & 
 & 
$-2.523^{+0.037}_{-0.034}$ & 
$0.10$ & 
$266$ \\
$9.4$ & 
$-2.055^{+0.014}_{-0.013}$ & 
$0.050$ & 
$1874$ & 
 & 
$-2.211^{+0.014}_{-0.013}$ & 
$0.040$ & 
$1594$ & 
 & 
$-2.576^{+0.033}_{-0.031}$ & 
$0.096$ & 
$280$ \\
$9.5$ & 
$-2.106^{+0.012}_{-0.012}$ & 
$0.053$ & 
$2106$ & 
 & 
$-2.272^{+0.012}_{-0.012}$ & 
$0.044$ & 
$1790$ & 
 & 
$-2.603^{+0.030}_{-0.028}$ & 
$0.090$ & 
$316$ \\
$9.6$ & 
$-2.144^{+0.012}_{-0.011}$ & 
$0.046$ & 
$2465$ & 
 & 
$-2.313^{+0.012}_{-0.012}$ & 
$0.040$ & 
$2079$ & 
 & 
$-2.634^{+0.026}_{-0.025}$ & 
$0.070$ & 
$386$ \\
$9.7$ & 
$-2.179^{+0.012}_{-0.012}$ & 
$0.051$ & 
$2820$ & 
 & 
$-2.362^{+0.011}_{-0.011}$ & 
$0.043$ & 
$2385$ & 
 & 
$-2.642^{+0.028}_{-0.026}$ & 
$0.072$ & 
$435$ \\
$9.8$ & 
$-2.188^{+0.010}_{-0.010}$ & 
$0.046$ & 
$3434$ & 
 & 
$-2.371^{+0.011}_{-0.011}$ & 
$0.040$ & 
$2886$ & 
 & 
$-2.652^{+0.021}_{-0.020}$ & 
$0.062$ & 
$548$ \\
$9.9$ & 
$-2.2160^{+0.0086}_{-0.0084}$ & 
$0.048$ & 
$3971$ & 
 & 
$-2.4120^{+0.0092}_{-0.0090}$ & 
$0.039$ & 
$3255$ & 
 & 
$-2.655^{+0.018}_{-0.017}$ & 
$0.065$ & 
$716$ \\
$10.0$ & 
$-2.2340^{+0.0080}_{-0.0078}$ & 
$0.047$ & 
$4667$ & 
 & 
$-2.4450^{+0.0090}_{-0.0088}$ & 
$0.041$ & 
$3710$ & 
 & 
$-2.649^{+0.015}_{-0.015}$ & 
$0.056$ & 
$957$ \\
$10.1$ & 
$-2.2350^{+0.0069}_{-0.0068}$ & 
$0.045$ & 
$5631$ & 
 & 
$-2.4700^{+0.0079}_{-0.0078}$ & 
$0.040$ & 
$4247$ & 
 & 
$-2.614^{+0.013}_{-0.012}$ & 
$0.051$ & 
$1384$ \\
$10.2$ & 
$-2.2620^{+0.0063}_{-0.0062}$ & 
$0.046$ & 
$6601$ & 
 & 
$-2.5240^{+0.0074}_{-0.0072}$ & 
$0.041$ & 
$4746$ & 
 & 
$-2.607^{+0.011}_{-0.011}$ & 
$0.048$ & 
$1855$ \\
$10.3$ & 
$-2.2520^{+0.0056}_{-0.0056}$ & 
$0.049$ & 
$8096$ & 
 & 
$-2.5410^{+0.0071}_{-0.0070}$ & 
$0.042$ & 
$5423$ & 
 & 
$-2.5640^{+0.0089}_{-0.0087}$ & 
$0.050$ & 
$2673$ \\
$10.4$ & 
$-2.2850^{+0.0051}_{-0.0051}$ & 
$0.045$ & 
$9341$ & 
 & 
$-2.6090^{+0.0066}_{-0.0065}$ & 
$0.042$ & 
$5831$ & 
 & 
$-2.5640^{+0.0077}_{-0.0076}$ & 
$0.043$ & 
$3510$ \\
$10.5$ & 
$-2.3170^{+0.0047}_{-0.0046}$ & 
$0.046$ & 
$10901$ & 
 & 
$-2.6600^{+0.0063}_{-0.0062}$ & 
$0.041$ & 
$6441$ & 
 & 
$-2.5800^{+0.0069}_{-0.0068}$ & 
$0.047$ & 
$4460$ \\
$10.6$ & 
$-2.3650^{+0.0044}_{-0.0044}$ & 
$0.049$ & 
$12177$ & 
 & 
$-2.7370^{+0.0062}_{-0.0061}$ & 
$0.043$ & 
$6706$ & 
 & 
$-2.6050^{+0.0062}_{-0.0061}$ & 
$0.049$ & 
$5471$ \\
$10.7$ & 
$-2.4190^{+0.0041}_{-0.0041}$ & 
$0.049$ & 
$13594$ & 
 & 
$-2.8110^{+0.0059}_{-0.0059}$ & 
$0.044$ & 
$7001$ & 
 & 
$-2.6450^{+0.0057}_{-0.0056}$ & 
$0.050$ & 
$6593$ \\
$10.8$ & 
$-2.5040^{+0.0040}_{-0.0040}$ & 
$0.047$ & 
$14172$ & 
 & 
$-2.9340^{+0.0061}_{-0.0060}$ & 
$0.040$ & 
$6580$ & 
 & 
$-2.7050^{+0.0053}_{-0.0052}$ & 
$0.049$ & 
$7592$ \\
$10.9$ & 
$-2.6070^{+0.0039}_{-0.0039}$ & 
$0.046$ & 
$14148$ & 
 & 
$-3.0770^{+0.0064}_{-0.0063}$ & 
$0.041$ & 
$5829$ & 
 & 
$-2.7860^{+0.0050}_{-0.0050}$ & 
$0.046$ & 
$8319$ \\
$11.0$ & 
$-2.7280^{+0.0040}_{-0.0040}$ & 
$0.046$ & 
$13361$ & 
 & 
$-3.2500^{+0.0071}_{-0.0070}$ & 
$0.043$ & 
$4715$ & 
 & 
$-2.8840^{+0.0049}_{-0.0049}$ & 
$0.045$ & 
$8646$ \\
$11.1$ & 
$-2.8880^{+0.0043}_{-0.0043}$ & 
$0.043$ & 
$11592$ & 
 & 
$-3.4720^{+0.0085}_{-0.0084}$ & 
$0.041$ & 
$3306$ & 
 & 
$-3.0190^{+0.0050}_{-0.0050}$ & 
$0.041$ & 
$8286$ \\
$11.2$ & 
$-3.1040^{+0.0049}_{-0.0048}$ & 
$0.041$ & 
$8682$ & 
 & 
$-3.769^{+0.011}_{-0.010}$ & 
$0.044$ & 
$1918$ & 
 & 
$-3.2090^{+0.0055}_{-0.0054}$ & 
$0.038$ & 
$6764$ \\
$11.3$ & 
$-3.3320^{+0.0059}_{-0.0059}$ & 
$0.042$ & 
$5717$ & 
 & 
$-4.102^{+0.016}_{-0.015}$ & 
$0.049$ & 
$936$ & 
 & 
$-3.4130^{+0.0065}_{-0.0064}$ & 
$0.038$ & 
$4781$ \\
$11.4$ & 
$-3.6060^{+0.0080}_{-0.0079}$ & 
$0.042$ & 
$3119$ & 
 & 
$-4.487^{+0.024}_{-0.023}$ & 
$0.052$ & 
$391$ & 
 & 
$-3.6670^{+0.0085}_{-0.0084}$ & 
$0.037$ & 
$2728$ \\
$11.5$ & 
$-3.953^{+0.012}_{-0.012}$ & 
$0.047$ & 
$1398$ & 
 & 
$-4.930^{+0.042}_{-0.038}$ & 
$0.077$ & 
$140$ & 
 & 
$-4.002^{+0.013}_{-0.012}$ & 
$0.041$ & 
$1258$ \\
$11.6$ & 
$-4.363^{+0.020}_{-0.019}$ & 
$0.050$ & 
$535$ & 
 & 
$-5.437^{+0.079}_{-0.067}$ & 
$0.072$ & 
$43$ & 
 & 
$-4.401^{+0.021}_{-0.020}$ & 
$0.046$ & 
$492$ \\
$11.7$ & 
$-4.778^{+0.033}_{-0.031}$ & 
$0.057$ & 
$201$ & 
 & 
$-5.98^{+0.20}_{-0.10}$ & 
$0.10$ & 
$12$ & 
 & 
$-4.806^{+0.034}_{-0.032}$ & 
$0.055$ & 
$189$ \\
$11.8$ & 
$-5.255^{+0.060}_{-0.053}$ & 
$0.066$ & 
$67$ & 
 & 
$-6.30^{+0.30}_{-0.20}$ & 
$0.20$ & 
$6$ & 
 & 
$-5.296^{+0.063}_{-0.056}$ & 
$0.059$ & 
$61$ \\
$11.9$ & 
$-5.87^{+0.10}_{-0.10}$ & 
$0.10$ & 
$16$ & 
 & 
$-6.77^{+0.60}_{-0.30}$ & 
$0.30$ & 
$2$ & 
 & 
$-5.93^{+0.10}_{-0.10}$ & 
$0.10$ & 
$14$ \\
$12.0$ & 
$-6.49^{+0.30}_{-0.20}$ & 
$0.20$ & 
$4$ & 
 & 
$-7.09^{+1.00}_{-0.40}$ & 
$0.40$ & 
$1$ & 
 & 
$-6.61^{+0.40}_{-0.20}$ & 
$0.30$ & 
$3$
\enddata
\end{deluxetable*}

\clearpage
\LongTables
\setlength{\tabcolsep}{0.005in}
\begin{deluxetable}{cccccccccccccccccccccccc}
\tabletypesize{\large}
\tablecaption{PRIMUS Stellar Mass Function for All, Quiescent, and Star-Forming Galaxies\label{table:mfevol}}
\tablewidth{0pt}
\tablehead{
\colhead{$\log\,\mass$} & 
\multicolumn{3}{c}{$0.20<z<0.30$} & 
\colhead{} & 
\multicolumn{3}{c}{$0.30<z<0.40$} & 
\colhead{} & 
\multicolumn{3}{c}{$0.40<z<0.50$} & 
\colhead{} & 
\multicolumn{3}{c}{$0.50<z<0.65$} & 
\colhead{} & 
\multicolumn{3}{c}{$0.65<z<0.80$} & 
\colhead{} & 
\multicolumn{3}{c}{$0.80<z<1.00$} \\
\cline{2-4}\cline{6-8}\cline{10-12}\cline{14-16}\cline{18-20}\cline{22-24}
\colhead{$(\msun)$} & 
\colhead{$\Phi$} & 
\colhead{$\sigma_{\rm cv}$} & 
\colhead{$N$} & 
\colhead{} & 
\colhead{$\Phi$} & 
\colhead{$\sigma_{\rm cv}$} & 
\colhead{$N$} & 
\colhead{} & 
\colhead{$\Phi$} & 
\colhead{$\sigma_{\rm cv}$} & 
\colhead{$N$} & 
\colhead{} & 
\colhead{$\Phi$} & 
\colhead{$\sigma_{\rm cv}$} & 
\colhead{$N$} & 
\colhead{} & 
\colhead{$\Phi$} & 
\colhead{$\sigma_{\rm cv}$} & 
\colhead{$N$} & 
\colhead{} & 
\colhead{$\Phi$} & 
\colhead{$\sigma_{\rm cv}$} & 
\colhead{$N$} 
}
\startdata
\cutinhead{All}
$  8.8$ & 
$-2.009^{+0.054}_{-0.048}$ & 
$0.20$ & 
$172$ & 
 & 
\nodata & 
\nodata & 
\nodata & 
 & 
\nodata & 
\nodata & 
\nodata & 
 & 
\nodata & 
\nodata & 
\nodata & 
 & 
\nodata & 
\nodata & 
\nodata & 
 & 
\nodata & 
\nodata & 
\nodata \\
$  8.9$ & 
$-2.039^{+0.056}_{-0.050}$ & 
$0.20$ & 
$157$ & 
 & 
\nodata & 
\nodata & 
\nodata & 
 & 
\nodata & 
\nodata & 
\nodata & 
 & 
\nodata & 
\nodata & 
\nodata & 
 & 
\nodata & 
\nodata & 
\nodata & 
 & 
\nodata & 
\nodata & 
\nodata \\
$  9.0$ & 
$-2.160^{+0.045}_{-0.041}$ & 
$0.10$ & 
$146$ & 
 & 
\nodata & 
\nodata & 
\nodata & 
 & 
\nodata & 
\nodata & 
\nodata & 
 & 
\nodata & 
\nodata & 
\nodata & 
 & 
\nodata & 
\nodata & 
\nodata & 
 & 
\nodata & 
\nodata & 
\nodata \\
$  9.1$ & 
$-2.185^{+0.050}_{-0.045}$ & 
$0.20$ & 
$140$ & 
 & 
\nodata & 
\nodata & 
\nodata & 
 & 
\nodata & 
\nodata & 
\nodata & 
 & 
\nodata & 
\nodata & 
\nodata & 
 & 
\nodata & 
\nodata & 
\nodata & 
 & 
\nodata & 
\nodata & 
\nodata \\
$  9.2$ & 
$-2.078^{+0.066}_{-0.057}$ & 
$0.10$ & 
$160$ & 
 & 
$-2.132^{+0.043}_{-0.040}$ & 
$0.088$ & 
$240$ & 
 & 
\nodata & 
\nodata & 
\nodata & 
 & 
\nodata & 
\nodata & 
\nodata & 
 & 
\nodata & 
\nodata & 
\nodata & 
 & 
\nodata & 
\nodata & 
\nodata \\
$  9.3$ & 
$-2.085^{+0.061}_{-0.054}$ & 
$0.099$ & 
$170$ & 
 & 
$-2.210^{+0.042}_{-0.038}$ & 
$0.080$ & 
$218$ & 
 & 
\nodata & 
\nodata & 
\nodata & 
 & 
\nodata & 
\nodata & 
\nodata & 
 & 
\nodata & 
\nodata & 
\nodata & 
 & 
\nodata & 
\nodata & 
\nodata \\
$  9.4$ & 
$-2.142^{+0.038}_{-0.035}$ & 
$0.089$ & 
$190$ & 
 & 
$-2.190^{+0.079}_{-0.068}$ & 
$0.20$ & 
$200$ & 
 & 
\nodata & 
\nodata & 
\nodata & 
 & 
\nodata & 
\nodata & 
\nodata & 
 & 
\nodata & 
\nodata & 
\nodata & 
 & 
\nodata & 
\nodata & 
\nodata \\
$  9.5$ & 
$-2.155^{+0.036}_{-0.033}$ & 
$0.042$ & 
$202$ & 
 & 
$-2.183^{+0.055}_{-0.049}$ & 
$0.20$ & 
$238$ & 
 & 
\nodata & 
\nodata & 
\nodata & 
 & 
\nodata & 
\nodata & 
\nodata & 
 & 
\nodata & 
\nodata & 
\nodata & 
 & 
\nodata & 
\nodata & 
\nodata \\
$  9.6$ & 
$-2.124^{+0.036}_{-0.033}$ & 
$0.045$ & 
$227$ & 
 & 
$-2.282^{+0.032}_{-0.030}$ & 
$0.10$ & 
$237$ & 
 & 
$-2.292^{+0.072}_{-0.063}$ & 
$0.20$ & 
$243$ & 
 & 
\nodata & 
\nodata & 
\nodata & 
 & 
\nodata & 
\nodata & 
\nodata & 
 & 
\nodata & 
\nodata & 
\nodata \\
$  9.7$ & 
$-2.200^{+0.037}_{-0.034}$ & 
$0.044$ & 
$206$ & 
 & 
$-2.258^{+0.030}_{-0.028}$ & 
$0.094$ & 
$263$ & 
 & 
$-2.347^{+0.031}_{-0.029}$ & 
$0.20$ & 
$268$ & 
 & 
\nodata & 
\nodata & 
\nodata & 
 & 
\nodata & 
\nodata & 
\nodata & 
 & 
\nodata & 
\nodata & 
\nodata \\
$  9.8$ & 
$-2.212^{+0.034}_{-0.031}$ & 
$0.040$ & 
$207$ & 
 & 
$-2.235^{+0.038}_{-0.035}$ & 
$0.078$ & 
$288$ & 
 & 
$-2.289^{+0.030}_{-0.028}$ & 
$0.10$ & 
$311$ & 
 & 
\nodata & 
\nodata & 
\nodata & 
 & 
\nodata & 
\nodata & 
\nodata & 
 & 
\nodata & 
\nodata & 
\nodata \\
$  9.9$ & 
$-2.242^{+0.034}_{-0.032}$ & 
$0.035$ & 
$197$ & 
 & 
$-2.241^{+0.029}_{-0.027}$ & 
$0.080$ & 
$290$ & 
 & 
$-2.308^{+0.040}_{-0.036}$ & 
$0.097$ & 
$308$ & 
 & 
\nodata & 
\nodata & 
\nodata & 
 & 
\nodata & 
\nodata & 
\nodata & 
 & 
\nodata & 
\nodata & 
\nodata \\
$ 10.0$ & 
$-2.215^{+0.038}_{-0.035}$ & 
$0.052$ & 
$199$ & 
 & 
$-2.208^{+0.031}_{-0.029}$ & 
$0.031$ & 
$327$ & 
 & 
$-2.325^{+0.028}_{-0.027}$ & 
$0.060$ & 
$328$ & 
 & 
$-2.419^{+0.036}_{-0.033}$ & 
$0.074$ & 
$459$ & 
 & 
\nodata & 
\nodata & 
\nodata & 
 & 
\nodata & 
\nodata & 
\nodata \\
$ 10.1$ & 
$-2.320^{+0.037}_{-0.034}$ & 
$0.033$ & 
$169$ & 
 & 
$-2.288^{+0.033}_{-0.030}$ & 
$0.071$ & 
$288$ & 
 & 
$-2.253^{+0.087}_{-0.073}$ & 
$0.023$ & 
$342$ & 
 & 
$-2.394^{+0.027}_{-0.026}$ & 
$0.058$ & 
$499$ & 
 & 
\nodata & 
\nodata & 
\nodata & 
 & 
\nodata & 
\nodata & 
\nodata \\
$ 10.2$ & 
$-2.285^{+0.035}_{-0.033}$ & 
$0.073$ & 
$184$ & 
 & 
$-2.241^{+0.026}_{-0.024}$ & 
$0.041$ & 
$338$ & 
 & 
$-2.342^{+0.030}_{-0.028}$ & 
$0.093$ & 
$322$ & 
 & 
$-2.371^{+0.022}_{-0.021}$ & 
$0.048$ & 
$550$ & 
 & 
\nodata & 
\nodata & 
\nodata & 
 & 
\nodata & 
\nodata & 
\nodata \\
$ 10.3$ & 
$-2.330^{+0.038}_{-0.035}$ & 
$0.039$ & 
$166$ & 
 & 
$-2.233^{+0.027}_{-0.025}$ & 
$0.017$ & 
$341$ & 
 & 
$-2.372^{+0.027}_{-0.025}$ & 
$0.055$ & 
$326$ & 
 & 
$-2.388^{+0.027}_{-0.025}$ & 
$0.036$ & 
$550$ & 
 & 
\nodata & 
\nodata & 
\nodata & 
 & 
\nodata & 
\nodata & 
\nodata \\
$ 10.4$ & 
$-2.350^{+0.038}_{-0.035}$ & 
$0.039$ & 
$158$ & 
 & 
$-2.290^{+0.027}_{-0.025}$ & 
$0.043$ & 
$316$ & 
 & 
$-2.327^{+0.036}_{-0.033}$ & 
$0.026$ & 
$357$ & 
 & 
$-2.382^{+0.021}_{-0.020}$ & 
$0.050$ & 
$588$ & 
 & 
$-2.387^{+0.022}_{-0.021}$ & 
$0.047$ & 
$593$ & 
 & 
\nodata & 
\nodata & 
\nodata \\
$ 10.5$ & 
$-2.380^{+0.039}_{-0.036}$ & 
$0.062$ & 
$154$ & 
 & 
$-2.283^{+0.027}_{-0.025}$ & 
$0.050$ & 
$309$ & 
 & 
$-2.332^{+0.030}_{-0.028}$ & 
$0.072$ & 
$369$ & 
 & 
$-2.346^{+0.020}_{-0.019}$ & 
$0.035$ & 
$667$ & 
 & 
$-2.320^{+0.027}_{-0.025}$ & 
$0.037$ & 
$672$ & 
 & 
\nodata & 
\nodata & 
\nodata \\
$ 10.6$ & 
$-2.396^{+0.041}_{-0.037}$ & 
$0.067$ & 
$145$ & 
 & 
$-2.332^{+0.028}_{-0.027}$ & 
$0.082$ & 
$278$ & 
 & 
$-2.384^{+0.028}_{-0.026}$ & 
$0.065$ & 
$336$ & 
 & 
$-2.408^{+0.020}_{-0.019}$ & 
$0.031$ & 
$614$ & 
 & 
$-2.353^{+0.033}_{-0.031}$ & 
$0.055$ & 
$673$ & 
 & 
\nodata & 
\nodata & 
\nodata \\
$ 10.7$ & 
$-2.422^{+0.043}_{-0.039}$ & 
$0.040$ & 
$132$ & 
 & 
$-2.407^{+0.031}_{-0.029}$ & 
$0.060$ & 
$238$ & 
 & 
$-2.360^{+0.033}_{-0.031}$ & 
$0.049$ & 
$331$ & 
 & 
$-2.431^{+0.020}_{-0.019}$ & 
$0.039$ & 
$609$ & 
 & 
$-2.387^{+0.029}_{-0.027}$ & 
$0.053$ & 
$673$ & 
 & 
\nodata & 
\nodata & 
\nodata \\
$ 10.8$ & 
$-2.542^{+0.052}_{-0.047}$ & 
$0.033$ & 
$94$ & 
 & 
$-2.472^{+0.034}_{-0.032}$ & 
$0.070$ & 
$199$ & 
 & 
$-2.493^{+0.029}_{-0.028}$ & 
$0.086$ & 
$265$ & 
 & 
$-2.502^{+0.021}_{-0.020}$ & 
$0.038$ & 
$525$ & 
 & 
$-2.443^{+0.020}_{-0.019}$ & 
$0.045$ & 
$667$ & 
 & 
$-2.583^{+0.028}_{-0.026}$ & 
$0.084$ & 
$521$ \\
$ 10.9$ & 
$-2.642^{+0.063}_{-0.055}$ & 
$0.074$ & 
$69$ & 
 & 
$-2.579^{+0.042}_{-0.039}$ & 
$0.083$ & 
$150$ & 
 & 
$-2.644^{+0.039}_{-0.036}$ & 
$0.041$ & 
$187$ & 
 & 
$-2.602^{+0.025}_{-0.023}$ & 
$0.031$ & 
$428$ & 
 & 
$-2.487^{+0.023}_{-0.022}$ & 
$0.054$ & 
$626$ & 
 & 
$-2.658^{+0.023}_{-0.022}$ & 
$0.052$ & 
$517$ \\
$ 11.0$ & 
$-2.784^{+0.089}_{-0.075}$ & 
$0.10$ & 
$39$ & 
 & 
$-2.709^{+0.046}_{-0.042}$ & 
$0.048$ & 
$117$ & 
 & 
$-2.734^{+0.039}_{-0.036}$ & 
$0.058$ & 
$159$ & 
 & 
$-2.729^{+0.027}_{-0.025}$ & 
$0.023$ & 
$329$ & 
 & 
$-2.599^{+0.033}_{-0.031}$ & 
$0.052$ & 
$473$ & 
 & 
$-2.701^{+0.028}_{-0.026}$ & 
$0.061$ & 
$499$ \\
$ 11.1$ & 
$-2.83^{+0.10}_{-0.084}$ & 
$0.30$ & 
$30$ & 
 & 
$-2.819^{+0.052}_{-0.047}$ & 
$0.037$ & 
$90$ & 
 & 
$-2.978^{+0.052}_{-0.047}$ & 
$0.084$ & 
$93$ & 
 & 
$-2.921^{+0.033}_{-0.031}$ & 
$0.058$ & 
$212$ & 
 & 
$-2.772^{+0.028}_{-0.026}$ & 
$0.040$ & 
$352$ & 
 & 
$-2.842^{+0.028}_{-0.026}$ & 
$0.059$ & 
$398$ \\
$ 11.2$ & 
$-3.17^{+0.40}_{-0.20}$ & 
$0.073$ & 
$8$ & 
 & 
$-3.109^{+0.076}_{-0.065}$ & 
$0.059$ & 
$46$ & 
 & 
$-3.114^{+0.066}_{-0.057}$ & 
$0.10$ & 
$64$ & 
 & 
$-3.118^{+0.042}_{-0.039}$ & 
$0.065$ & 
$139$ & 
 & 
$-2.919^{+0.042}_{-0.039}$ & 
$0.091$ & 
$235$ & 
 & 
$-3.039^{+0.035}_{-0.032}$ & 
$0.036$ & 
$274$ \\
$ 11.3$ & 
$-3.54^{+0.30}_{-0.20}$ & 
$0.030$ & 
$6$ & 
 & 
$-3.34^{+0.10}_{-0.086}$ & 
$0.090$ & 
$27$ & 
 & 
$-3.46^{+0.10}_{-0.083}$ & 
$0.10$ & 
$30$ & 
 & 
$-3.311^{+0.059}_{-0.052}$ & 
$0.076$ & 
$82$ & 
 & 
$-3.233^{+0.045}_{-0.041}$ & 
$0.065$ & 
$133$ & 
 & 
$-3.296^{+0.039}_{-0.036}$ & 
$0.077$ & 
$169$ \\
$ 11.4$ & 
\nodata & 
\nodata & 
\nodata & 
 & 
$-3.58^{+0.20}_{-0.10}$ & 
$0.079$ & 
$12$ & 
 & 
$-3.67^{+0.10}_{-0.10}$ & 
$0.20$ & 
$17$ & 
 & 
$-3.649^{+0.083}_{-0.071}$ & 
$0.083$ & 
$42$ & 
 & 
$-3.470^{+0.056}_{-0.050}$ & 
$0.10$ & 
$81$ & 
 & 
$-3.453^{+0.079}_{-0.068}$ & 
$0.063$ & 
$100$ \\
$ 11.5$ & 
\nodata & 
\nodata & 
\nodata & 
 & 
$-4.34^{+0.40}_{-0.20}$ & 
$0.044$ & 
$3$ & 
 & 
$-4.12^{+0.30}_{-0.20}$ & 
$0.047$ & 
$6$ & 
 & 
$-3.80^{+0.10}_{-0.10}$ & 
$0.094$ & 
$23$ & 
 & 
$-3.93^{+0.10}_{-0.097}$ & 
$0.20$ & 
$22$ & 
 & 
$-3.77^{+0.10}_{-0.090}$ & 
$0.084$ & 
$50$ \\
$ 11.6$ & 
\nodata & 
\nodata & 
\nodata & 
 & 
\nodata & 
\nodata & 
\nodata & 
 & 
$-4.35^{+0.40}_{-0.20}$ & 
$0.091$ & 
$4$ & 
 & 
$-4.52^{+0.20}_{-0.20}$ & 
$0.10$ & 
$7$ & 
 & 
$-4.22^{+0.20}_{-0.10}$ & 
$0.088$ & 
$11$ & 
 & 
$-4.32^{+0.10}_{-0.10}$ & 
$0.087$ & 
$19$ \\
$ 11.7$ & 
\nodata & 
\nodata & 
\nodata & 
 & 
\nodata & 
\nodata & 
\nodata & 
 & 
$-5.09^{+1.00}_{-0.40}$ & 
$0.50$ & 
$1$ & 
 & 
$-4.21^{+0.20}_{-0.20}$ & 
$0.10$ & 
$8$ & 
 & 
$-4.60^{+0.30}_{-0.20}$ & 
$0.20$ & 
$5$ & 
 & 
$-4.44^{+0.20}_{-0.20}$ & 
$0.10$ & 
$11$ \\
$ 11.8$ & 
\nodata & 
\nodata & 
\nodata & 
 & 
\nodata & 
\nodata & 
\nodata & 
 & 
$-5.05^{+1.00}_{-0.40}$ & 
$0.50$ & 
$1$ & 
 & 
\nodata & 
\nodata & 
\nodata & 
 & 
$-4.94^{+1.00}_{-0.40}$ & 
$0.40$ & 
$1$ & 
 & 
$-5.07^{+0.40}_{-0.20}$ & 
$0.088$ & 
$3$ \\
$ 11.9$ & 
\nodata & 
\nodata & 
\nodata & 
 & 
\nodata & 
\nodata & 
\nodata & 
 & 
\nodata & 
\nodata & 
\nodata & 
 & 
\nodata & 
\nodata & 
\nodata & 
 & 
$-4.48^{+1.00}_{-0.40}$ & 
$0.10$ & 
$1$ & 
 & 
\nodata & 
\nodata & 
\nodata \\
$ 12.0$ & 
\nodata & 
\nodata & 
\nodata & 
 & 
\nodata & 
\nodata & 
\nodata & 
 & 
\nodata & 
\nodata & 
\nodata & 
 & 
\nodata & 
\nodata & 
\nodata & 
 & 
$-5.48^{+1.00}_{-0.40}$ & 
$1.4$ & 
$1$ & 
 & 
$-5.71^{+1.00}_{-0.40}$ & 
$0.40$ & 
$1$ \\
\cutinhead{Star-Forming}
$  8.8$ & 
$-2.014^{+0.054}_{-0.049}$ & 
$0.20$ & 
$171$ & 
 & 
\nodata & 
\nodata & 
\nodata & 
 & 
\nodata & 
\nodata & 
\nodata & 
 & 
\nodata & 
\nodata & 
\nodata & 
 & 
\nodata & 
\nodata & 
\nodata & 
 & 
\nodata & 
\nodata & 
\nodata \\
$  8.9$ & 
$-2.125^{+0.043}_{-0.040}$ & 
$0.084$ & 
$149$ & 
 & 
\nodata & 
\nodata & 
\nodata & 
 & 
\nodata & 
\nodata & 
\nodata & 
 & 
\nodata & 
\nodata & 
\nodata & 
 & 
\nodata & 
\nodata & 
\nodata & 
 & 
\nodata & 
\nodata & 
\nodata \\
$  9.0$ & 
$-2.205^{+0.044}_{-0.040}$ & 
$0.10$ & 
$140$ & 
 & 
\nodata & 
\nodata & 
\nodata & 
 & 
\nodata & 
\nodata & 
\nodata & 
 & 
\nodata & 
\nodata & 
\nodata & 
 & 
\nodata & 
\nodata & 
\nodata & 
 & 
\nodata & 
\nodata & 
\nodata \\
$  9.1$ & 
$-2.250^{+0.045}_{-0.041}$ & 
$0.20$ & 
$133$ & 
 & 
$-2.156^{+0.057}_{-0.051}$ & 
$0.097$ & 
$196$ & 
 & 
\nodata & 
\nodata & 
\nodata & 
 & 
\nodata & 
\nodata & 
\nodata & 
 & 
\nodata & 
\nodata & 
\nodata & 
 & 
\nodata & 
\nodata & 
\nodata \\
$  9.2$ & 
$-2.133^{+0.071}_{-0.062}$ & 
$0.10$ & 
$147$ & 
 & 
$-2.196^{+0.032}_{-0.030}$ & 
$0.085$ & 
$235$ & 
 & 
\nodata & 
\nodata & 
\nodata & 
 & 
\nodata & 
\nodata & 
\nodata & 
 & 
\nodata & 
\nodata & 
\nodata & 
 & 
\nodata & 
\nodata & 
\nodata \\
$  9.3$ & 
$-2.114^{+0.064}_{-0.056}$ & 
$0.087$ & 
$160$ & 
 & 
$-2.290^{+0.034}_{-0.031}$ & 
$0.073$ & 
$209$ & 
 & 
\nodata & 
\nodata & 
\nodata & 
 & 
\nodata & 
\nodata & 
\nodata & 
 & 
\nodata & 
\nodata & 
\nodata & 
 & 
\nodata & 
\nodata & 
\nodata \\
$  9.4$ & 
$-2.216^{+0.038}_{-0.035}$ & 
$0.099$ & 
$171$ & 
 & 
$-2.316^{+0.041}_{-0.037}$ & 
$0.20$ & 
$188$ & 
 & 
$-2.325^{+0.034}_{-0.032}$ & 
$0.10$ & 
$228$ & 
 & 
\nodata & 
\nodata & 
\nodata & 
 & 
\nodata & 
\nodata & 
\nodata & 
 & 
\nodata & 
\nodata & 
\nodata \\
$  9.5$ & 
$-2.233^{+0.037}_{-0.034}$ & 
$0.052$ & 
$180$ & 
 & 
$-2.294^{+0.034}_{-0.031}$ & 
$0.10$ & 
$220$ & 
 & 
$-2.335^{+0.034}_{-0.032}$ & 
$0.10$ & 
$248$ & 
 & 
\nodata & 
\nodata & 
\nodata & 
 & 
\nodata & 
\nodata & 
\nodata & 
 & 
\nodata & 
\nodata & 
\nodata \\
$  9.6$ & 
$-2.231^{+0.035}_{-0.033}$ & 
$0.053$ & 
$195$ & 
 & 
$-2.364^{+0.034}_{-0.031}$ & 
$0.088$ & 
$208$ & 
 & 
$-2.318^{+0.077}_{-0.066}$ & 
$0.20$ & 
$233$ & 
 & 
\nodata & 
\nodata & 
\nodata & 
 & 
\nodata & 
\nodata & 
\nodata & 
 & 
\nodata & 
\nodata & 
\nodata \\
$  9.7$ & 
$-2.289^{+0.041}_{-0.038}$ & 
$0.047$ & 
$170$ & 
 & 
$-2.340^{+0.032}_{-0.030}$ & 
$0.075$ & 
$229$ & 
 & 
$-2.369^{+0.032}_{-0.030}$ & 
$0.20$ & 
$261$ & 
 & 
$-2.438^{+0.029}_{-0.027}$ & 
$0.20$ & 
$383$ & 
 & 
\nodata & 
\nodata & 
\nodata & 
 & 
\nodata & 
\nodata & 
\nodata \\
$  9.8$ & 
$-2.336^{+0.039}_{-0.036}$ & 
$0.036$ & 
$159$ & 
 & 
$-2.346^{+0.030}_{-0.029}$ & 
$0.054$ & 
$248$ & 
 & 
$-2.364^{+0.032}_{-0.029}$ & 
$0.10$ & 
$280$ & 
 & 
$-2.528^{+0.027}_{-0.025}$ & 
$0.20$ & 
$354$ & 
 & 
\nodata & 
\nodata & 
\nodata & 
 & 
\nodata & 
\nodata & 
\nodata \\
$  9.9$ & 
$-2.347^{+0.039}_{-0.036}$ & 
$0.042$ & 
$154$ & 
 & 
$-2.354^{+0.033}_{-0.031}$ & 
$0.074$ & 
$233$ & 
 & 
$-2.402^{+0.033}_{-0.030}$ & 
$0.087$ & 
$275$ & 
 & 
$-2.493^{+0.033}_{-0.031}$ & 
$0.10$ & 
$390$ & 
 & 
\nodata & 
\nodata & 
\nodata & 
 & 
\nodata & 
\nodata & 
\nodata \\
$ 10.0$ & 
$-2.364^{+0.046}_{-0.042}$ & 
$0.032$ & 
$143$ & 
 & 
$-2.354^{+0.038}_{-0.035}$ & 
$0.030$ & 
$236$ & 
 & 
$-2.430^{+0.031}_{-0.029}$ & 
$0.033$ & 
$276$ & 
 & 
$-2.480^{+0.039}_{-0.036}$ & 
$0.052$ & 
$421$ & 
 & 
$-2.642^{+0.029}_{-0.027}$ & 
$0.20$ & 
$329$ & 
 & 
\nodata & 
\nodata & 
\nodata \\
$ 10.1$ & 
$-2.509^{+0.047}_{-0.043}$ & 
$0.035$ & 
$109$ & 
 & 
$-2.452^{+0.042}_{-0.039}$ & 
$0.081$ & 
$200$ & 
 & 
$-2.36^{+0.10}_{-0.093}$ & 
$0.048$ & 
$268$ & 
 & 
$-2.516^{+0.024}_{-0.022}$ & 
$0.037$ & 
$434$ & 
 & 
$-2.637^{+0.032}_{-0.030}$ & 
$0.082$ & 
$329$ & 
 & 
\nodata & 
\nodata & 
\nodata \\
$ 10.2$ & 
$-2.526^{+0.047}_{-0.043}$ & 
$0.077$ & 
$108$ & 
 & 
$-2.426^{+0.032}_{-0.030}$ & 
$0.038$ & 
$224$ & 
 & 
$-2.502^{+0.036}_{-0.033}$ & 
$0.090$ & 
$235$ & 
 & 
$-2.491^{+0.025}_{-0.024}$ & 
$0.018$ & 
$443$ & 
 & 
$-2.564^{+0.030}_{-0.028}$ & 
$0.055$ & 
$406$ & 
 & 
\nodata & 
\nodata & 
\nodata \\
$ 10.3$ & 
$-2.576^{+0.050}_{-0.045}$ & 
$0.039$ & 
$97$ & 
 & 
$-2.430^{+0.032}_{-0.030}$ & 
$0.026$ & 
$223$ & 
 & 
$-2.530^{+0.032}_{-0.030}$ & 
$0.050$ & 
$234$ & 
 & 
$-2.593^{+0.025}_{-0.023}$ & 
$0.046$ & 
$384$ & 
 & 
$-2.527^{+0.031}_{-0.029}$ & 
$0.041$ & 
$429$ & 
 & 
$-2.639^{+0.066}_{-0.058}$ & 
$0.064$ & 
$340$ \\
$ 10.4$ & 
$-2.715^{+0.058}_{-0.051}$ & 
$0.10$ & 
$74$ & 
 & 
$-2.515^{+0.035}_{-0.033}$ & 
$0.054$ & 
$188$ & 
 & 
$-2.545^{+0.032}_{-0.029}$ & 
$0.055$ & 
$231$ & 
 & 
$-2.593^{+0.026}_{-0.025}$ & 
$0.056$ & 
$380$ & 
 & 
$-2.561^{+0.025}_{-0.024}$ & 
$0.038$ & 
$438$ & 
 & 
$-2.697^{+0.057}_{-0.051}$ & 
$0.10$ & 
$323$ \\
$ 10.5$ & 
$-2.614^{+0.052}_{-0.046}$ & 
$0.065$ & 
$92$ & 
 & 
$-2.542^{+0.037}_{-0.034}$ & 
$0.045$ & 
$172$ & 
 & 
$-2.556^{+0.036}_{-0.033}$ & 
$0.063$ & 
$227$ & 
 & 
$-2.566^{+0.028}_{-0.026}$ & 
$0.026$ & 
$412$ & 
 & 
$-2.575^{+0.025}_{-0.024}$ & 
$0.063$ & 
$441$ & 
 & 
$-2.783^{+0.044}_{-0.040}$ & 
$0.081$ & 
$304$ \\
$ 10.6$ & 
$-2.691^{+0.059}_{-0.053}$ & 
$0.083$ & 
$73$ & 
 & 
$-2.640^{+0.041}_{-0.038}$ & 
$0.056$ & 
$139$ & 
 & 
$-2.617^{+0.038}_{-0.035}$ & 
$0.044$ & 
$197$ & 
 & 
$-2.646^{+0.028}_{-0.026}$ & 
$0.044$ & 
$357$ & 
 & 
$-2.579^{+0.050}_{-0.045}$ & 
$0.075$ & 
$401$ & 
 & 
$-2.873^{+0.036}_{-0.034}$ & 
$0.038$ & 
$301$ \\
$ 10.7$ & 
$-2.769^{+0.065}_{-0.057}$ & 
$0.066$ & 
$61$ & 
 & 
$-2.736^{+0.046}_{-0.041}$ & 
$0.056$ & 
$116$ & 
 & 
$-2.615^{+0.036}_{-0.033}$ & 
$0.074$ & 
$195$ & 
 & 
$-2.686^{+0.028}_{-0.026}$ & 
$0.048$ & 
$332$ & 
 & 
$-2.612^{+0.045}_{-0.041}$ & 
$0.082$ & 
$390$ & 
 & 
$-2.944^{+0.034}_{-0.031}$ & 
$0.066$ & 
$289$ \\
$ 10.8$ & 
$-2.894^{+0.087}_{-0.073}$ & 
$0.096$ & 
$39$ & 
 & 
$-2.865^{+0.055}_{-0.049}$ & 
$0.071$ & 
$83$ & 
 & 
$-2.762^{+0.041}_{-0.037}$ & 
$0.069$ & 
$146$ & 
 & 
$-2.815^{+0.031}_{-0.029}$ & 
$0.045$ & 
$252$ & 
 & 
$-2.744^{+0.029}_{-0.027}$ & 
$0.053$ & 
$333$ & 
 & 
$-2.908^{+0.041}_{-0.038}$ & 
$0.10$ & 
$280$ \\
$ 10.9$ & 
$-3.18^{+0.10}_{-0.10}$ & 
$0.028$ & 
$19$ & 
 & 
$-2.905^{+0.059}_{-0.052}$ & 
$0.063$ & 
$74$ & 
 & 
$-2.978^{+0.067}_{-0.059}$ & 
$0.071$ & 
$84$ & 
 & 
$-2.980^{+0.035}_{-0.033}$ & 
$0.033$ & 
$189$ & 
 & 
$-2.788^{+0.037}_{-0.034}$ & 
$0.079$ & 
$304$ & 
 & 
$-3.011^{+0.036}_{-0.033}$ & 
$0.079$ & 
$242$ \\
$ 11.0$ & 
$-3.38^{+0.20}_{-0.20}$ & 
$0.20$ & 
$7$ & 
 & 
$-3.113^{+0.078}_{-0.067}$ & 
$0.054$ & 
$46$ & 
 & 
$-3.117^{+0.061}_{-0.054}$ & 
$0.092$ & 
$69$ & 
 & 
$-3.139^{+0.043}_{-0.039}$ & 
$0.023$ & 
$130$ & 
 & 
$-3.000^{+0.042}_{-0.038}$ & 
$0.049$ & 
$193$ & 
 & 
$-3.113^{+0.043}_{-0.039}$ & 
$0.10$ & 
$203$ \\
$ 11.1$ & 
$-3.37^{+0.20}_{-0.10}$ & 
$0.094$ & 
$10$ & 
 & 
$-3.43^{+0.10}_{-0.090}$ & 
$0.086$ & 
$24$ & 
 & 
$-3.347^{+0.083}_{-0.070}$ & 
$0.061$ & 
$40$ & 
 & 
$-3.368^{+0.060}_{-0.053}$ & 
$0.090$ & 
$73$ & 
 & 
$-3.148^{+0.045}_{-0.041}$ & 
$0.097$ & 
$139$ & 
 & 
$-3.279^{+0.044}_{-0.040}$ & 
$0.10$ & 
$148$ \\
$ 11.2$ & 
\nodata & 
\nodata & 
\nodata & 
 & 
$-3.67^{+0.20}_{-0.10}$ & 
$0.076$ & 
$13$ & 
 & 
$-3.54^{+0.10}_{-0.093}$ & 
$0.10$ & 
$24$ & 
 & 
$-3.539^{+0.069}_{-0.060}$ & 
$0.084$ & 
$55$ & 
 & 
$-3.286^{+0.059}_{-0.052}$ & 
$0.10$ & 
$107$ & 
 & 
$-3.405^{+0.063}_{-0.055}$ & 
$0.065$ & 
$114$ \\
$ 11.3$ & 
$-4.60^{+1.00}_{-0.40}$ & 
$1.6$ & 
$1$ & 
 & 
$-4.04^{+0.40}_{-0.20}$ & 
$0.20$ & 
$4$ & 
 & 
$-3.83^{+0.20}_{-0.10}$ & 
$0.090$ & 
$13$ & 
 & 
$-3.93^{+0.10}_{-0.100}$ & 
$0.10$ & 
$20$ & 
 & 
$-3.671^{+0.072}_{-0.063}$ & 
$0.10$ & 
$51$ & 
 & 
$-3.630^{+0.063}_{-0.056}$ & 
$0.095$ & 
$72$ \\
$ 11.4$ & 
\nodata & 
\nodata & 
\nodata & 
 & 
$-4.15^{+0.40}_{-0.20}$ & 
$0.20$ & 
$3$ & 
 & 
$-4.23^{+0.40}_{-0.20}$ & 
$0.20$ & 
$4$ & 
 & 
$-4.08^{+0.20}_{-0.10}$ & 
$0.10$ & 
$15$ & 
 & 
$-4.04^{+0.10}_{-0.092}$ & 
$0.096$ & 
$23$ & 
 & 
$-3.92^{+0.10}_{-0.087}$ & 
$0.10$ & 
$37$ \\
$ 11.5$ & 
\nodata & 
\nodata & 
\nodata & 
 & 
$-4.57^{+0.60}_{-0.30}$ & 
$0.60$ & 
$2$ & 
 & 
$-4.73^{+1.00}_{-0.40}$ & 
$0.70$ & 
$1$ & 
 & 
$-4.66^{+0.30}_{-0.20}$ & 
$0.50$ & 
$5$ & 
 & 
$-4.43^{+0.20}_{-0.20}$ & 
$0.20$ & 
$7$ & 
 & 
$-4.22^{+0.10}_{-0.10}$ & 
$0.046$ & 
$19$ \\
$ 11.6$ & 
\nodata & 
\nodata & 
\nodata & 
 & 
\nodata & 
\nodata & 
\nodata & 
 & 
$-5.01^{+1.00}_{-0.40}$ & 
$1.4$ & 
$1$ & 
 & 
$-5.34^{+1.00}_{-0.40}$ & 
$0.0089$ & 
$1$ & 
 & 
$-5.06^{+0.60}_{-0.30}$ & 
$1.0$ & 
$2$ & 
 & 
$-5.05^{+0.40}_{-0.20}$ & 
$0.10$ & 
$3$ \\
$ 11.7$ & 
\nodata & 
\nodata & 
\nodata & 
 & 
\nodata & 
\nodata & 
\nodata & 
 & 
\nodata & 
\nodata & 
\nodata & 
 & 
$-4.81^{+0.40}_{-0.20}$ & 
$0.0026$ & 
$3$ & 
 & 
$-4.75^{+0.50}_{-0.30}$ & 
$0.50$ & 
$3$ & 
 & 
$-4.81^{+0.60}_{-0.30}$ & 
$0.20$ & 
$3$ \\
$ 11.8$ & 
\nodata & 
\nodata & 
\nodata & 
 & 
\nodata & 
\nodata & 
\nodata & 
 & 
\nodata & 
\nodata & 
\nodata & 
 & 
\nodata & 
\nodata & 
\nodata & 
 & 
$-4.94^{+1.00}_{-0.40}$ & 
$0.80$ & 
$1$ & 
 & 
$-5.42^{+1.00}_{-0.40}$ & 
$0.60$ & 
$1$ \\
$ 12.0$ & 
\nodata & 
\nodata & 
\nodata & 
 & 
\nodata & 
\nodata & 
\nodata & 
 & 
\nodata & 
\nodata & 
\nodata & 
 & 
\nodata & 
\nodata & 
\nodata & 
 & 
$-5.48^{+1.00}_{-0.40}$ & 
$2.7$ & 
$1$ & 
 & 
\nodata & 
\nodata & 
\nodata \\
\cutinhead{Quiescent}
$  9.3$ & 
$-3.27^{+0.20}_{-0.10}$ & 
$0.20$ & 
$10$ & 
 & 
\nodata & 
\nodata & 
\nodata & 
 & 
\nodata & 
\nodata & 
\nodata & 
 & 
\nodata & 
\nodata & 
\nodata & 
 & 
\nodata & 
\nodata & 
\nodata & 
 & 
\nodata & 
\nodata & 
\nodata \\
$  9.4$ & 
$-2.95^{+0.20}_{-0.10}$ & 
$0.052$ & 
$19$ & 
 & 
\nodata & 
\nodata & 
\nodata & 
 & 
\nodata & 
\nodata & 
\nodata & 
 & 
\nodata & 
\nodata & 
\nodata & 
 & 
\nodata & 
\nodata & 
\nodata & 
 & 
\nodata & 
\nodata & 
\nodata \\
$  9.5$ & 
$-2.94^{+0.10}_{-0.10}$ & 
$0.047$ & 
$22$ & 
 & 
\nodata & 
\nodata & 
\nodata & 
 & 
\nodata & 
\nodata & 
\nodata & 
 & 
\nodata & 
\nodata & 
\nodata & 
 & 
\nodata & 
\nodata & 
\nodata & 
 & 
\nodata & 
\nodata & 
\nodata \\
$  9.6$ & 
$-2.79^{+0.10}_{-0.098}$ & 
$0.079$ & 
$32$ & 
 & 
$-3.043^{+0.100}_{-0.083}$ & 
$0.047$ & 
$29$ & 
 & 
\nodata & 
\nodata & 
\nodata & 
 & 
\nodata & 
\nodata & 
\nodata & 
 & 
\nodata & 
\nodata & 
\nodata & 
 & 
\nodata & 
\nodata & 
\nodata \\
$  9.7$ & 
$-2.931^{+0.088}_{-0.074}$ & 
$0.076$ & 
$36$ & 
 & 
$-3.021^{+0.095}_{-0.079}$ & 
$0.10$ & 
$34$ & 
 & 
\nodata & 
\nodata & 
\nodata & 
 & 
\nodata & 
\nodata & 
\nodata & 
 & 
\nodata & 
\nodata & 
\nodata & 
 & 
\nodata & 
\nodata & 
\nodata \\
$  9.8$ & 
$-2.818^{+0.075}_{-0.064}$ & 
$0.096$ & 
$48$ & 
 & 
$-2.88^{+0.20}_{-0.10}$ & 
$0.20$ & 
$40$ & 
 & 
\nodata & 
\nodata & 
\nodata & 
 & 
\nodata & 
\nodata & 
\nodata & 
 & 
\nodata & 
\nodata & 
\nodata & 
 & 
\nodata & 
\nodata & 
\nodata \\
$  9.9$ & 
$-2.910^{+0.079}_{-0.067}$ & 
$0.020$ & 
$43$ & 
 & 
$-2.880^{+0.070}_{-0.061}$ & 
$0.10$ & 
$57$ & 
 & 
$-3.02^{+0.20}_{-0.10}$ & 
$0.20$ & 
$33$ & 
 & 
\nodata & 
\nodata & 
\nodata & 
 & 
\nodata & 
\nodata & 
\nodata & 
 & 
\nodata & 
\nodata & 
\nodata \\
$ 10.0$ & 
$-2.754^{+0.072}_{-0.062}$ & 
$0.10$ & 
$56$ & 
 & 
$-2.753^{+0.052}_{-0.047}$ & 
$0.041$ & 
$91$ & 
 & 
$-2.991^{+0.076}_{-0.065}$ & 
$0.20$ & 
$52$ & 
 & 
\nodata & 
\nodata & 
\nodata & 
 & 
\nodata & 
\nodata & 
\nodata & 
 & 
\nodata & 
\nodata & 
\nodata \\
$ 10.1$ & 
$-2.773^{+0.066}_{-0.058}$ & 
$0.040$ & 
$60$ & 
 & 
$-2.791^{+0.053}_{-0.047}$ & 
$0.064$ & 
$88$ & 
 & 
$-2.908^{+0.061}_{-0.054}$ & 
$0.10$ & 
$74$ & 
 & 
\nodata & 
\nodata & 
\nodata & 
 & 
\nodata & 
\nodata & 
\nodata & 
 & 
\nodata & 
\nodata & 
\nodata \\
$ 10.2$ & 
$-2.655^{+0.057}_{-0.051}$ & 
$0.077$ & 
$76$ & 
 & 
$-2.700^{+0.046}_{-0.041}$ & 
$0.059$ & 
$114$ & 
 & 
$-2.854^{+0.056}_{-0.050}$ & 
$0.10$ & 
$87$ & 
 & 
$-2.988^{+0.052}_{-0.047}$ & 
$0.10$ & 
$107$ & 
 & 
\nodata & 
\nodata & 
\nodata & 
 & 
\nodata & 
\nodata & 
\nodata \\
$ 10.3$ & 
$-2.694^{+0.062}_{-0.054}$ & 
$0.082$ & 
$69$ & 
 & 
$-2.672^{+0.051}_{-0.046}$ & 
$0.037$ & 
$118$ & 
 & 
$-2.889^{+0.052}_{-0.047}$ & 
$0.070$ & 
$92$ & 
 & 
$-2.814^{+0.061}_{-0.054}$ & 
$0.049$ & 
$166$ & 
 & 
\nodata & 
\nodata & 
\nodata & 
 & 
\nodata & 
\nodata & 
\nodata \\
$ 10.4$ & 
$-2.595^{+0.054}_{-0.048}$ & 
$0.067$ & 
$84$ & 
 & 
$-2.682^{+0.043}_{-0.039}$ & 
$0.027$ & 
$128$ & 
 & 
$-2.730^{+0.083}_{-0.070}$ & 
$0.044$ & 
$126$ & 
 & 
$-2.795^{+0.035}_{-0.032}$ & 
$0.042$ & 
$208$ & 
 & 
\nodata & 
\nodata & 
\nodata & 
 & 
\nodata & 
\nodata & 
\nodata \\
$ 10.5$ & 
$-2.760^{+0.064}_{-0.057}$ & 
$0.067$ & 
$62$ & 
 & 
$-2.631^{+0.041}_{-0.038}$ & 
$0.068$ & 
$137$ & 
 & 
$-2.726^{+0.056}_{-0.050}$ & 
$0.090$ & 
$142$ & 
 & 
$-2.746^{+0.031}_{-0.029}$ & 
$0.050$ & 
$255$ & 
 & 
$-2.672^{+0.054}_{-0.049}$ & 
$0.049$ & 
$231$ & 
 & 
\nodata & 
\nodata & 
\nodata \\
$ 10.6$ & 
$-2.702^{+0.060}_{-0.053}$ & 
$0.059$ & 
$72$ & 
 & 
$-2.628^{+0.041}_{-0.038}$ & 
$0.10$ & 
$139$ & 
 & 
$-2.765^{+0.041}_{-0.038}$ & 
$0.10$ & 
$139$ & 
 & 
$-2.783^{+0.031}_{-0.029}$ & 
$0.032$ & 
$257$ & 
 & 
$-2.745^{+0.040}_{-0.037}$ & 
$0.036$ & 
$272$ & 
 & 
\nodata & 
\nodata & 
\nodata \\
$ 10.7$ & 
$-2.681^{+0.060}_{-0.053}$ & 
$0.077$ & 
$71$ & 
 & 
$-2.682^{+0.044}_{-0.040}$ & 
$0.077$ & 
$122$ & 
 & 
$-2.713^{+0.063}_{-0.056}$ & 
$0.047$ & 
$136$ & 
 & 
$-2.784^{+0.029}_{-0.027}$ & 
$0.031$ & 
$277$ & 
 & 
$-2.781^{+0.030}_{-0.029}$ & 
$0.062$ & 
$283$ & 
 & 
\nodata & 
\nodata & 
\nodata \\
$ 10.8$ & 
$-2.799^{+0.069}_{-0.060}$ & 
$0.096$ & 
$55$ & 
 & 
$-2.698^{+0.046}_{-0.041}$ & 
$0.072$ & 
$116$ & 
 & 
$-2.830^{+0.045}_{-0.041}$ & 
$0.10$ & 
$119$ & 
 & 
$-2.792^{+0.030}_{-0.028}$ & 
$0.040$ & 
$273$ & 
 & 
$-2.745^{+0.028}_{-0.026}$ & 
$0.045$ & 
$334$ & 
 & 
$-2.861^{+0.040}_{-0.037}$ & 
$0.061$ & 
$241$ \\
$ 10.9$ & 
$-2.790^{+0.074}_{-0.064}$ & 
$0.076$ & 
$50$ & 
 & 
$-2.856^{+0.065}_{-0.057}$ & 
$0.10$ & 
$76$ & 
 & 
$-2.915^{+0.049}_{-0.044}$ & 
$0.031$ & 
$103$ & 
 & 
$-2.837^{+0.035}_{-0.032}$ & 
$0.063$ & 
$239$ & 
 & 
$-2.788^{+0.028}_{-0.026}$ & 
$0.032$ & 
$322$ & 
 & 
$-2.913^{+0.031}_{-0.029}$ & 
$0.041$ & 
$275$ \\
$ 11.0$ & 
$-2.913^{+0.098}_{-0.081}$ & 
$0.082$ & 
$32$ & 
 & 
$-2.927^{+0.060}_{-0.053}$ & 
$0.079$ & 
$71$ & 
 & 
$-2.967^{+0.053}_{-0.047}$ & 
$0.052$ & 
$90$ & 
 & 
$-2.942^{+0.035}_{-0.032}$ & 
$0.027$ & 
$199$ & 
 & 
$-2.819^{+0.049}_{-0.045}$ & 
$0.056$ & 
$280$ & 
 & 
$-2.914^{+0.037}_{-0.034}$ & 
$0.042$ & 
$296$ \\
$ 11.1$ & 
$-2.97^{+0.10}_{-0.10}$ & 
$0.30$ & 
$20$ & 
 & 
$-2.942^{+0.062}_{-0.055}$ & 
$0.049$ & 
$66$ & 
 & 
$-3.220^{+0.072}_{-0.062}$ & 
$0.10$ & 
$53$ & 
 & 
$-3.114^{+0.041}_{-0.038}$ & 
$0.058$ & 
$139$ & 
 & 
$-3.008^{+0.037}_{-0.034}$ & 
$0.022$ & 
$213$ & 
 & 
$-3.040^{+0.038}_{-0.035}$ & 
$0.034$ & 
$250$ \\
$ 11.2$ & 
$-3.17^{+0.40}_{-0.20}$ & 
$0.073$ & 
$8$ & 
 & 
$-3.248^{+0.092}_{-0.077}$ & 
$0.081$ & 
$33$ & 
 & 
$-3.317^{+0.086}_{-0.073}$ & 
$0.20$ & 
$40$ & 
 & 
$-3.325^{+0.057}_{-0.050}$ & 
$0.063$ & 
$84$ & 
 & 
$-3.162^{+0.063}_{-0.055}$ & 
$0.091$ & 
$128$ & 
 & 
$-3.283^{+0.042}_{-0.038}$ & 
$0.034$ & 
$160$ \\
$ 11.3$ & 
$-3.58^{+0.30}_{-0.20}$ & 
$0.065$ & 
$5$ & 
 & 
$-3.44^{+0.10}_{-0.092}$ & 
$0.20$ & 
$23$ & 
 & 
$-3.71^{+0.10}_{-0.10}$ & 
$0.10$ & 
$17$ & 
 & 
$-3.431^{+0.070}_{-0.061}$ & 
$0.081$ & 
$62$ & 
 & 
$-3.430^{+0.060}_{-0.053}$ & 
$0.076$ & 
$82$ & 
 & 
$-3.566^{+0.050}_{-0.045}$ & 
$0.085$ & 
$97$ \\
$ 11.4$ & 
\nodata & 
\nodata & 
\nodata & 
 & 
$-3.72^{+0.20}_{-0.10}$ & 
$0.10$ & 
$9$ & 
 & 
$-3.81^{+0.20}_{-0.10}$ & 
$0.10$ & 
$13$ & 
 & 
$-3.85^{+0.10}_{-0.088}$ & 
$0.065$ & 
$27$ & 
 & 
$-3.606^{+0.068}_{-0.059}$ & 
$0.10$ & 
$58$ & 
 & 
$-3.63^{+0.10}_{-0.092}$ & 
$0.087$ & 
$63$ \\
$ 11.5$ & 
\nodata & 
\nodata & 
\nodata & 
 & 
$-4.71^{+1.00}_{-0.40}$ & 
$1.2$ & 
$1$ & 
 & 
$-4.25^{+0.30}_{-0.20}$ & 
$0.30$ & 
$5$ & 
 & 
$-3.86^{+0.10}_{-0.10}$ & 
$0.084$ & 
$18$ & 
 & 
$-4.09^{+0.20}_{-0.10}$ & 
$0.10$ & 
$15$ & 
 & 
$-3.95^{+0.20}_{-0.10}$ & 
$0.10$ & 
$31$ \\
$ 11.6$ & 
\nodata & 
\nodata & 
\nodata & 
 & 
\nodata & 
\nodata & 
\nodata & 
 & 
$-4.46^{+0.40}_{-0.20}$ & 
$0.045$ & 
$3$ & 
 & 
$-4.59^{+0.30}_{-0.20}$ & 
$0.50$ & 
$6$ & 
 & 
$-4.29^{+0.20}_{-0.20}$ & 
$0.10$ & 
$9$ & 
 & 
$-4.41^{+0.10}_{-0.10}$ & 
$0.089$ & 
$16$ \\
$ 11.7$ & 
\nodata & 
\nodata & 
\nodata & 
 & 
\nodata & 
\nodata & 
\nodata & 
 & 
$-5.09^{+1.00}_{-0.40}$ & 
$0.20$ & 
$1$ & 
 & 
$-4.33^{+0.30}_{-0.20}$ & 
$0.10$ & 
$5$ & 
 & 
$-5.12^{+0.60}_{-0.30}$ & 
$0.80$ & 
$2$ & 
 & 
$-4.68^{+0.20}_{-0.20}$ & 
$0.099$ & 
$8$ \\
$ 11.8$ & 
\nodata & 
\nodata & 
\nodata & 
 & 
\nodata & 
\nodata & 
\nodata & 
 & 
$-5.05^{+1.00}_{-0.40}$ & 
$0.20$ & 
$1$ & 
 & 
\nodata & 
\nodata & 
\nodata & 
 & 
\nodata & 
\nodata & 
\nodata & 
 & 
$-5.33^{+0.60}_{-0.30}$ & 
$0.40$ & 
$2$ \\
$ 11.9$ & 
\nodata & 
\nodata & 
\nodata & 
 & 
\nodata & 
\nodata & 
\nodata & 
 & 
\nodata & 
\nodata & 
\nodata & 
 & 
\nodata & 
\nodata & 
\nodata & 
 & 
$-4.48^{+1.00}_{-0.40}$ & 
$0.20$ & 
$1$ & 
 & 
\nodata & 
\nodata & 
\nodata \\
$ 12.0$ & 
\nodata & 
\nodata & 
\nodata & 
 & 
\nodata & 
\nodata & 
\nodata & 
 & 
\nodata & 
\nodata & 
\nodata & 
 & 
\nodata & 
\nodata & 
\nodata & 
 & 
\nodata & 
\nodata & 
\nodata & 
 & 
$-5.71^{+1.00}_{-0.40}$ & 
$1.1$ & 
$1$
\enddata
\tablecomments{PRIMUS stellar mass function at $z=0.2-1$ for all, star-forming, and quiescent galaxies.  The units of $\Phi$ are $10^{-4}\,{\rm Mpc}^{-3}\, {\rm dex}^{-1}$, $\sigma_{\rm cv}$ is the estimated $1\sigma$ uncertainty in $\Phi$ due to sample variance, and $N$ is the number of galaxies in each stellar mass bin.}
\end{deluxetable}

\clearpage
\setlength{\tabcolsep}{5pt}
\begin{deluxetable*}{cccccccccccc}
\tablecaption{Cumulative Number and Stellar Mass Density of All Galaxies\label{table:numden_all}}
\tablewidth{0pt}
\tabletypesize{\small}
\tablehead{
\colhead{$\langle z\rangle$} & 
\colhead{$\log\,(n)$\tablenotemark{a}} & 
\colhead{$\log\,(\rho)$\tablenotemark{b}} & 
\colhead{} & 
\colhead{$\log\,(n)$} & 
\colhead{$\log\,(\rho)$} & 
\colhead{} & 
\colhead{$\log\,(n)$} & 
\colhead{$\log\,(\rho)$} & 
\colhead{} & 
\colhead{$\log\,(n)$} & 
\colhead{$\log\,(\rho)$} 
}
\startdata
\colhead{} & 
\multicolumn{2}{c}{$\log\,(\mass/\msun)>9.5$} & 
\colhead{} & 
\multicolumn{2}{c}{$\log\,(\mass/\msun)>10$} & 
\colhead{} & 
\multicolumn{2}{c}{$\log\,(\mass/\msun)>10.5$} & 
\colhead{} & 
\multicolumn{2}{c}{$\log\,(\mass/\msun)>11$} \\
\cline{2-3}\cline{5-6}\cline{8-9}\cline{11-12}
$0.100$ & 
$-2.09\pm0.05$ & $ 8.35\pm0.05$ & & 
$-2.32\pm0.05$ & $ 8.31\pm0.05$ & & 
$-2.68\pm0.05$ & $ 8.18\pm0.05$ & & 
$-3.41\pm0.05$ & $ 7.78\pm0.04$ \\
$0.250$ & 
$-2.12\pm0.05$ & $ 8.32\pm0.09$ & & 
$-2.36\pm0.06$ & $ 8.28\pm0.10$ & & 
$-2.70\pm0.09$ & $ 8.16\pm0.13$ & & 
$-3.41\pm0.29$ & $ 7.78\pm0.29$ \\
$0.350$ & 
$-2.11\pm0.06$ & $ 8.35\pm0.06$ & & 
$-2.30\pm0.05$ & $ 8.32\pm0.06$ & & 
$-2.65\pm0.07$ & $ 8.20\pm0.07$ & & 
$-3.39\pm0.10$ & $ 7.77\pm0.11$ \\
$0.450$ & 
$-2.16\pm0.09$ & $ 8.31\pm0.08$ & & 
$-2.35\pm0.07$ & $ 8.27\pm0.08$ & & 
$-2.69\pm0.08$ & $ 8.16\pm0.09$ & & 
$-3.47\pm0.14$ & $ 7.70\pm0.16$ \\
$0.575$ & 
$>-2.22$ & $> 8.30$ & & 
$-2.39\pm0.04$ & $ 8.28\pm0.05$ & & 
$-2.69\pm0.04$ & $ 8.18\pm0.05$ & & 
$-3.42\pm0.08$ & $ 7.78\pm0.10$ \\
$0.725$ & 
\nodata & \nodata & & 
$>-2.37$ & $> 8.35$ & & 
$-2.61\pm0.05$ & $ 8.27\pm0.05$ & & 
$-3.28\pm0.07$ & $ 7.90\pm0.08$ \\
$0.900$ & 
\nodata & \nodata & & 
\nodata & \nodata & & 
$>-2.79$ & $> 8.15$ & & 
$-3.35\pm0.07$ & $ 7.85\pm0.07$
\enddata
\tablenotetext{a}{Number density in $h_{70}^{3}$~Mpc$^{-3}$.}
\tablenotetext{b}{Stellar mass density in $h_{70}$~\msun~Mpc$^{-3}$.}
\end{deluxetable*}

\begin{deluxetable*}{cccccccc}
\tablecaption{Number and Stellar Mass Density of All, Quiescent, and Star-Forming Galaxies\label{table:numden_rhoden_bymass}}
\tablewidth{0pt}
\tablehead{
\colhead{} & 
\multicolumn{3}{c}{$\log\,(n)$} & 
\colhead{} & 
\multicolumn{3}{c}{$\log\,(\rho)$} \\
\colhead{$\langle z\rangle$} & 
\multicolumn{3}{c}{$(h_{70}^{3}$~Mpc$^{-3})$} & 
\colhead{} & 
\multicolumn{3}{c}{$(h_{70}$~\msun~Mpc$^{-3})$} 
}
\startdata
\colhead{} & \multicolumn{7}{c}{$9.5<\log\,(\mass/\msun)<10$} \\
\cline{1-8}
\colhead{} &  \colhead{All} &  \colhead{Q\tablenotemark{a}} & 
\colhead{SF\tablenotemark{a}} &  \colhead{} &  \colhead{All} &  \colhead{Q} & 
\colhead{SF} \\
\cline{2-4} \cline{6-8}
$0.100$ & 
$-2.48\pm0.05$ & 
$-2.94\pm0.07$ & 
$-2.66\pm0.04$ & & 
$ 7.28\pm0.05$ & 
$ 6.83\pm0.07$ & 
$ 7.10\pm0.04$ \\
$0.250$ & 
$-2.49\pm0.03$ & 
$-3.15\pm0.04$ & 
$-2.60\pm0.03$ & & 
$ 7.27\pm0.03$ & 
$ 6.63\pm0.05$ & 
$ 7.16\pm0.03$ \\
$0.350$ & 
$-2.54\pm0.09$ & 
$-3.21\pm0.16$ & 
$-2.65\pm0.07$ & & 
$ 7.24\pm0.08$ & 
$ 6.58\pm0.13$ & 
$ 7.13\pm0.06$ \\
$0.450$ & 
$-2.61\pm0.15$ & 
$>-3.49$ & 
$-2.67\pm0.14$ & & 
$ 7.16\pm0.13$ & 
$> 6.33$ & 
$ 7.10\pm0.12$ \\
$0.575$ & 
$>-2.71$ & 
\nodata & 
$>-2.74$ & & 
$> 7.06$ & 
\nodata & 
$> 7.02$ \\
\cline{1-8}
\colhead{} & \multicolumn{7}{c}{$10<\log\,(\mass/\msun)<10.5$} \\
\cline{1-8}
\colhead{} &  \colhead{All} &  \colhead{Q} &  \colhead{SF} &  \colhead{} & 
\colhead{All} &  \colhead{Q} &  \colhead{SF} \\
\cline{2-4} \cline{6-8}
$0.100$ & 
$-2.56\pm0.05$ & 
$-2.89\pm0.05$ & 
$-2.84\pm0.04$ & & 
$ 7.71\pm0.05$ & 
$ 7.39\pm0.05$ & 
$ 7.42\pm0.04$ \\
$0.250$ & 
$-2.62\pm0.03$ & 
$-2.99\pm0.05$ & 
$-2.85\pm0.04$ & & 
$ 7.65\pm0.03$ & 
$ 7.29\pm0.04$ & 
$ 7.39\pm0.04$ \\
$0.350$ & 
$-2.56\pm0.04$ & 
$-3.01\pm0.04$ & 
$-2.75\pm0.04$ & & 
$ 7.71\pm0.03$ & 
$ 7.28\pm0.04$ & 
$ 7.51\pm0.04$ \\
$0.450$ & 
$-2.62\pm0.04$ & 
$-3.14\pm0.07$ & 
$-2.78\pm0.03$ & & 
$ 7.65\pm0.04$ & 
$ 7.16\pm0.05$ & 
$ 7.47\pm0.04$ \\
$0.575$ & 
$-2.68\pm0.05$ & 
$>-3.20$ & 
$-2.84\pm0.04$ & & 
$ 7.60\pm0.05$ & 
$> 7.12$ & 
$ 7.42\pm0.04$ \\
$0.725$ & 
$>-2.74$ & 
\nodata & 
$>-2.88$ & & 
$> 7.56$ & 
\nodata & 
$> 7.40$ \\
\cline{1-8}
\colhead{} & \multicolumn{7}{c}{$10.5<\log\,(\mass/\msun)<11$} \\
\cline{1-8}
\colhead{} &  \colhead{All} &  \colhead{Q} &  \colhead{SF} &  \colhead{} & 
\colhead{All} &  \colhead{Q} &  \colhead{SF} \\
\cline{2-4} \cline{6-8}
$0.100$ & 
$-2.77\pm0.05$ & 
$-2.99\pm0.05$ & 
$-3.17\pm0.04$ & & 
$ 7.97\pm0.05$ & 
$ 7.76\pm0.05$ & 
$ 7.55\pm0.04$ \\
$0.250$ & 
$-2.80\pm0.04$ & 
$-3.06\pm0.04$ & 
$-3.15\pm0.07$ & & 
$ 7.93\pm0.04$ & 
$ 7.70\pm0.04$ & 
$ 7.55\pm0.09$ \\
$0.350$ & 
$-2.74\pm0.06$ & 
$-3.02\pm0.09$ & 
$-3.07\pm0.04$ & & 
$ 7.99\pm0.06$ & 
$ 7.73\pm0.08$ & 
$ 7.66\pm0.04$ \\
$0.450$ & 
$-2.76\pm0.06$ & 
$-3.11\pm0.06$ & 
$-3.03\pm0.06$ & & 
$ 7.97\pm0.06$ & 
$ 7.64\pm0.06$ & 
$ 7.69\pm0.06$ \\
$0.575$ & 
$-2.78\pm0.03$ & 
$-3.11\pm0.04$ & 
$-3.06\pm0.03$ & & 
$ 7.96\pm0.03$ & 
$ 7.65\pm0.04$ & 
$ 7.66\pm0.03$ \\
$0.725$ & 
$-2.72\pm0.05$ & 
$-3.06\pm0.04$ & 
$-2.99\pm0.07$ & & 
$ 8.03\pm0.05$ & 
$ 7.70\pm0.04$ & 
$ 7.75\pm0.07$ \\
$0.900$ & 
$>-2.93$ & 
$>-3.23$ & 
$-3.23\pm0.07$ & & 
$> 7.84$ & 
$> 7.56$ & 
$ 7.52\pm0.08$ \\
\cline{1-8}
\colhead{} & \multicolumn{7}{c}{$11<\log\,(\mass/\msun)<11.5$} \\
\cline{1-8}
\colhead{} &  \colhead{All} &  \colhead{Q} &  \colhead{SF} &  \colhead{} & 
\colhead{All} &  \colhead{Q} &  \colhead{SF} \\
\cline{2-4} \cline{6-8}
$0.100$ & 
$-3.42\pm0.04$ & 
$-3.54\pm0.04$ & 
$-4.04\pm0.04$ & & 
$ 7.74\pm0.04$ & 
$ 7.63\pm0.04$ & 
$ 7.09\pm0.04$ \\
$0.250$ & 
$-3.42\pm0.20$ & 
$-3.55\pm0.24$ & 
$-4.10\pm0.17$ & & 
$ 7.75\pm0.21$ & 
$ 7.60\pm0.27$ & 
$ 7.03\pm0.28$ \\
$0.350$ & 
$-3.40\pm0.04$ & 
$-3.55\pm0.06$ & 
$-3.94\pm0.03$ & & 
$ 7.76\pm0.04$ & 
$ 7.62\pm0.07$ & 
$ 7.20\pm0.05$ \\
$0.450$ & 
$-3.48\pm0.10$ & 
$-3.71\pm0.11$ & 
$-3.88\pm0.09$ & & 
$ 7.68\pm0.12$ & 
$ 7.45\pm0.12$ & 
$ 7.27\pm0.10$ \\
$0.575$ & 
$-3.43\pm0.04$ & 
$-3.62\pm0.04$ & 
$-3.90\pm0.07$ & & 
$ 7.74\pm0.05$ & 
$ 7.56\pm0.04$ & 
$ 7.25\pm0.08$ \\
$0.725$ & 
$-3.29\pm0.05$ & 
$-3.51\pm0.05$ & 
$-3.69\pm0.05$ & & 
$ 7.88\pm0.05$ & 
$ 7.66\pm0.06$ & 
$ 7.47\pm0.05$ \\
$0.900$ & 
$-3.37\pm0.05$ & 
$-3.58\pm0.03$ & 
$-3.77\pm0.08$ & & 
$ 7.81\pm0.05$ & 
$ 7.59\pm0.04$ & 
$ 7.40\pm0.07$ 
\enddata
\tablenotetext{a}{Q = quiescent; SF = star-forming.}
\end{deluxetable*}

\clearpage
\begin{deluxetable*}{lccccc}
\tablecaption{Power-Law Fits to the Number and Stellar Mass Density Evolution\tablenotemark{a}\label{table:numden_bymass_coeff}}
\tablewidth{0pt}
\tablehead{
\colhead{Sample\tablenotemark{a}} & 
\colhead{$\log\,(n_{0})$\tablenotemark{b}} & 
\colhead{$\gamma$\tablenotemark{b}} & 
\colhead{} & 
\colhead{$\log\,(\rho_{0})$\tablenotemark{c}} & 
\colhead{$\beta$\tablenotemark{c}} 
}
\startdata
\colhead{} & \multicolumn{5}{c}{$9.5<\log\,(\mass/\msun)<10$} \\
\cline{2-6}
All & $-2.441\pm  0.03$ & $ -0.729\pm  0.328$ &  & $ 7.319\pm  0.03$ & $ -0.663\pm  0.294$  \\
Q & $-2.802\pm  0.03$ & $ -3.444\pm  0.441$ &  & $ 6.957\pm  0.04$ & $ -3.150\pm  0.514$  \\
SF & $-2.666\pm  0.04$ & $  0.317\pm  0.445$ &  & $ 7.091\pm  0.03$ & $  0.351\pm  0.400$  \\
\cline{1-6}
\colhead{} & \multicolumn{5}{c}{$10<\log\,(\mass/\msun)<10.5$} \\
\cline{2-6}
All & $-2.527\pm  0.04$ & $ -0.644\pm  0.318$ &  & $ 7.735\pm  0.04$ & $ -0.576\pm  0.313$  \\
Q & $-2.821\pm  0.03$ & $ -1.668\pm  0.363$ &  & $ 7.456\pm  0.03$ & $ -1.551\pm  0.326$  \\
SF & $-2.830\pm  0.04$ & $  0.090\pm  0.326$ &  & $ 7.419\pm  0.04$ & $  0.168\pm  0.339$  \\
\cline{1-6}
\colhead{} & \multicolumn{5}{c}{$10.5<\log\,(\mass/\msun)<11$} \\
\cline{2-6}
All & $-2.766\pm  0.02$ & $ -0.062\pm  0.145$ &  & $ 7.968\pm  0.02$ & $ -0.037\pm  0.130$  \\
Q & $-2.956\pm  0.02$ & $ -0.775\pm  0.144$ &  & $ 7.786\pm  0.02$ & $ -0.699\pm  0.132$  \\
SF & $-3.150\pm  0.06$ & $  0.245\pm  0.390$ &  & $ 7.559\pm  0.05$ & $  0.344\pm  0.347$  \\
\cline{1-6}
\colhead{} & \multicolumn{5}{c}{$11<\log\,(\mass/\msun)<11.5$} \\
\cline{2-6}
All & $-3.433\pm  0.02$ & $  0.148\pm  0.124$ &  & $ 7.730\pm  0.02$ & $  0.189\pm  0.123$  \\
Q & $-3.534\pm  0.02$ & $ -0.285\pm  0.168$ &  & $ 7.641\pm  0.02$ & $ -0.297\pm  0.174$  \\
SF & $-4.090\pm  0.01$ & $  1.114\pm  0.100$ &  & $ 7.036\pm  0.01$ & $  1.274\pm  0.109$ 
\enddata
\tablenotetext{a}{Q = quiescent; SF = star-forming.}
\tablenotetext{b}{Model given by $n(z) = n_{0}(1+z)^{\gamma}$ with $n$ in $h_{70}^{3}$~Mpc$^{-3}$.}
\tablenotetext{c}{Model given by $\rho(z) = \rho_{0}(1+z)^{\beta}$ with $\rho$ in $h_{70}$~\msun~Mpc$^{-3}$.}
\end{deluxetable*}

\end{document}